\documentclass[12pt, centerh1]{article} 

\usepackage{graphicx,colonequals}
\usepackage{natbib}
\usepackage{amsmath, amsthm, amssymb, amsfonts}
\usepackage{supertabular}
\usepackage{longtable, multirow}
\usepackage{rotating}
\usepackage{subfigure}
\usepackage{epsfig}
\usepackage{color}

\newcommand{\vecmu}{\mbox{\boldmath$\mu$}}

\newcommand{\noisev}{\mathbf\Psi}

\newcommand{\vecw}{\mathbf{w}}
\newcommand{\veca}{\mathbf{a}}
\newcommand{\vecx}{\mathbf{x}}

\newcommand{\vecX}{\mathbf{X}}

\newcommand{\vecV}{\mathbf{V}}
\newcommand{\vecv}{\mathbf{v}}
\newcommand{\vecz}{\mathbf{z}}

\newcommand{\gam}{\mathbf\Gamma}

\newcommand{\matsig}{\mathbf\Sigma}

\newcommand{\vecGam}{\mathbf\Gamma}
\newcommand{\veclambda}{\mbox{\boldmath$\lambda$}}

\newcommand{\vecpi}{\mbox{\boldmath$\pi$}}
\newcommand{\vectheta}{\mbox{\boldmath$\theta$}}

\newcommand{\vecr}{\mathbf{r}}
\newcommand{\Beta}{\mbox{\boldmath$\beta$}}
\newcommand{\zig}{\hat{z}_{ig}}
\newcommand{\varthet}{\mbox{\boldmath$\vartheta$}}
\newcommand{\vecR}{\mathbf{R}}

\newcommand{\vecomega}{\mbox{\boldmath$\omega$}}

\newcommand{\diag}{\,\mbox{diag}}
\newcommand{\tr}{\,\mbox{tr}}

\newcommand{\vecomegaP}{\mbox{\boldmath$\omega$}}
\newcommand{\vecA}{\mathbf{A}}
\newcommand{\vecY}{\mathbf{Y}}
\newcommand{\vecS}{\mathbf{S}}

\newcommand{\matw}{\mathbf{W}}

\newcommand{\ident}{\mathbf{I}}
\newcommand{\del}{\mathbf\Delta}

\newcommand{\vecalpha}{\mbox{\boldmath$\alpha$}}

\newcommand{\vecnu}{\mbox{\boldmath$\nu$}}

\newcommand{\vecDelta}{\mathbf\Delta}
\newcommand{\vecPhi}{\mathbf\Phi}
\newcommand{\prev}{\mbox{\tiny prev}}

\newtheorem{definition}{Definition}
\newtheorem{lemma}{Lemma}
\newtheorem{theorem}{Theorem}

\title{A Mixture of Coalesced Generalized\\ Hyperbolic Distributions}  
  \author{Cristina Tortora$^{*}$, Brian C.\ Franczak$^{**}$, Ryan P.\ Browne$^{\dagger}$, Paul D.\ McNicholas$^{\dagger\dagger}$}
  \date{\small $^{*}$Department of Mathematics \& Statistics, San Jos\'{e} State University, CA, USA.\\
 $^{**}$Department of Mathematics \& Statistics, MacEwan University, Edmonton, AB, Canada.\\
 $^{\dagger}$Department of Statistics and Actuarial Sciences, University of Waterloo, ON, Canada.\\
 $^{\dagger\dagger}$Department of Mathematics and Statistics, McMaster University, ON, Canada.}
 
\begin{document}

\maketitle 

\begin{abstract}
A mixture of multiple scaled generalized hyperbolic distributions (MMSGHDs) is introduced. Then, a coalesced generalized hyperbolic distribution (CGHD) is developed by joining a generalized hyperbolic 
distribution with a multiple scaled generalized hyperbolic distribution. After detailing the development of the MMSGHDs, 
which arises via implementation of a multi-dimensional weight function, the density of the mixture of CGHDs is developed. A parameter estimation scheme is developed using the ever-expanding class of MM algorithms and the Bayesian information criterion is used for model selection. The issue of cluster convexity is examined and a special case of the MMSGHDs is developed that is guaranteed to have convex clusters. These approaches are illustrated and compared using simulated and real data. The identifiability of the MMSGHDs and the mixture of CGHDs is discussed in an appendix.\\

\end{abstract}

\section{Introduction}

Finite mixture models have been linked with clustering since the idea of defining a cluster in terms of a component in finite mixture model was put forth more than 60 years ago \citep[see][Section~2.1]{mcnicholas16}. Nowadays, mixture model-based clustering is a popular approach to clustering. A random vector $\vecX$ arises from a finite mixture model if, for all $\vecx \subset \vecX$, its density can be written $$f(\vecx\mid\varthet)= \sum_{g=1}^G \pi_g f_g(\vecx\mid\vectheta_g),$$ where $\pi_g >0$ such that $\sum_{g=1}^G \pi_g = 1$ are the mixing proportions, $f_g(\vecx\mid\vectheta_g)$ is the $g$th component density, and $\varthet=(\vecpi,\vectheta_1,\ldots,\vectheta_G)$ denotes the vector of parameters with $\vecpi=(\pi_1,\ldots,\pi_G)$. The component densities $f_1(\vecx\mid\vectheta_1),\ldots,f_G(\vecx\mid\vectheta_G)$ are typically taken to be of the same type, most commonly multivariate Gaussian. In fact, until a few years after the turn of the century, almost all work on clustering and classification using mixture models had been based on Gaussian mixture models  \citep[e.g.,][]{banfield93,celeux95,ghahramani97,tipping99b,mclachlan00a,fraley02a}. 

Early work on non-Gaussian mixtures was on mixtures of multivariate $t$-distributions \cite[e.g.,][]{peel00}. A little beyond the turn of the century, work on $t$-mixtures burgeoned into a substantial subfield of mixture model-based classification \citep[e.g.,][]{mclachlan07,andrews11a,andrews11c,andrews12,baek11,steane12,lin14,pesevski18}. 
Around the same time, work on mixtures of skewed distributions took off, including work on skew-normal mixtures \citep[e.g.,][]{lin09}, skew-$t$ mixtures \citep[e.g.,][]{lin10,vrbik12,vrbik14,lee13b,lee13,murray14b}, Laplace mixtures \citep[e.g.,][]{franczak14}, variance-gamma mixtures \citep{smcnicholas17}, 
generalized hyperbolic mixtures \citep{browne15}, and other non-elliptically contoured distributions \cite[e.g.,][]{karlis09,murray17,tang18}. 
A thorough review of work on model-based clustering is given by \cite{mcnicholas16b}.

More recently, mixtures of multiple scaled distributions have been considered (Section~\ref{sec:multscaledists}). In the present manuscript, a multiple scaled generalized hyperbolic distribution is introduced (Section~\ref{sec:MSGHD}). Because the (multivariate) generalized hyperbolic distribution is not a special case of the multiple scaled generalized hyperbolic distribution, a coalesced generalized hyperbolic distribution is also developed (Section~\ref{sec:MCGHD}). The issue of cluster convexity, which has essentially been ignored in work to date on multiple scaled distributions is discussed and a special case of the multiple scaled generalized hyperbolic distribution is developed to guarantee that the components, i.e., the clusters, are convex (Section~\ref{sec:cMSGHD}).

\section{Multiple Scaled Distributions}\label{sec:multscaledists}
The distribution of a $p$-dimensional random variable $\vecX$ is said to be a normal variance-mean mixture if its density can be written in the form
\begin{equation}\label{eq:nvmm_dens}
f(\vecx\mid\vecmu,\matsig,\vecalpha,\vectheta)=
\int_{0}^{\infty}{\phi_p\left(\vecx\mid\vecmu+w\vecalpha,w\matsig\right)h\left(w\mid\vectheta\right)dw},\end{equation}
where $\phi_p\left(\vecx\mid\vecmu+w\vecalpha,{w}\matsig\right)$ is the density of a $p$-dimensional Gaussian distribution with mean $\vecmu+w\vecalpha$ and covariance matrix $w\matsig$, and $h\left(w\mid\vectheta\right)$ is the density of a univariate random variable $W>0$ that has the role of a weight function \cite[see][]{barndorff82,gneiting97}. This weight function can take on many forms, some of which lead to density representations for well-known non-Gaussian distributions, e.g., if $h\left(w\mid\vectheta\right)$ is the density of an inverse-gamma random variable with parameters $(\nu/2,\nu/2)$, then \eqref{eq:nvmm_dens} is a representation of the skew-$t$ distribution with $\nu$ degrees of freedom \citep[see][]{demarta05,murray14b}. 
Further details on, and examples of, normal variance-mean mixtures are given by \cite{barndorff78b}, \cite{kotz01}, and \cite{kotz04}, amongst others. 

Now, the density of $W>0$ from an inverse-gamma distribution with parameters $(\alpha,\beta)$ is given by
\begin{equation}\label{eqn:gaminess1}
h\left(w\mid\alpha,\beta\right)=w^{-\alpha-1}\frac{\beta^{\alpha}}{\Gamma\left(\alpha\right)}\exp\left\{-\frac{\beta}{ w}\right\},
\end{equation}%
where $\Gamma(\cdot)$ is the Gamma function. Setting $\alpha=\beta=\nu/2$ in \eqref{eqn:gaminess1} gives
\begin{equation}\label{eqn:gaminess}
h\left(w\mid\nu/2,\nu/2\right)=w^{-\nu/2-1}\frac{(\nu/2)^{\nu/2}}{\Gamma\left(\nu/2\right)}\exp\left\{-\frac{\nu}{2w}\right\},
\end{equation}%
Setting $\vecalpha = \mathbf{0}$ in \eqref{eq:nvmm_dens}, and using \eqref{eqn:gaminess} for $h\left(w\mid\vectheta\right)$, it follows that the density of the multivariate $t$-distribution with $\nu$ degrees of freedom can be written 
\begin{align}\label{eq:nvmm_t}
f_t(\vecx\mid\vecmu,\matsig,\nu)
&=\int_{0}^{\infty}{\phi_p\left(\vecx\mid\vecmu,w \matsig\right)h\left(w\mid\nu/2,\nu/2\right)dw}\nonumber\\
&=\frac{\Gamma \left( [{\nu+p}]/{2} \right) |\matsig|^{-{1}/{2}}}{(\pi \nu)^{{p}/{2}} \Gamma \left({\nu}/{2} \right) \left[1 + {\delta(\vecx, \vecmu~|~\matsig)}/{\nu} \right]^{{(\nu + p)}/{2}}}
\end{align}
where $\delta\left(\vecx,\vecmu\mid\matsig\right)$ is the squared Mahalanobis distance between $\vecx$ and $\vecmu$.
\cite{forbes14} show that a  multi-dimensional weight variable $$\del_\matw = \diag\left(w_1^{-1},\dots,w_p^{-1}\right)$$%
can be incorporated into \eqref{eq:nvmm_dens} via an eigen-decomposition of the symmetric positive-definite matrix~$\matsig$. Specifically, they set $\matsig=\gam\vecPhi\gam',$ where $\gam$ is a $p\times p$ matrix of eigenvectors and $\vecPhi$ is a $p\times p$ diagonal matrix containing the eigenvalues of~$\matsig$. It follows that the density of $\vecX$ becomes
\begin{equation}\begin{split}\label{eq:nvmm_mdens}
f&\left(\vecx\mid\vecmu,\gam , \vecPhi,\vecalpha,\vectheta\right)=\\
&\int_{0}^{\infty}\cdots\int_{0}^{\infty}\phi_p\left(\vecx\mid\vecmu+\del_\matw\vecalpha, \gam\del_\matw\vecPhi\gam' \right) h_\matw\left(w_1,\dots,w_p\mid\vectheta\right)dw_1\dots dw_p,
\end{split}\end{equation}
where $$h_{\matw}\left(w_1,\dots,w_p\mid\vectheta\right) = h\left(w_1\mid\vectheta_1\right)\times\dots\times h\left(w_p\mid\vectheta_p\right)$$ is a $p$-dimensional density such that the random variables $W_1,\ldots,W_p$ are independent, i.e., the weights are independent. 
The density given in~\eqref{eq:nvmm_mdens} adds flexibility to normal variance-mean mixtures because the parameters $\vectheta_1,\dots,\vectheta_p$ are free to vary in each dimension.  Using the density in~\eqref{eq:nvmm_mdens}, \cite{forbes14} derive the density of a multiple scaled multivariate-$t$ distribution, \cite{wraith15} derive a multiple scaled normal-inverse Gaussian distribution, and \cite{franczak15} develop a multiple scaled shifted asymmetric Laplace distribution.

Setting $\matsig=\gam\vecPhi\gam'$, it follows from~\eqref{eq:nvmm_mdens} that the density of a multiple scaled analogue of~\eqref{eq:nvmm_t} can be written
\begin{equation}\begin{split}\label{eq:nvmm_dens2}
&f_{t\text{\tiny{MS}}}(\vecx\mid\vecmu,\gam,\vecPhi,\vecnu)=
\int_{0}^{\infty}\cdots\int_{0}^{\infty}\phi_p\left(\vecx\mid\vecmu,\gam\vecPhi\del_\matw\gam'\right)h_{\matw}\left(w_1,\dots,w_p\mid\vecnu\right)dw_1\dots dw_p,
\end{split}\end{equation}
where $\del_\matw = \diag\left(w_1^{-1},\dots,w_p^{-1}\right)$ and the weight function $$h_{\matw}\left(w_1,\dots,w_p\mid\vecnu\right) = h\left(w_1\mid\nu_1/2,\nu_1/2\right)\times\dots\times h\left(w_p\mid\nu_p/2,\nu_p/2\right)$$ is a $p$-dimensional Gamma density, where $h\left(w_j\mid\nu_j/2,\nu_j/2\right)$ is given by \eqref{eqn:gaminess}. 
Note that the scaled Gaussian density in~\eqref{eq:nvmm_dens2} can be written
\begin{align}\label{eq:scale_gauss}
\phi_p\left(\vecx\mid\vecmu,\gam\vecPhi\del_\matw\gam'\right) 
&= \prod_{j=1}^p{\phi_1\left([\gam'\vecx]_j \mid [\gam'\vecmu]_j, \Phi_jw_j^{-1}\right)}
= \prod_{j=1}^p{\phi_1\left([\gam'(\vecx-\vecmu)]_j\mid0,\Phi_jw_j^{-1} \right) },
\end{align}
where $\phi_1\left([\gam'(\vecx-\vecmu)]_j\mid0,\Phi_jw_j^{-1} \right) $ is the density of a univariate Gaussian distribution with mean $0$ and variance $\Phi_jw_j^{-1}$, $[\gam'(\vecx-\vecmu)]_j$ is the $j$th element of $\gam'(\vecx-\vecmu)$, and $\Phi_j$ is the 
$j$th eigenvalue of $\vecPhi$, i.e., the $j$th diagonal element of the matrix $\vecPhi$. It follows that~\eqref{eq:nvmm_dens2} can be written
\begin{equation}\label{eq:nvmm_dens3}
\begin{split}
f_{t\text{\tiny{MS}}}(\vecx\mid\vecmu,&\gam,\vecPhi,\vecnu)=\prod_{j=1}^p\int_0^{\infty}{\phi_1\left([\gam'(\vecx-\vecmu)]_j\mid 0, \Phi_jw_j^{-1} \right)h\left(w_j\mid\nu_j/2,\nu_j/2\right)dw_j}.
\end{split}\end{equation}
%
Solving the integral in~\eqref{eq:nvmm_dens3} gives the density of a multiple scaled multivariate-$t$ distribution,
\begin{equation}\label{eq:mst_dens}
f_{t\text{\tiny{MS}}}(\vecx\mid\vecmu,\gam,\vecPhi,\vecnu)=\prod_{j=1}^p{\frac{\Gamma([\nu_j+1]/2)}{\Gamma(\nu_j/2)(\Phi_j\nu_j\pi)^{1/2} }\left[1+\frac{[\gam'(\vecx-\vecmu)]_j^2}{\Phi_j\nu_j}\right]^{-(\nu_j+1)/2}},
\end{equation}
where $\Phi_j$ is the $j$th eigenvalue of $\vecPhi$, $\gam$ is a matrix of eigenvectors, $\vecmu$ is a location parameter, and $[\gam'(\vecx-\vecmu)]_j^2/\Phi_j$ can be regarded as the squared Mahalanobis distance between $\vecx$ and $\vecmu$.

The main difference between the traditional multivariate-$t$ density given in~\eqref{eq:nvmm_t} and the multiple scaled multivariate-$t$ density given in~\eqref{eq:mst_dens} is that the degrees of freedom can now be parameterized separately in each dimension~$j$. Therefore, unlike the standard multivariate-$t$ distribution, the multiple scaled density in \eqref{eq:mst_dens} can account for different tail weight in each dimension  \citep{forbes14}.

\section{Mixture of Multiple Scaled Generalized Hyperbolic Distributions}\label{sec:MSGHD}

There are different ways to formulate the density of a generalized hyperbolic distribution \cite[GHD; see][for example]{mcneil05}. \cite{browne15} use a mixture of GHDs (MGHDs) for clustering and they use the following formulation for the density of a $p$-dimensional random vector $\vecX$ from a GHD:
\begin{equation} \label{dist ghy}
\begin{split}
f_{\text{GH}}(\vecx\mid\vectheta) =& 
\left[ \frac{ \omega + \delta\left(\vecx, \vecmu| \matsig\right) }{ \omega+ \vecalpha'\matsig^{-1}\vecalpha} \right]^{(\lambda-{p}/{2})/2}\frac{  K_{\lambda - {p}/{2}}\Big(\sqrt{\big[ \omega+ \vecalpha'\matsig^{-1}\vecalpha\big]\big[\omega +\delta\left(\vecx, \vecmu| \matsig\right)\big]}\Big)}{ \left(2\pi\right)^{{p}/{2}} \left| \matsig \right|^{{1}/{2}} K_{\lambda}\left( \omega\right)\exp\big\{-\left(\vecx-\vecmu\right)'\matsig^{-1}\vecalpha\big\}},
\end{split}\end{equation}
where $\vecmu\in\mathbb{R}^p$ is the location parameter, $\vecalpha\in\mathbb{R}^p$ is the skewness parameter, $\matsig\in\mathbb{R}^{p\times p}$ is the scale matrix, $\lambda\in\mathbb{R}$ is the index parameter,  $\omega\in\mathbb{R}^+$ is the concentration parameter, and $K_{\lambda}$ is the modified Bessel function of the third kind with index~$\lambda$.
Now, $\vecX\mid w \backsim \text{N}(\vecmu+w\vecalpha,w\matsig)$ and $W\mid\vecx\sim\text{GIG}(\omega+ \vecalpha'\matsig^{-1}\vecalpha,\omega+\delta\left(\vecx, \vecmu| \matsig\right),\lambda - {p}/{2}),$
where $W\sim\text{GIG}(a,b,\lambda)$ denotes that $W$ follows a generalized inverse Gaussian (GIG) distribution with density formulated as
\begin{equation}\label{gig}
q(w~|~a,b,\lambda) = \frac{(a/b)^{\lambda/2}w^{\lambda-1}}{2K_{\lambda}(\sqrt{ab})} \exp\left\{-\frac{aw+b/w}{2}\right\},
\end{equation}
for $w>0$, where $a,b\in\mathbb{R}^+$, and  $\lambda\in\mathbb{R}$.  The GIG distribution has some attractive properties including the tractability of the following expected values:
\begin{eqnarray}
\label{eqn:exp_vals}
&&\mathbb{E}\left[ W \right] =
\sqrt{\frac{b}{a}}\frac{K_{\lambda+1}\big(\sqrt{ab}\big)}{K_{\lambda}(\sqrt{ab})},
\qquad\qquad\mathbb{E}\left[{1}/{W}\right] = 
\sqrt{\frac{a}{b}}\frac{K_{\lambda+1}\big(\sqrt{ab}\big)}{K_{\lambda}\big(\sqrt{ab}\big)} -\frac{2\lambda}{b},\label{eqn:gigex2}\\
&&\mathbb{E}[\log W] =\log\left(\sqrt{\frac{b}{a}}\right)+\frac{1}{K_\lambda\big(\sqrt{ab}\big)}\frac{\partial}{\partial\lambda}K_\lambda\big(\sqrt{ab}\big).\label{eqn:gigex3}
\end{eqnarray}
Write $\vecX\sim\text{GHD}(\vecmu,\matsig,\vecalpha,\omega,\lambda)$ to denote that the $p$-dimensional random variable $\vecX$ has the density in \eqref{dist ghy}. 

Now, note that the formulation of the GHD in \eqref{dist ghy} can be written as a normal variance-mean mixture where the univariate density is GIG, i.e., 
\begin{equation}\label{eqn:hypstorep2}
\vecX = \vecmu + W\vecalpha+ \sqrt{W} \vecV,
\end{equation}
where $\vecV\sim\text{N}(\mathbf{0}, \matsig)$ and $W$ has density 
\begin{equation}\label{eq:gig_density}
h(w\mid\omega,1,\lambda)=\frac{w^{\lambda-1}}{2 K_\lambda(\omega)}\exp{\left\{-\frac{\omega}{2}\left(w + \frac{1}{w}\right)\right\}},
\end{equation}
for $w>0$, where $\omega$ and $\lambda$ are as previously defined. Note that \eqref{eq:gig_density} is just an alternative parameterization of the GIG distribution.
From \eqref{eqn:hypstorep2} and \eqref{eq:gig_density}, it follows that the generalized hyperbolic density can be written
\begin{equation}\label{eq:ghvmm_dens}
f(\vecx\mid\vecmu,\matsig,\vecalpha,\omega,\lambda)=
\int_{0}^{\infty}{\phi_p\left(\vecx\mid\vecmu+w\vecalpha,{w}\matsig\right)h(w\mid\omega,1,\lambda)dw}.\end{equation}

We can use \eqref{eq:nvmm_mdens}, \eqref{eq:ghvmm_dens} and an alternative parameterization to write the density of a multiple scaled generalized hyperbolic distribution (MSGHD) as
\begin{equation}\begin{split}\label{eq:GHNMVM2}
f_{\text{\tiny{MSGHD}}}(&\vecx\mid\vecmu,\gam,\vecPhi,\vecalpha,\vecomega,\veclambda)=\\&\int_{0}^{\infty}\dots\int_0^{\infty}\phi_p\left(\gam' \vecx - \vecmu-\vecDelta_\vecw \vecalpha \mid \mathbf{0}, \vecDelta_\vecw \vecPhi  \right) h_\vecw(w_1,\dots,w_p\mid\vecomegaP,\mathbf{1},\veclambda)dw_1\dots dw_p,
\end{split}\end{equation}
where $\vecomegaP=(\omega_1,\ldots,\omega_p)'$, $\veclambda=(\lambda_1,\ldots,\lambda_p)'$, $\mathbf{1}$ is a $p$-vector of $1$s, and
$$h_\matw(w_1,\dots,w_p\mid\vecomegaP,\mathbf{1},\veclambda) = h(w_1\mid\omega_1,1,\lambda_1)\times\dots\times h(w_p\mid\omega_p,1,\lambda_p).$$
From~\eqref{eq:nvmm_dens2} and \eqref{eq:scale_gauss}, it follows that \eqref{eq:GHNMVM2} can be written
\begin{align}\label{eq:msghd_dens}
&f_{\text{\tiny{MSGHD}}}\left(\vecx\mid\vecmu,\gam,\vecPhi,\vecalpha,\vecomegaP,\veclambda\right)\nonumber
=\prod_{j=1}^p\int_0^{\infty}{\phi_1\big(\left[\gam' \vecx-\vecmu-\vecDelta_\vecw \vecalpha \right]_j\mid 0,\Phi_jw_j \big)h_{W}\left(w_j\mid\omega_j,1,\lambda_j\right)dw_j}\nonumber\\
&=\prod_{j=1}^p\left\{\left[\frac{\omega_j+ \Phi_j^{-1}\big(\left[\gam'\vecx\right]_j-\mu_j\big)^{2}}{\omega_j+ \alpha_j^2 {\Phi_j}^{-1}} \right]^{\frac{\lambda_j-{1}/{2}}{2}}\frac{K_{\lambda_j-{1}/{2}}\bigg(\sqrt {\left[\omega_j+\alpha_j^2 {\Phi_j}^{-1}\right]\left[\omega_j+ \Phi_j^{-1}\big(\left[\gam'\vecx\right]_j-\mu_j\big)^{2}\right]}\bigg)}{(2\pi)^{{1}/{2}}{\Phi_j}^{{1}/{2}}K_{\lambda_j}(\omega_j)\exp{\left\{-\big(\left[\gam'\vecx\right]_j-\mu_j\big){\Phi_j^{-1}\alpha_j}\right\}}}\right\},\nonumber
\end{align}
%
%
where $\left[\gam'\vecx\right]_j$ is the $j${th} element of the vector $\gam'\vecx$, $\mu_j$ is the $j${th} element of the location parameter $\vecmu$, $\alpha_j$ is the $j${th} element of the skewness parameter $\vecalpha$, $\gam$ is a $p\times p$ matrix of eigenvectors, $\Phi_j$ is the $j$th eigenvalue of the diagonal matrix $\vecPhi$, $\vecomegaP= (\omega_1, \ldots, \omega_p)'$ controls the concentration in each dimension~$p$, and $\veclambda=(\lambda_1, \ldots, \lambda_p)'$ is a $p$-dimensional index parameter. 
Write $\vecX\backsim\text{MSGHD}(\vecmu,\gam,\vecPhi,\vecalpha,\vecomegaP,\veclambda)$ to indicate that the random vector $\vecX$ follows an MSGHD with density $f_{\text{\tiny{MSGHD}}}(\vecx\mid\vecmu,\gam,\vecPhi,\vecalpha,\vecomegaP,\veclambda)$. 
Then, a mixture of MSGHDs (MMSGHDs) has density
\begin{equation}
f(\vecx\mid\varthet)= \sum_{g=1}^G \pi_g f_{\text{\tiny{MSGHD}}}\left(\vecx\mid\vecmu_g,\gam_g, \vecPhi_g, \vecalpha_g,\vecomegaP_{g},\veclambda_{g}\right).
\end{equation}
The identifiability of the MMSGHD is discussed in Appendix~\ref{sec:PE}.

\section{Mixture of Coalesced Generalized Hyperbolic Distributions}\label{sec:MCGHD}

Note that the generalized hyperbolic distribution is not a special or limiting case of the MSGHD under any parameterization with $p>1$. Motivated by this, consider a coalesced generalized hyperbolic distribution (CGHD) that contains both the generalized hyperbolic distribution and MSGHD as limiting cases. The CGHD arises through the introduction of a random vector 
\begin{equation}\label{eq:def_R}
\vecR=U\vecX+(1-U)\vecS,
\end{equation}
where $\vecX=\gam\vecY$, $\vecY\backsim\text{GHD}\left(\vecmu,\matsig,\vecalpha,\omega_0,\lambda_0\right)$, $\matsig=\gam \vecPhi\gam'$, $\vecS\backsim\text{MSGHD}\left(\vecmu,\gam,\vecPhi,\vecalpha,\vecomegaP,\veclambda\right)$, and  $U$ is an indicator variable such that 
$$U=\begin{cases} 1 & \text{ if } \vecR \text{ follows a generalized hyperbolic distribution, and}\\ 0 & \text{ if } \vecR \text{ follows a MSGHD.}\end{cases}$$
It follows that
$\vecX = \gam\vecmu + W\gam\vecalpha + \sqrt{W}\gam\vecV,$
where $\gam\vecV\backsim\text{N}_p\left(\mathbf{0},\gam\vecPhi\gam'\right)$, 
$\vecS=\gam\vecmu + \gam\vecalpha\vecDelta_{\vecw} + \gam\vecA,$ where 
$\gam\vecA\backsim\text{N}_p\left(\mathbf{0},\gam\vecDelta_{\vecw}\vecPhi\gam'\right)$, 
and the density of $\vecR$ can be written
\begin{equation}\label{eq:cghd_dens}
\begin{split}
f_{\text{\tiny{CGHD}}}(\vecr\mid\vecmu,&\gam, \vecPhi,\vecalpha,\vecomegaP,\veclambda,\omega_{0},\lambda_{0},\varpi)\\ &= \varpi f_{\text{\tiny{GHD}}}\left(\vecr\mid\vecmu,\gam \vecPhi\gam' ,\vecalpha,\omega_0,\lambda_0\right)+ (1-\varpi)f_{\text{\tiny{MSGHD}}}\left(\vecr\mid\vecmu,\gam,\vecPhi,\vecalpha,\vecomegaP,\veclambda\right),
\end{split}\end{equation}
where $f_{\text{\tiny{GHD}}}(\cdot)$ is the density of a generalized hyperbolic random variable, $f_{\text{\tiny{MSGHD}}}(\cdot)$ is the density of a MSGHD random variable, and $\varpi\in(0,1)$ is a mixing proportion. Note that the random vector $\vecR$ would be distributed generalized hyperbolic if $\varpi=1$ and would be distributed MSGHD if $\varpi=0$. 
However, we must restrict $\varpi\in(0,1)$ for identifiability reasons. Specifically, if $\varpi=0$ then the value of $f_{\text{\tiny{CGHD}}}(\vecr\mid\vecmu,\gam, \vecPhi,\vecalpha,\vecomegaP,\veclambda,\omega_{0},\lambda_{0},\varpi)$ will be the same for any $\omega_0$ and $\lambda_0$ whereas, if $\varpi=1$, then the value of $f_{\text{\tiny{CGHD}}}(\vecr\mid\vecmu,\gam, \vecPhi,\vecalpha,\vecomegaP,\veclambda,\omega_{0},\lambda_{0},\varpi)$ will be the same for any $\vecomega$ and $\veclambda$.
The parameters $\vecmu$, $\vecalpha$, $\gam$, and $\vecPhi$ are the same for both densities, the parameters $\omega_0$ and $\lambda_0$ are univariate values peculiar to the generalized hyperbolic distribution, and the $p$-dimensional parameters $\vecomega$ and $\veclambda$ are peculiar to the MSGHD. 
Write $\vecR\backsim\text{CGHD}(\vecmu,\gam,\vecPhi,\vecalpha,\vecomegaP,\veclambda,\omega_0,\lambda_0,\varpi)$ to indicate that the random vector $\vecR$ follows a CGHD with density in \eqref{eq:cghd_dens}. 

A mixture of CGHDs (MCGHDs) has density
\begin{equation*}\label{eq:mcghd_dens}
f(\vecx\mid\varthet)= \sum_{g=1}^G \pi_gf_{\text{\tiny{CGHD}}}\left(\vecx\mid\vecmu_g,\gam_g,\vecPhi_g,\vecalpha_g,\vecomegaP_{g},\veclambda_{g},\omega_{0g},\lambda_{0g},\varpi_g\right).
\end{equation*}
Parameter estimation can be carried out via a generalized expectation-maximization (GEM) algorithm \citep{dempster77}. 
There are four sources of missing data: the latent $w_{0ig}$, the multi-dimensional weights $\vecDelta_{\vecw ig}=\diag(w_{1ig}, \ldots, w_{pig} )$, the component membership labels $z_{ig}$, and the inner component labels $u_{ig}$, for~$i=1,\ldots,n$ and $g=1,\ldots,G$. 
As usual, $z_{ig} = 1$ if observation~$i$ belongs to component~$g$ and $z_{ig}=0$ otherwise. Similarly, $u_{ig} = 1$ if observation~$i$, in component~$g$, is distributed generalized hyperbolic and $u_{ig} = 0$ if observation~$i$, in component~$g$, is distributed MSGHD. It follows that the complete-data log-likelihood for the MCGHD is
\begin{equation*}\begin{split}
l_{\text{c}}& = \sum_{i=1}^n  \sum_{g=1}^G \bigg\{ z_{ig} \log \pi_g   + z_{ig}{u_{ig} } \log \varpi_g  + z_{ig}(1 - u_{ig}) \log (1-\varpi_g)
+   z_{ig}{u_{ig} } \log h \left(  w_{0ig}~|~\omega_{0g}, 1,   \lambda_{0g}  \right)\\& +  z_{ig}(1 - u_{ig})  \sum_{j=1}^p\log h \left(  w_{jig}~|~\omega_{jg}, 1,   \lambda_{jg}  \right) + z_{ig}{u_{ig} } \log  \phi_p \left(\gam_g'\vecx_i~|~\vecmu_g + w_{0ig} \vecalpha_g , w_{0ig} \vecPhi \right)\\ 
&+ z_{ig}(1 - u_{ig})  \sum_{j=1}^p \log \phi_1 \left( [\gam_g'\vecx_i ]_j~|~\mu_{jg} + w_{jig} \alpha_{jg} ,     \omega_{jg} \phi_{jg} \right)\bigg\}.
\end{split}\end{equation*}
Further details on parameter estimation (Appendix~\ref{sec:ParaEst}) 
 and the identifiability of the MCGHD (Appendix~\ref{sec:PE}) are discussed in appendices.

%


\section{Cluster Convexity}\label{sec:cMSGHD}

The definition of clusters has been discussed quite extensively in the literature. Recently, \cite{hennig15} provides a very interesting discussion of some potential characteristics that clusters may have. Even though they cannot always be observed --- and, in some situations, desired characteristics may even conflict --- such characteristics provide important links and contrasts between different definitions that are used for a cluster. In fact, the idea of desirable characteristics is not a new one, e.g., \cite{cormack71} gives internal
cohesion and external isolation as two ``basic ideas" in this direction. In this paper, we follow the definition given by \cite{mcnicholas16}: ``a cluster is a unimodal component within an appropriate finite mixture model''. The term appropriate means that the model has the flexibility to fit the data and, in many cases, this also means that each cluster is convex \cite[see][Section~9.1]{mcnicholas16}. The development of flexible models outlined herein may make it a little easier to find an ``appropriate" component density.

\begin{figure}[!t]
\centering\includegraphics[width=0.425\textwidth]{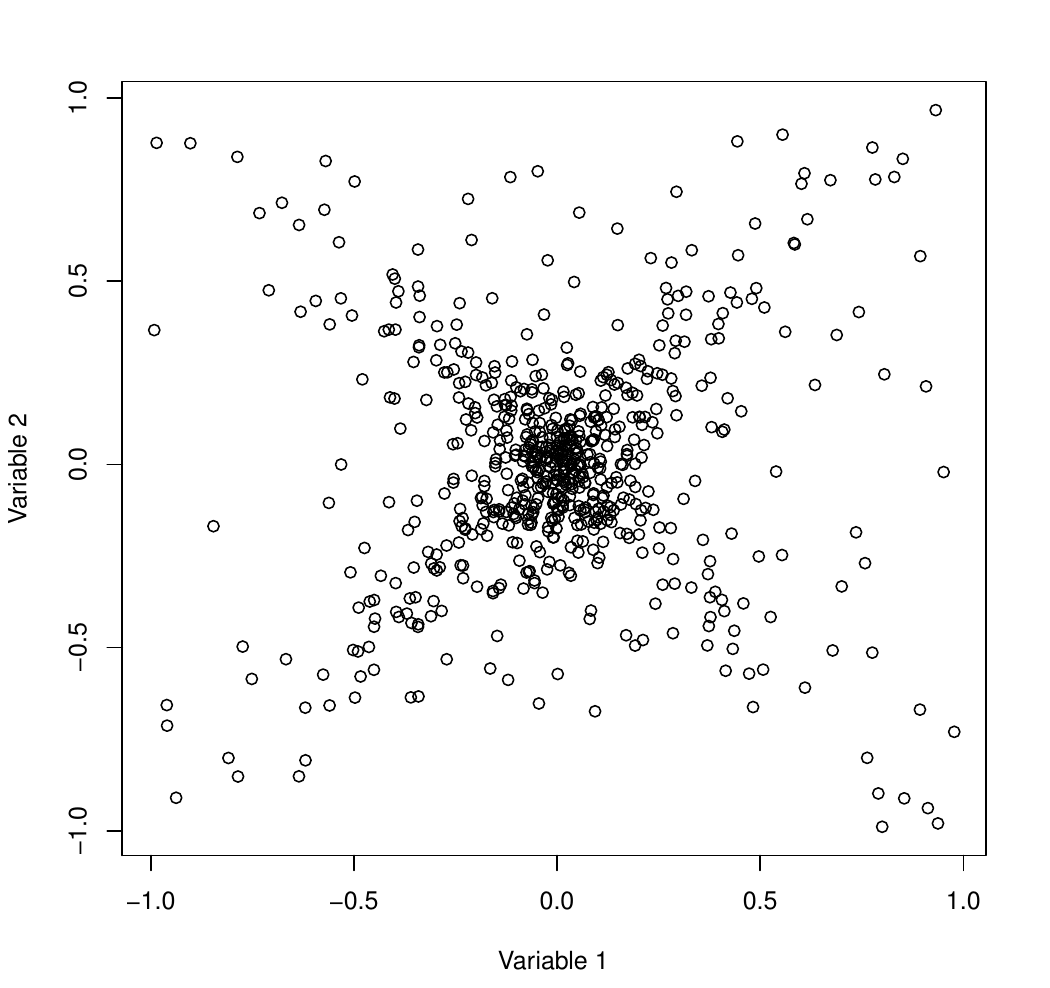} \
\includegraphics[width=0.425\textwidth]{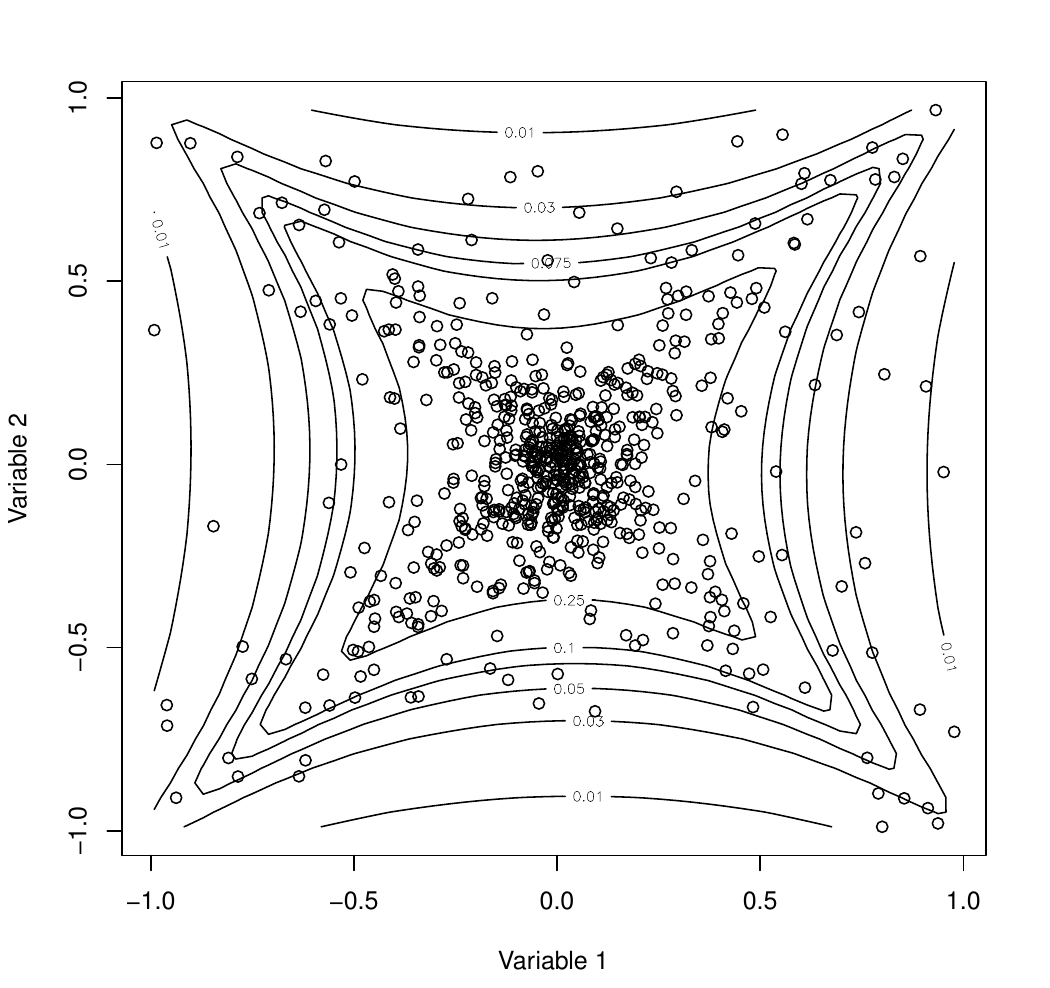}\\
\includegraphics[width=0.425\textwidth]{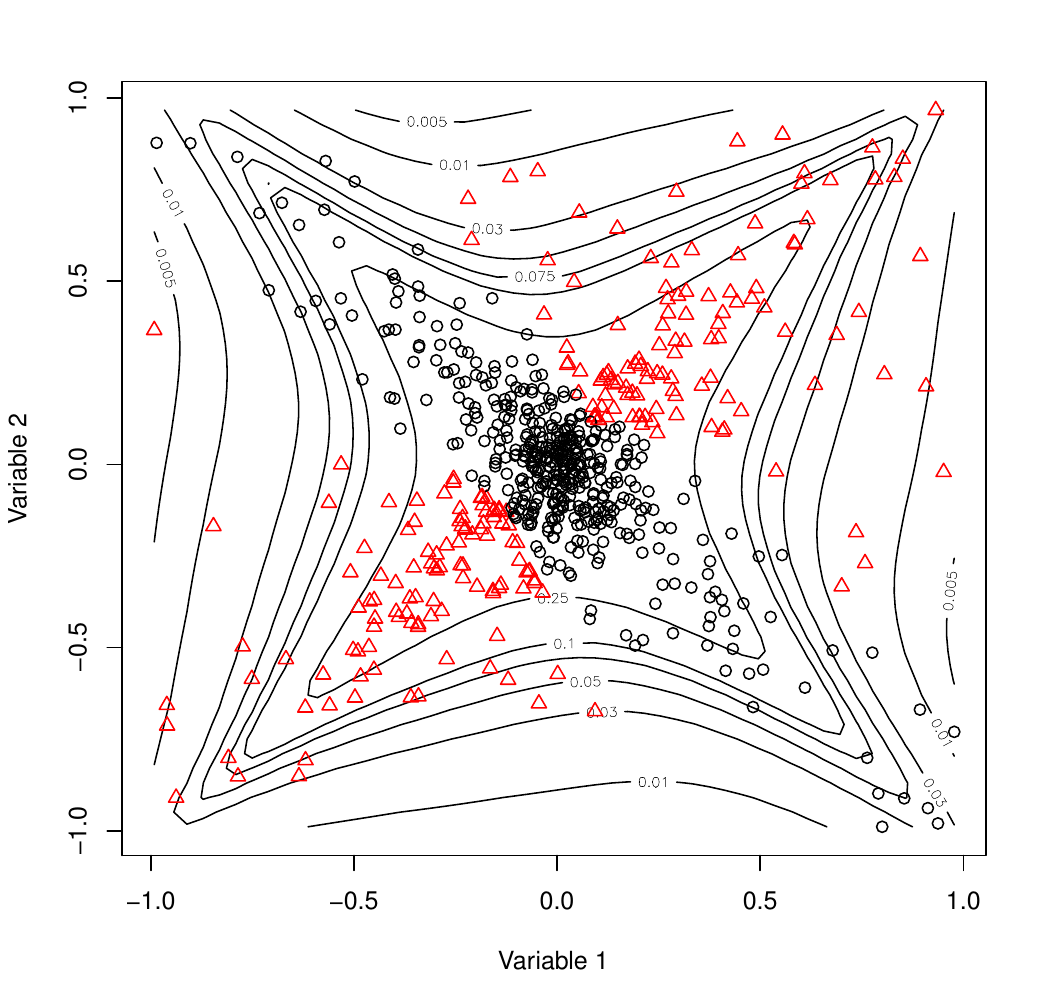} \
\includegraphics[width=0.425\textwidth]{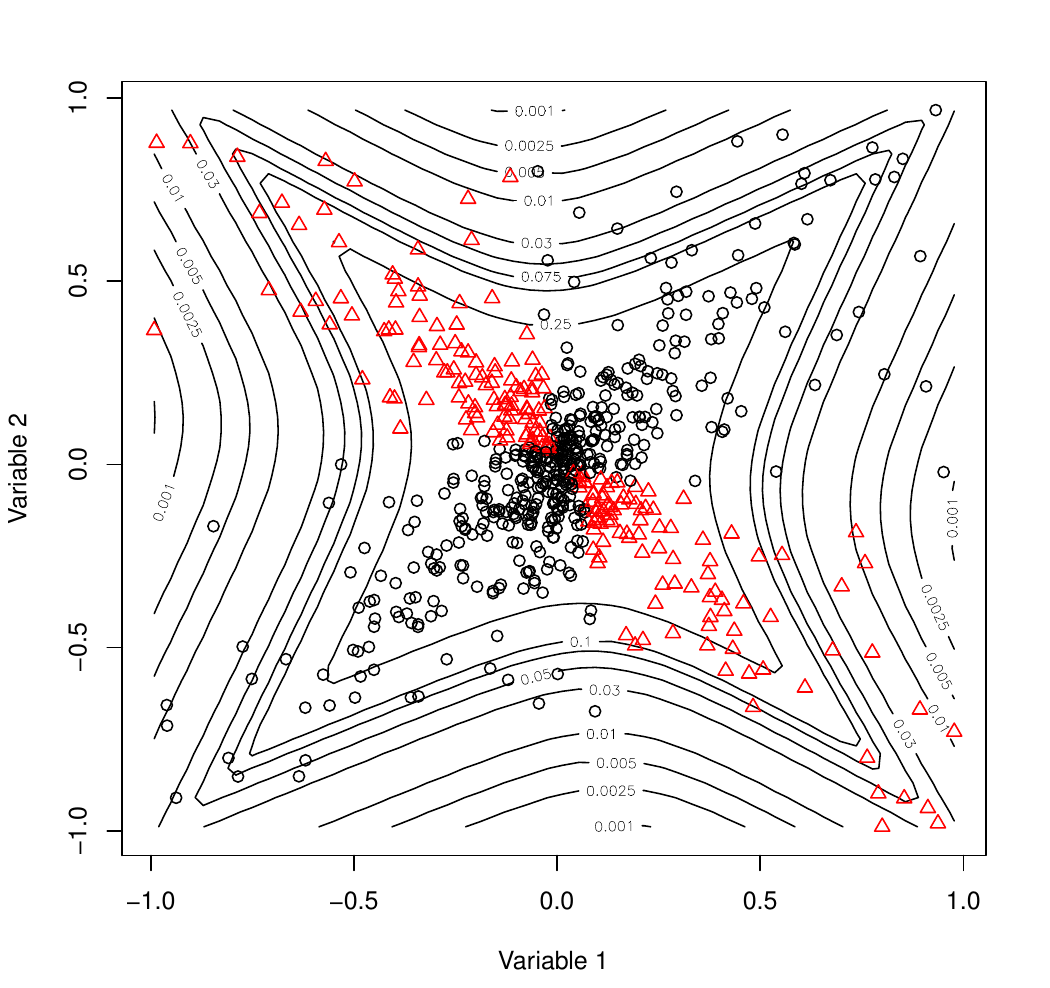}
\caption{\label{fig:xplot1} Scatter plot of data generated from a two-component $t$-mixture (top-left) along with contours from fitted $G=1$ component MMSGHD (top-right), $G=2$ component MMSGHD (bottom-left), and $G=2$ component McMSGHD (bottom-right) models, where plotting symbol and colour represent predicted classifications.}
\end{figure}

The MSGHD is more flexible than the GHD; however, similar to the multiple scaled multivariate $t$-distribution of \cite{forbes14}, the MSGHD can have contours that are not convex. Accordingly, the MMSGHD can have components that are non-convex, leading to non-convex clusters. 
%
%
Consider the data in Figure~\ref{fig:xplot1}. How many clusters are there? 
The most plausible answer to this question is two overlapping clusters: one with positive correlation between the variables and another with negative correlation between the variables. There may also be an argument for four or five. Note that the data are generated from a $G=2$ component mixture of multivariate $t$-distributions. 

\begin{figure}[!t]
\centering\includegraphics[width=0.4\textwidth]{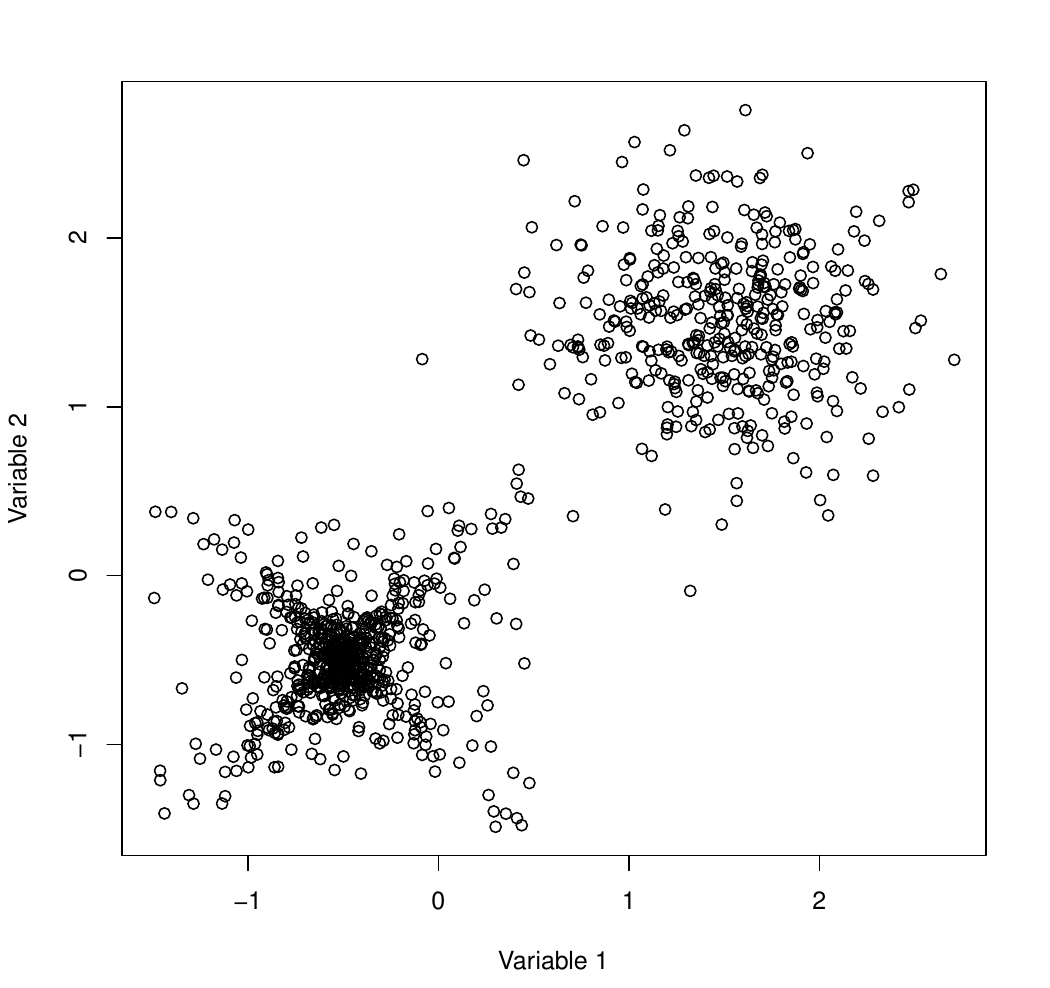}\ 
\includegraphics[width=0.4\textwidth]{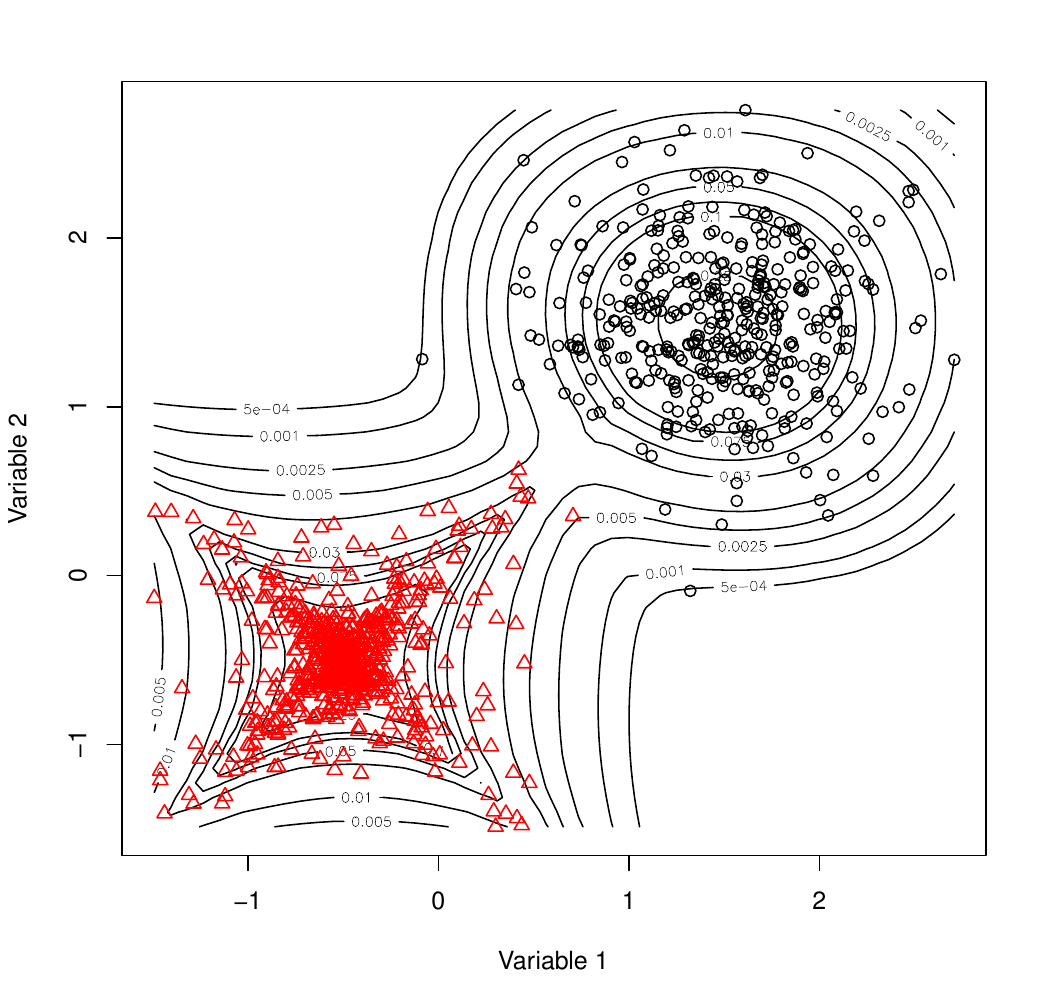}\\
\includegraphics[width=0.4\textwidth]{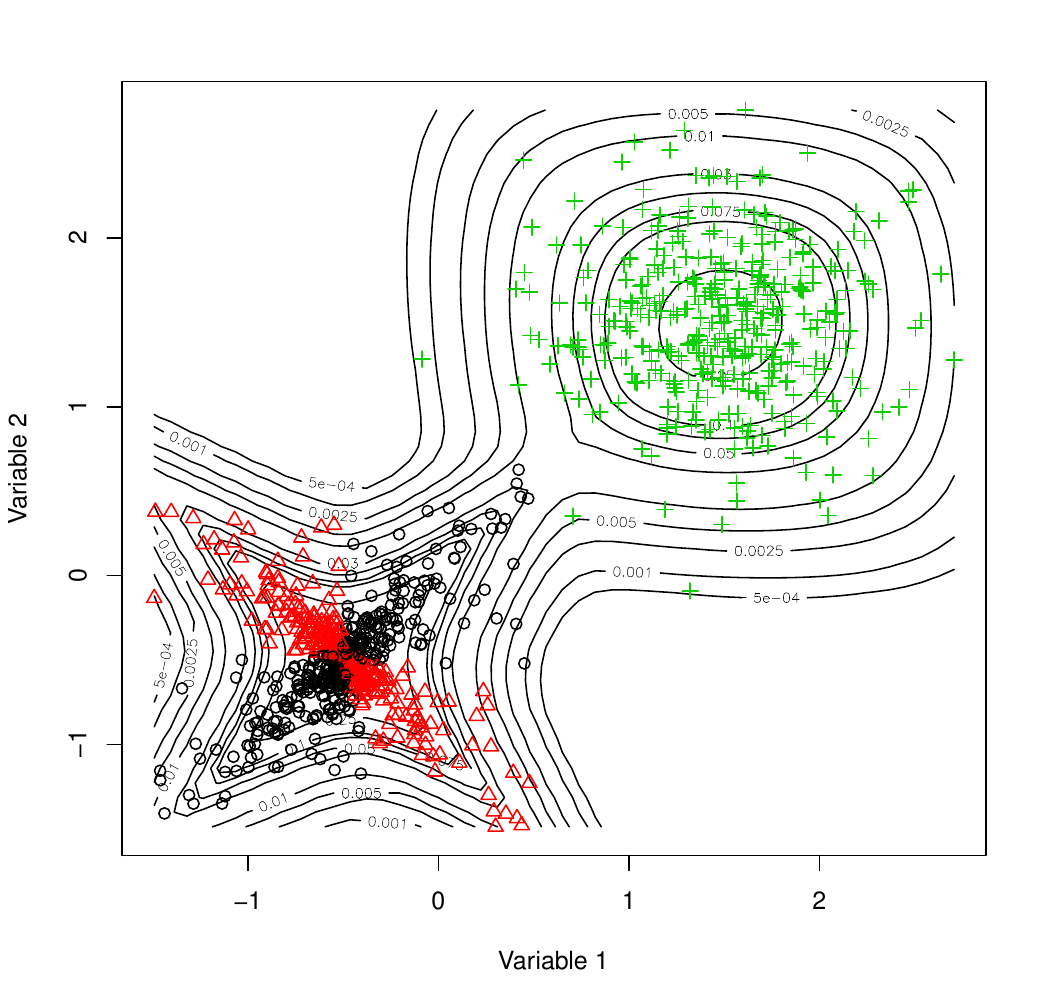}
\caption{\label{fig:xoplot} Scatter plots for model-based clustering results on data simulated from a three-component $t$-mixture (top-left), with contours from the selected MMSGHD (top-right) and McMSGHD (bottom) models, respectively, where plotting symbol and colour represent predicted classifications.}
\end{figure}

The MMSGHD is fitted to these data for $G=1,\ldots,5$ and, as is common in model-based clustering applications, the Bayesian information criterion \citep[BIC;][]{schwarz78} is used to select $G$. Note that the BIC is given by
$$\text{BIC}=2l(\hat\varthet)-\rho\log n,$$
where $l(\hat\varthet)$ is the maximized log-likelihood, $\rho$ is the number of free parameters, and $n$ is the number of observations. For the MMSGHD, the BIC selects a $G=1$ component model (Figure~\ref{fig:xplot1}). Furthermore, looking at the $G=2$ component MSGHD solution (Figure~\ref{fig:xplot1}) confirms that the problem is not just one of model selection; the $G=2$ component MSGHD solution selects one component that is roughy elliptical and another that is not convex. On the other hand, forcing the MSGHD to be convex --- which can be done by imposing the constraint $\lambda_j >1$, for $j=1,\ldots,p$ --- leads to what we call the convex MSGHD (cMSGHD). Fitting the corresponding mixture of cMSGHDs (McMSGHDs) ensures that, if each component is associated with a cluster, then convex clusters are guaranteed. Results in a $G=2$ component model being selected (Figure~\ref{fig:xplot1}). Formally, this amounts to insuring that the MSGHD is quasi-convex; see Appendix~\ref{app:convex}.
The general point here is that if convexity is not enforced, then the MSGHD can give components that contain multiple clusters. While it is easy to spot this in two dimensions, e.g., Figure~\ref{fig:xplot1}, this phenomenon may go unrecognized in higher dimensions, possibly resulting in greatly misleading results. Of course, the issue of non-convex clusters does not arise with most model-based approaches; however, when multiple scaled mixtures are considered, the issue can crop up. Another example in a similar vein is given in Figure~\ref{fig:xoplot}, where data are generated from a $G=3$ component mixture of multivariate $t$-distributions. The selected MMSGHD has $G=2$ components, including one clearly non-convex cluster, while the McMSGHD gives sensible clustering results. 

The intention behind the introduction of the McMSGHD is not that it should supplant the MMSGHD, but rather that it provides a convenient check on the number of components. In a higher dimensional application, where visualization is difficult or impossible, situations where the selected MMSGHD has fewer components than the selected McMSGHD will deserve special attention. Of course, this is not to say that the selected McMSGHD will always have more components in situations where the MSGHD has too few, but rather that it will help to avoid the sort of situations depicted in Figures~\ref{fig:xplot1} and~\ref{fig:xoplot}.
Note that parameter estimation for the McMSGHD is analogous to the algorithm for the MMSGHD algorithm but $\lambda_{kj}$ ($k=0,1,\ldots,G$, $j=1,\ldots,p$) is updated only if the value of its update exceeds 1.

\section{Illustrations}\label{sec:app}

\subsection{Implementation and Evaluation}
In the following illustrations, we fit the MGHDs, the MMSGHDs, the McMSGHDs, and the MCGHDs using the corresponding functions available in the \texttt{MixGHD} package \citep{tortora17} for {\sf R} \citep{R15}. We use $k$-means and $k$-medoids clustering to initialize the $\hat{z}_{ig}$. The adjusted Rand index \cite[ARI;][]{hubert85} is used to compare predicted classifications with true classes. The ARI corrects the Rand index \citep{rand71} for chance, its expected value under random classification is $0$, and it takes a value of $1$ when there is perfect class agreement. \cite{steinley04} gives guidelines for interpreting ARI values.

Several mixtures of skewed distributions have been proposed for model-based clustering and classification. Among them, three different formulations of the multivariate skew-{\it t} distribution have been used for clustering in the mixture setting. The formulation used by \cite{murray14b} is a special case of the generalized hyperbolic distributions. \cite{azzalini14} refers to the other two formulations as the classical and SDB formulations (for the initials of the authors' names), respectively. 
 \cite{lee13,lee14} compare the mixture of classical skew-{\it t} distributions to the mixture of SDB skew-{\it t} distributions (MSDBST) and to skew-normal analogues of both formulations; their results lead one to believe that the MSDBST is generally preferable \citep[see][for an alternative viewpoint]{azzalini14}. The MSDBST approach is implemented using the \texttt{EMMIXuskew} package \citep{lee13b} for {\sf R} and it is used for comparison herein.

\subsection{ Simulation Study}
\begin{figure}[!ht]
\centering\includegraphics[width=0.384\textwidth]{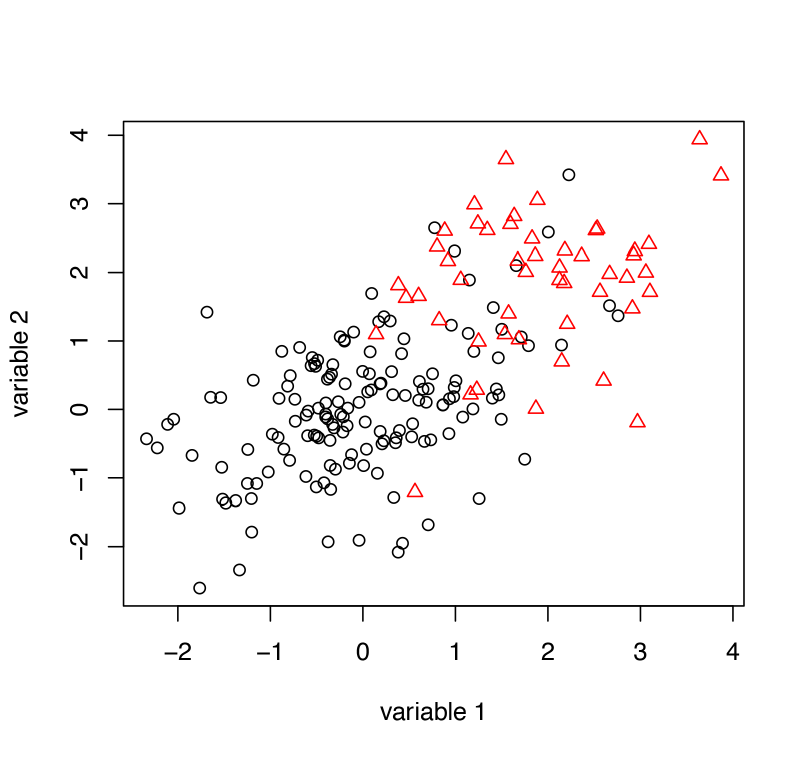}\
\includegraphics[width=0.384\textwidth]{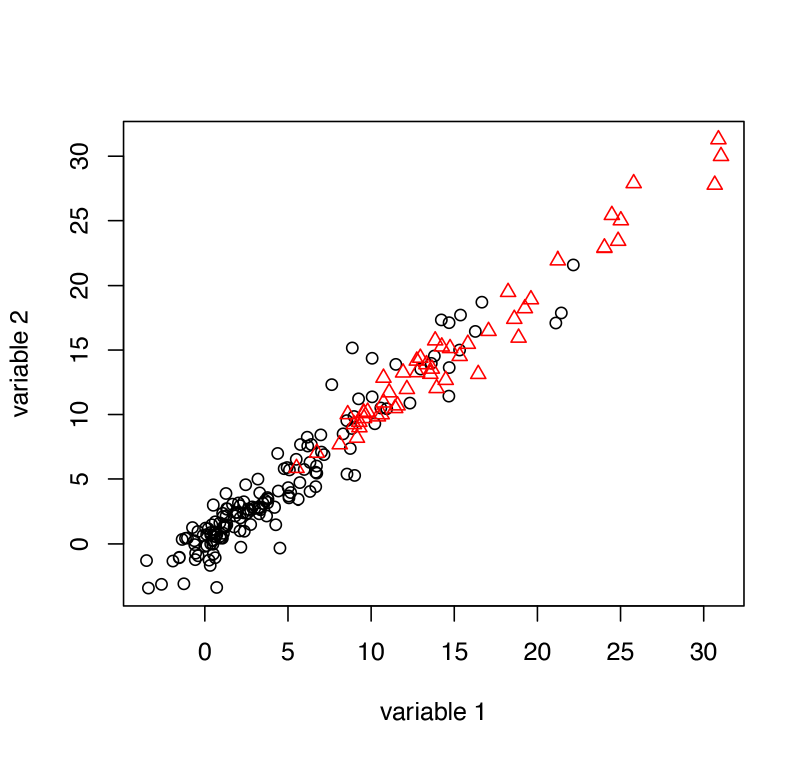}\\[-9pt]
\includegraphics[width=0.384\textwidth]{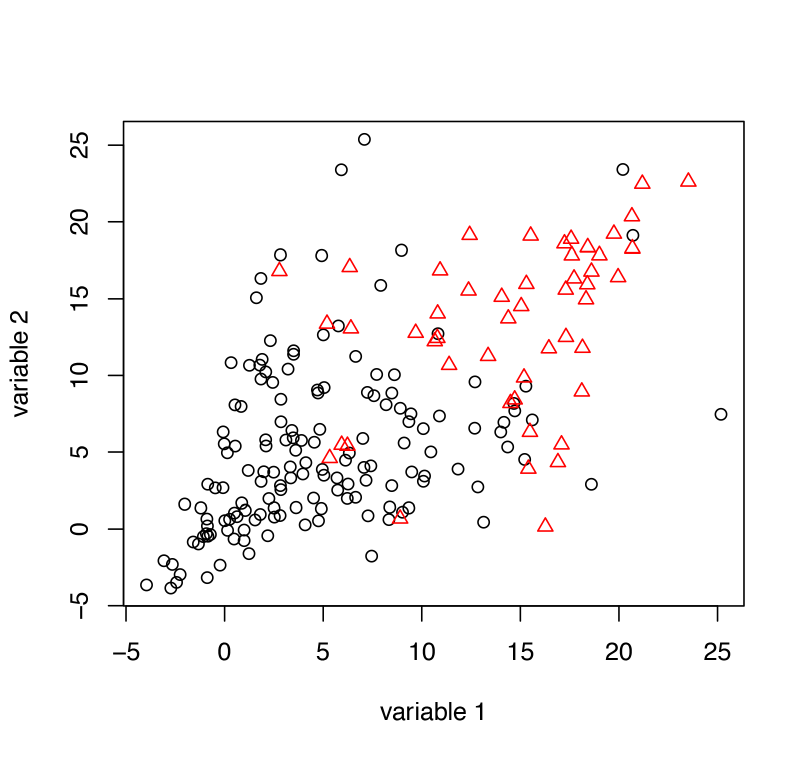}
\caption{\label{fig:simplot} Scatter plots for  data simulated from a two-component MGDs (top-left),  MGHDs (top-right) and MMSGHDs (bottom) models, respectively,  with high correlation, 50\% overlapping and clusters of different size, where plotting symbol and colour represent the true classifications.}
\end{figure}

We use a simulation study to measure the performance of the proposed methods. The interest is to observe the ARI under different circumstances, the data are generated using two-component mixtures of Gaussian (Scenario 1), generalized hyperbolic (Scenario 2), and multiple scaled generalized hyperbolic distributions (Scenario 3), using the {\sf R} function \texttt{mvrnorm} and  the stochastic relationships given in~\eqref{eqn:hypstorep2} and Section~\ref{sec:MCGHD}. For each scenario a three factors full factorial design was used, where the factors are: medium (M) or high (H)  correlation, 25\% or 50\% overlapping, and same (S), $n_1=n_2=100$,  or different (D), $n_1=150$ and $n_2=50$, number of elements per component. In all the scenarios $\vecmu_1 = \bf1$.
Table \ref{simGau} shows the average ARI, with standard deviation, obtained on 10 data sets generated from a Gaussian distribution with $p=10$, $G=2$, correlation equal to 0.33 for M and 0.66 for H, $\vecmu_2 = \bf2$ for 25\% overlapping and $\vecmu_2 = \bf3$ for 50\% overlapping. 
Table \ref{simGHD} shows the average ARI, with standard deviation,  obtained on 10 data sets generated from a MGHD distribution with $p=10$, $G=2$, $\vecalpha=\bf 1$, $\lambda=0.5$, $\omega=1$ and  correlation equal to 0.25 for M and 0.5 for H, $\vecmu_2 = \bf10$ for 25\% overlapping and $\vecmu_2 = \bf15$ for 50\% overlapping. 
Table \ref{simMSGHD} shows the average ARI, with 1 standard deviation, obtained on 10 data sets generated from a MMSGHD distribution with $p=10$, $G=2$,  $\vecalpha_1=\bf 1$, $\vecalpha_2=\bf -1$, $\veclambda=\bf 0.5$, $\vecomega=\bf 1$ and  correlation equal to 0.25 for M and 0.35 for H, $\vecmu_2 = \bf19$ for 25\% overlapping and $\vecmu_2 = \bf23$ for 50\% overlapping. 
For the data sets generated using the MGHD and the MMSGHD the values of the correlation had to be reduced in order to maintain 25\% and 50\% overlap, higher correlation lead to higher overlap. Figure \ref{fig:simplot} shows the scatterplots of the data simulated using the two-component MGDs, MGHDs, and MMSGHDs  with high correlation, 50\% overlapping and cluster of different size, where plotting symbol and colour represent the true classifications.
\begin{table}[!h]
\caption{\label{simGau} Average ARI values, with standard deviations, for each data set generated from MGDs, with different starting parameters.}
\begin{tabular*}{1.00\textwidth}{@{\extracolsep{\fill}}llrrrrr}
\hline
Correlation & Overlapping & $n_g$ &MCGHD&MGHD&MMSGHD&McMSGHD\\
\hline
M&25\%&S& $0.964$ $(0.020)$&$0.932$ $(0.056)$&$0.943$ $(0.025)$&$0.941$ $(0.026)$\\
H&25\%&S& $0.850$ $(0.044)$&$0.730$ $(0.132)$&$0.848$ $(0.050)$&$0.817$ $(0.062)$\\
M&50\%&S& $0.734$ $(0.064)$&$0.642$ $(0.085)$&$0.736$ $(0.062)$&$0.721$ $(0.084)$\\
H&50\%&S& $0.565$ $(0.116)$&$0.408$ $(0.179)$&$0.595$ $( 0.059)$&$0.553$ $(0.090)$\\
M&25\%&D& $0.981$ $(0.016)$&$0.979$ $( 0.017)$&$0.981$ $( 0.021)$&$0.975$ $(0.022)$\\
H&25\%&D& $0.831$ $(0.047)$&$0.771$ $(0.047)$&$0.780$ $(0.082)$&$0.787$ $(0.059)$\\
M&50\%&D&$ 0.715$ $( 0.053)$&$0.663$ $(0.064)$&$0.705$ $(0.069)$&$0.696$ $(0.073)$\\
H&50\%&D& $0.494$ $( 0.133)$&$0.354$ $(0.151)$&$0.446$ $(0.122)$&$0.447$ $(0.130)$\\
\hline
\end{tabular*}
\end{table}
\begin{table}[!h]
\caption{\label{simGHD} Average ARI values, with standard deviations, for each data set generated from MGHDs, with different starting parameters and $\vecalpha=\bf 1$.}
\begin{tabular*}{1.00\textwidth}{@{\extracolsep{\fill}}llrrrrr}
\hline
Correlation & Overlapping & $n_g$ &MCGHD&MGHD&MMSGHD&McMSGHD\\
\hline
M&25\%&S& $0.933$ $(0.045)$&$0.922$ $(0.041)$&$0.765$ $(0.228)$&$0.773$ $(0.211)$\\
H&25\%&S& $0.842$ $(0.040)$&$0.770$ $(0.119)$&$0.463$ $(0.348)$&$0.399$ $(0.371)$\\
M&50\%&S& $0.712$ $(0.093)$&$0.680$ $(0.107)$&$0.510$ $(0.229)$&$0.477$ $(0.221)$\\
H&50\%&S& $0.475$ $(0.176)$&$0.416$ $(0.104)$&$0.188$ $(0.180)$&$0.159$ $(0.113)$\\
M&25\%&D& $0.836$ $(0.058)$&$0.858$ $(0.058)$&$0.579$ $(0.233)$&$0.557$ $(0.233)$\\
H&25\%&D& $0.754$ $(0.106)$&$0.713$ $(0.124)$&$0.257$ $(0.250)$&$0.318$ $(0.283)$\\
M&50\%&D& $0.630$ $(0.092)$&$0.622$ $(0.098)$&$0.144$ $( 0.134)$&$0.194$ $(0.198)$\\
H&50\%&D& $0.427$ $(0.090)$&$0.360$ $(0.086)$&$0.121$ $(0.109)$&$0.127$ $(0.093)$\\
\hline
\end{tabular*}
\end{table}
\begin{table}[!h]
\caption{\label{simMSGHD} Average ARI, with standard deviations, for each data set generated from MMSGHDs, with different starting parameters and $\vecalpha=\bf 1$.}
\begin{tabular*}{1.00\textwidth}{@{\extracolsep{\fill}}llrrrrr}
\hline
Correlation & Overlapping & $n_g$ &MCGHD&MGHD&MMSGHD&McMSGHD\\
\hline
M&25\%&S& $0.903$ $(0.047)$&$0.801$ $(0.077)$&$0.912$ $(0.043)$&$0.922$ $(0.045)$\\%
H&25\%&S& $0.806$ $(0.085)$&$0.697$ $(0.069)$&$0.793$ $(0.073)$&$0.810$ $(0.060)$\\
M&50\%&S& $0.751$ $(0.080)$&$0.614$ $(0.081)$&$0.811$ $(0.053)$&$0.822$ $(0.044)$\\
H&50\%&S& $0.592$ $(0.062)$&$0.475$ $(0.121)$&$0.603$ $(0.090)$&$0.634$ $(0.078)$\\
M&25\%&D& $0.925$ $(0.042)$&$0.838$ $(0.108)$&$0.950$ $(0.042)$&$0.950$ $(0.037)$\\
H&25\%&D& $0.813$ $(0.049)$&$0.740$ $(0.088)$&$0.824$ $(0.066)$&$0.815$ $(0.070)$\\
M&50\%&D&$ 0.786$ $(0.090)$&$0.616$ $(0.104)$&$0.742$ $(0.126)$&$0.746$ $(0.120)$\\
H&50\%&D& $0.605$ $(0.111)$&$0.511$ $(0.118)$&$0.391$ $(0.231)$&$0.402$ $(0.224)$\\
\hline
\end{tabular*}
\end{table}

The MCGHDs perform at least comparably to the best of the MMSGHDs and MGHDs. As expected, the MGHDs over performs the MMSGHDs when the data are generated using a MGHDs model and \textit{vice versa} when the data are generated from a MMSGHDs. The two methods perform similarly on normally distributed clusters. The generated clusters are all convex and that explains the similarity of the results obtained using MMSGHDs and McMSGHDs. The level of overlapping has a notable impact on the ARI, as expected. The performance of all the methods are slightly better when the clusters are of the same size and when there is a lower correlation.

\subsection{Real Data Analysis}

To assess the 
classification performance of the MGHDs, the MMSGHDs, the McMSGHDs, the MCGHDs, and the MSDBST, we consider four real data sets that are commonly used within the model-based clustering literature (Table~\ref{datasets}). 
\begin{table}[!h]
\caption{\label{datasets} Summary details for four real data sets commonly used within the model-based clustering literature.} 
{\small\begin{tabular*}{1.01\textwidth}{@{\extracolsep{\fill}}lccclr}
\hline
&Classes &$n$& $p$&Source &Original source\\
\hline
Bankruptcy &2 &66&  2 &  {\sf R} package \texttt{MixGHD}&\cite{altman68}\\
Banknote &2&200& 7&  {\sf R} package  \texttt{MixGHD}&\cite{flury88}\\
Seeds&3&210& 7& UCI machine learning repository$^*$ &\cite{charytanowicz10}\\
AIS  &2&202& 11&  {\sf R} package \texttt{EMMIXuskew}&\cite{cook94}\\
\hline
\end{tabular*}}\\
$^*$\scriptsize{\tt http://archive.ics.uci.edu/ml/}
\end{table}

To compare the performance of the methods we set the number of components, $G$, equal to the true number of classes. In general, the BIC  is used to select the best starting criterion between $k$-means and $k$-medoids; however, because of the presence of outliers, only $k$-medoids starts are used for the bankruptcy data set. Table~\ref{appres} displays the classification performance. Note that MSDBST is not used on the AIS  data set because of the prohibitively high dimensionality ($p=11$). The MCGHD generally performs comparably to the best of the other four approaches. Specifically, the MCGHD gives the best classification performance --- either outright or jointly --- for the four data sets.   Interestingly, the MGHD gives very good classification performance on three of the four data sets; however, this must be taken in context with its very poor classification performance on the bankruptcy data set (ARI $\approx0$). It is also interesting to compare the classification performance of the McMSGHD to the MMSGHD as well as that of the MGHD to the MMSGHD. The McMSGHD and the MGHD approaches both  outperform MMSGHD for one data set, and give a similar performance on two of the other three data sets. 
This highlights the fact that a mixture of multiple scaled distributions may well not outperform its single scaled analogue, and underlines the need for an approach with both the MSGHD and MGHD models as special cases. The results for the bankruptcy data illustrate that the MCGHD approach can give very good classification performance in situations where neither the MGHD nor the MMSGHD perform well.
Finally, the MCGHD outperforms the MSDBST on two data sets and gives the same result on the third (recall that the MSDBST could not be fitted to the AIS data).\begin{table}[!h]
{\footnotesize
\caption{\label{appres} ARI values for the MCGHD, MGHD,  MMSGHD,  McMSGHD, and MSDBST approaches on four real data sets. }
\begin{tabular*}{1.00\textwidth}{@{\extracolsep{\fill}}lrrrrrr}
\hline
Data&$G$ &MCGHD&MGHD&MMSGHD&McMSGHD&MSDBST\\
\hline
Bankruptcy& 2&0.824&0.019&0.255&0.170&0.085\\
Bank note& 2&0.980&0.980&0.980&0.980&0.980\\
Seeds & 3&0.775&0.617&0.519&0.533&0.723\\
AIS& 2&0.903&0.884  &0.884&0.865&NA\\
\hline
\end{tabular*}
}
\end{table}

In this analysis, we took the number of components to be known. However, in a true clustering scenario, we would  not have \textit{a~priori} information about the number of groups. Therefore, the analysis was repeated without this assumption and, for all but the bankruptcy data set, all approaches were run for $G=1,\ldots,5$ components. 
Because of the small number of observations, the bankruptcy data were run for $G=1,2,3$. Table~\ref{index} gives the number of components selected by the BIC. 
\begin{table}[!h]
{\footnotesize
\caption{\label{index} Selected number of components using the BIC for each real data  set and corresponding ARI in brackets, where the correct number of components is highlighted in bold face font.}
\begin{tabular*}{1.00\textwidth}{@{\extracolsep{\fill}}lcrrrrr}
\hline
Data&Classes &MCGHD&MGHD&MMSGHD&McMSGHD& MSDBST\\
\hline
{Bankruptcy}&{$2$}&1(0.000)&1(0.000)&1(0.000)&{\bf 2}(0.170)&1(0.000)\\
{Bank note}&{$2$} &{\bf 2}(0.980)&{\bf 2}(0.980)&{\bf 2}(0.980)&{\bf 2}(0.980)&{\bf 2}(0.980)\\
{Seeds}&{$3$}&2(0.502)&4(0.484)&2(0.530)&2(0.530)&1(0.000)\\
{AIS}&{$2$} &4(0.435)&3(0.615)&1(0.000)&1(0.000)&NA\\
\hline
\end{tabular*}}
\end{table}

For the bank note data, the BIC selects the correct number of components for every method; however, for the bankruptcy data, it picks the right number of components only for McMSGHDs. For the seed and AIS data sets, the BIC does not select the correct number of components for any approach. Recall that the MSDBST could not be fitted to the AIS data; also, it could not be fitted to the seed data for $G=4$. These results illustrate that the BIC is not necessarily reliable for selecting the number of components in real data examples for any of the five approaches. 

\subsection{Computation time}
\begin{figure}[!ht]
\begin{center}
 \includegraphics[width=0.49\textwidth]{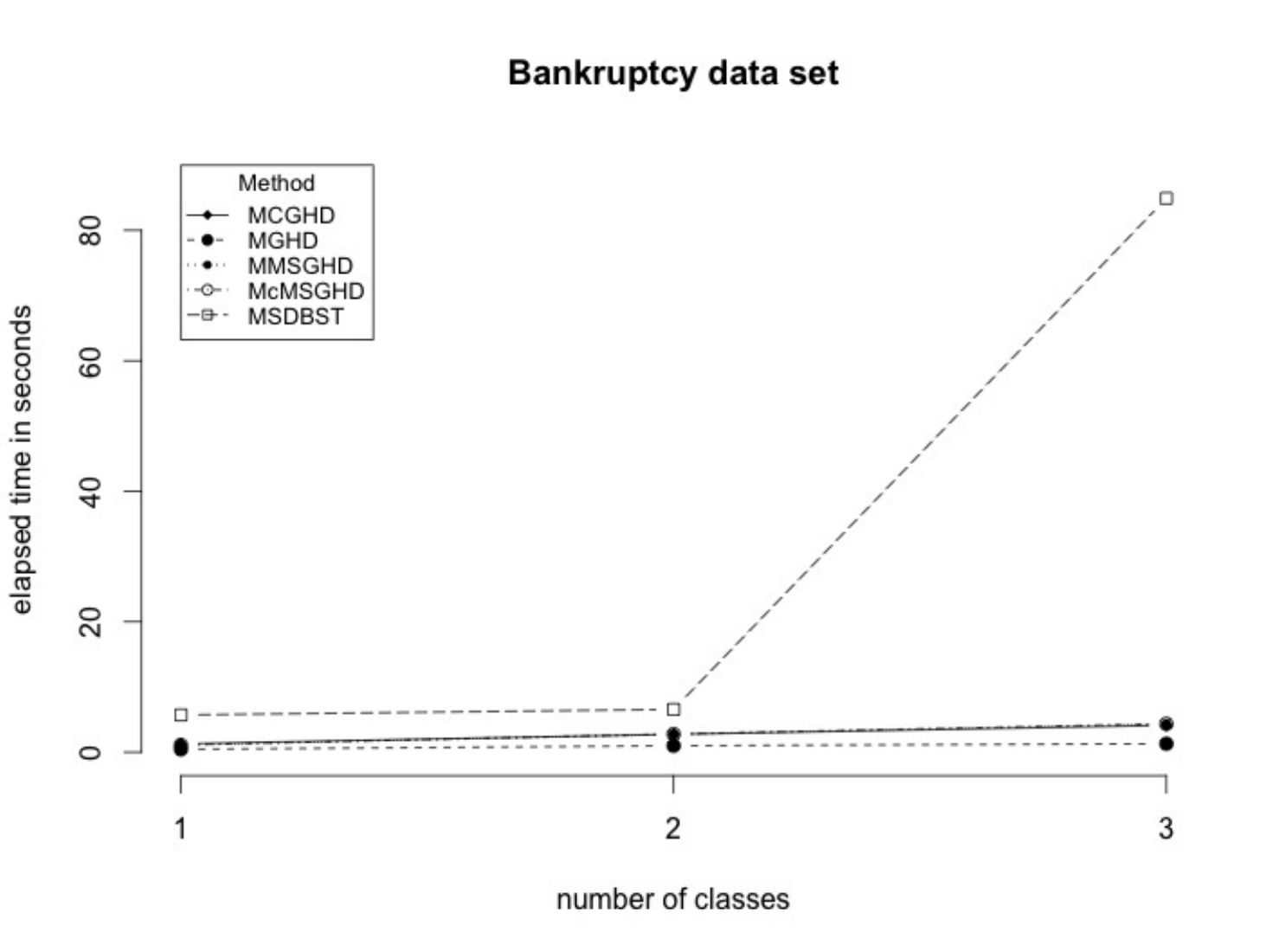} \
\includegraphics[width=0.49\textwidth]{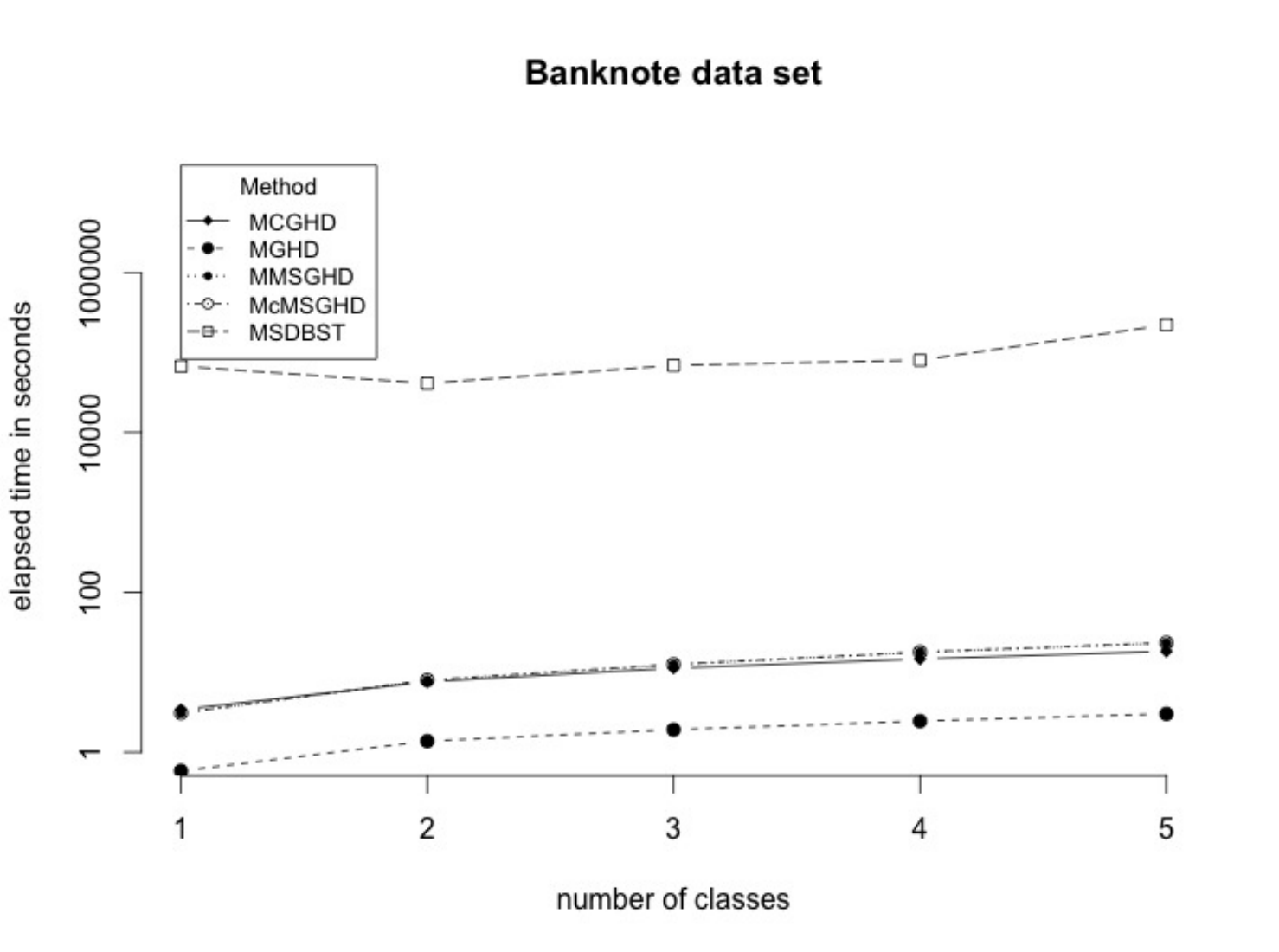}\\
\includegraphics[width=0.49\textwidth]{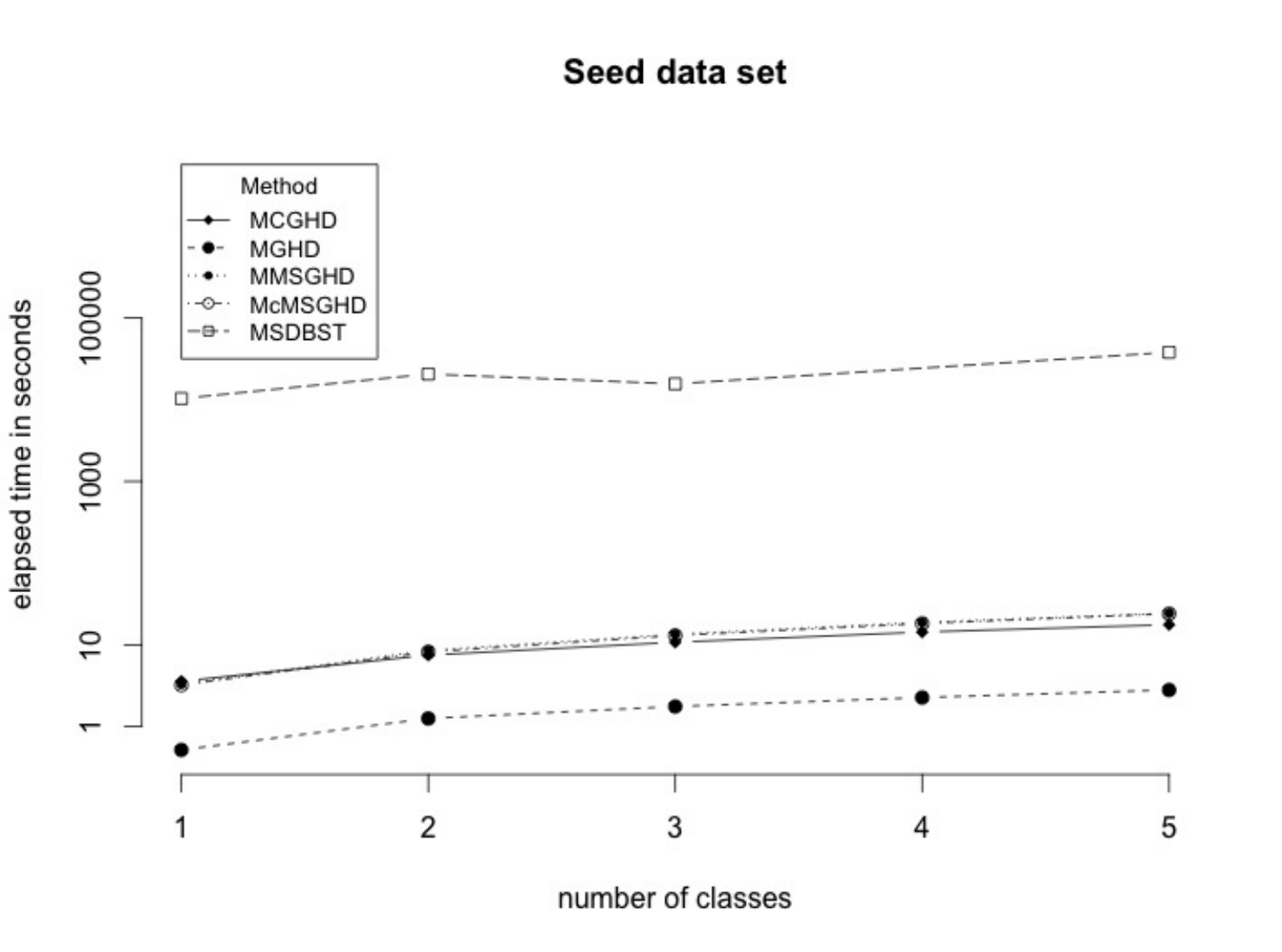}\
\includegraphics[width=0.49\textwidth]{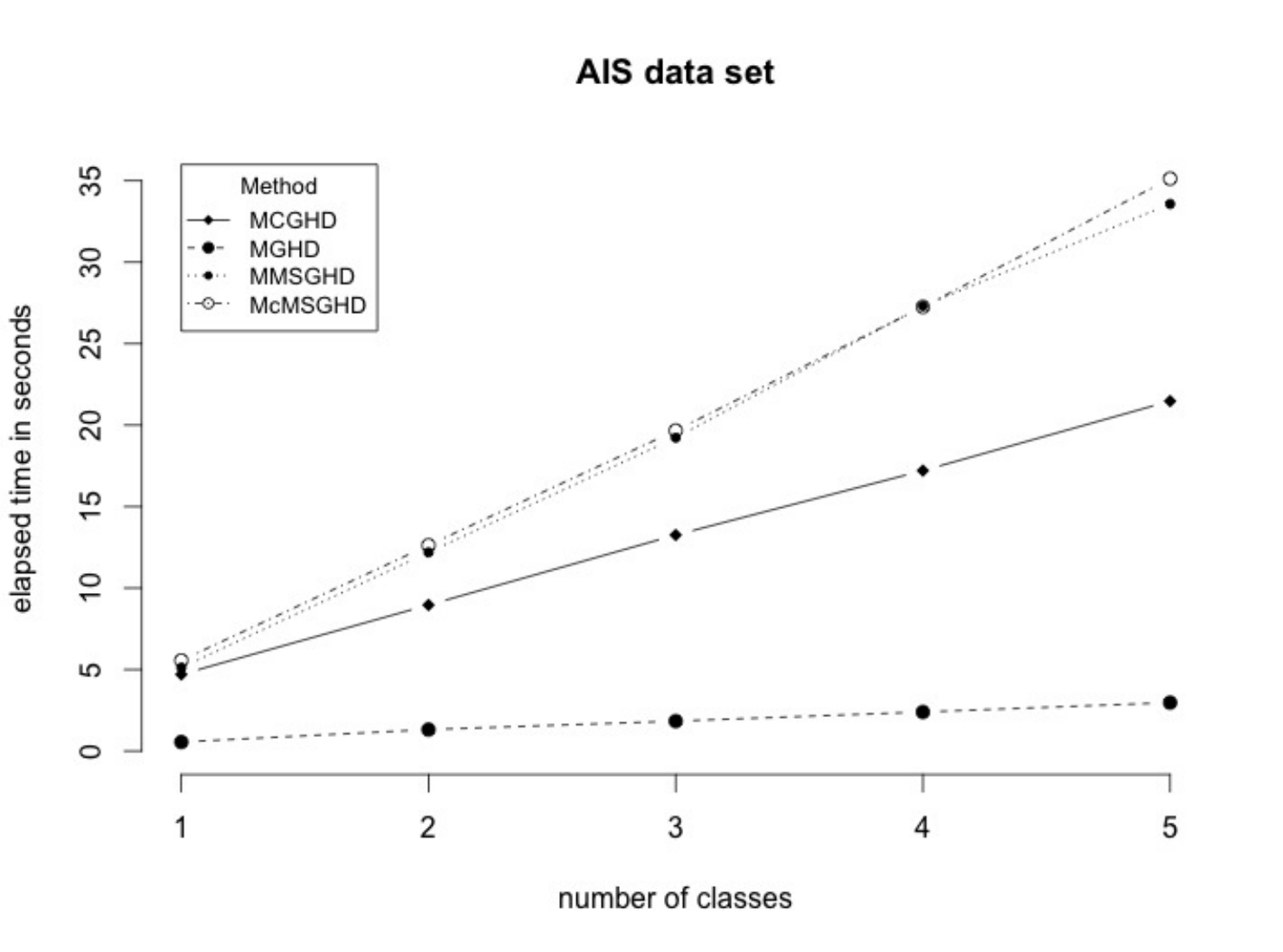}
\caption{\label{figTime2} The average elapsed time to perform 100 iterations of the (G)EM algorithm when varying the number of components for MCGHD, MGHD,  MMSGHD,  McMSGHD, and MSDBST models  on the  bankruptcy, banknote, seeds, and AIS data sets, respectively.}
\end{center}
\end{figure}

For the data sets listed in Table~\ref{datasets}, we measured the elapsed user times, in seconds, to perform 100 (G)EM iterations. Note that all code was run in {\sf R} version 3.0.2 on a 32-core Intel Xeon E5 server with 256GB RAM running 64-bit CentOS.
Figure~\ref{figTime2} displays the average elapsed time for two replications of each algorithm using a $k$-means and a $k$-medoids starting partition for $G=1,\ldots,5$ components.
The EM algorithm for the MSDBST is significantly slower than that of the hyperbolic-based approaches (Figure~\ref{figTime2}). In fact, on the banknote data set, it takes the MSDBST more than 11 hours to perform the required number of EM iterations when $G=2$ and about 63 hours when $G=5$. For the seeds data set, it takes the MSDBST more than 5 hours when $G=2$ and more than 10 hours when $G=5$, whereas the $G=5$ component MMSGHD, McMSGHD, and MCGHD approaches need less than 25 seconds. For the bankruptcy data set, the MSDBST requires approximately 80 seconds when $G=3$, whereas the hyperbolic distributions need less than 20 seconds each. 

\section{Discussion}\label{sec:discuss}

Novel MCGHDs, MMSGHDs, and McMSGHDs models have been introduced and applied for model-based clustering. The GHD is a flexible distribution, capable of handling skewness and heavy tails, and has many well known distributions as special or limiting cases. Furthermore, it is a normal variance-mean mixture, arising via a relationship between a multivariate Gaussian and an univariate GIG distribution. The MSGHD extends the GHD to include a multivariate GIG distribution, increasing the flexibility of the model. However, the GHD is not a special case of the MSGHD; hence, we created  MCGHDs, which has both the GHD and MSGHD as special cases. 
The McMSGHD approach was introduced as a convex version of the MMSGHD, and this point deserves some further discussion. The extension of the multivariate-$t$ distribution to multiple scale was carried out by \cite{forbes14}; as discussed in the Appendix~\ref{app:convex}, the multiple scaled multivariate $t$-distribution cannot be quasi-concave, i.e., the clusters associated with a mixture of multiple scaled multivariate $t$-distributions cannot be convex. We have seen examples where the MMSGHD can put multiple clusters into one component, and the McMSGHD has an important role in helping to prevent this; if both approaches are fitted and lead to different numbers of components, then further attention is warranted. 

Comparing the MGHD, MMSGHD, McMSGHD, and MCGHD approaches yielded some interesting results. Amongst them, we see that the MMSGHD does not necessarily outperform the MGHD; far from it, in fact, only the MGHD approach gave better clustering performance than the MMSGHD approach on one of the four real data sets we considered, as well as identical performance on a other two. This underlines the fact that a mixture of multiple scaled distributions may well not outperform its single scaled analogue, and highlights the benefit of approaches with both a multiple scaled distribution and its single scaled analogue as special cases. The MCGHDs represent one such approach, with the MSGHD and MGHD models as special cases. The approaches introduced herein, as well as the MGHDs, have been made freely available via the {\tt MixGHD} package for {\sf R}.

Future work will focus in several directions. For one, it will be interesting to study the performance of the approaches introduced herein within the fractionally-supervised classification framework \citep{vrbik15,gallaugher18b}. The extension of these approaches to the matrix-variate paradigm is also of interest and may proceed in an analogous fashion to the work of \cite{gallaugher18a,gallaugher18c}. Finally, the models introduced herein can be extended to account for missing data \citep[see][]{wei19}.

{\small
\subsection*{Acknowledgements}
This work was supported by a grant-in-aid from Compusense Inc., a Collaborative Research and Development Grant from the Natural Sciences and Engineering Research Council of Canada, and an Early Researcher Award from the Ontario Ministry of Research and Innovation.

\appendix


\section{Parameter estimation}\label{sec:ParaEst}

We use the EM algorithm to estimate the parameters of the MCGHDs. The EM algorithm belongs to a larger class of algorithms known as MM algorithms \citep{ortega70,hunter00} and is well-suited for problems involving missing data. `MM' stands for `minorize-maximize' or `majorize-minimize,' depending on the purpose of the algorithm; in the EM context, the minorizing function is the expected value of the complete-data log-likelihood. The EM algorithm iterates between two steps, an E-step and a M-step, and has been used to estimate the parameters of mixture models in many experiments \citep{mclachlan08}.
On each E-step, the expected value of the complete-data log-likelihood, $\mathcal{Q}$, is calculated and on each M-step is maximized with respect to $\pi_g, \vecmu_g,\vecPhi_g,\vecalpha_g,\vecomegaP_{g},\veclambda_{g},\omega_{0g},\lambda_{0g},\varpi_g$. However, in each M-step $\mathcal{Q}$ increases with respect to $\gam_g$ rather than maximize; accordingly, the algorithm is formally a generalized EM (GEM) algorithm.
For our MCGHDs, there are four sources of missing data: the latent variable $W_{0ig}$, the multi-dimensional weight variable $\vecDelta_{\vecw ig}$, the group component indicator labels $z_{ig}$, and inner component labels $u_{ig}$, for~$i=1,\ldots,n$ and $g=1,\ldots,G$. 
For each observation $i$, $z_{ig} = 1$ if observation $i$ is in component $g$ and $z_{ig}=0$ otherwise. Similarly, for each observation $i$, $u_{ig} = 1$ if observation $i$, in component $g$, is distributed generalized hyperbolic and $u_{ig} = 0$ if observation $i$, in component $g$, is distributed multiple scaled generalized hyperbolic. It follows that the complete-data log-likelihood for the MCGHDs is given by
\begin{equation*}\begin{split}
l_c& = \sum_{i=1}^n  \sum_{g=1}^G \bigg\{ z_{ig} \log \pi_g   + z_{ig}{u_{ig} } \log \varpi_g  + z_{ig}(1 - u_{ig}) \log (1-\varpi_g)+   z_{ig}{u_{ig} } \log h \left(  w_{0ig}~|~\omega_{0g}, 1,   \lambda_{0g}  \right)\\
& +  z_{ig}(1 - u_{ig})  \sum_{j=1}^p\log h \left(  w_{jig}~|~\omega_{jg}, 1,   \lambda_{jg}  \right)+ z_{ig}{u_{ig} } \log  \phi_p \left(\gam_g'\vecx_i~|~\vecmu_g + w_{0ig} \vecalpha_g , w_{0ig} \vecPhi \right)\\
&
+ z_{ig}(1 - u_{ig})  \sum_{j=1}^p \log \phi_1 \left( [\gam_g'\vecx_i ]_j~|~\mu_{jg} + w_{jig} \alpha_{jg} ,     \omega_{jg} \phi_{jg} \right)\bigg\},
\end{split}\end{equation*}
%
%
%
where  $ \phi_p(\cdot)$ represents a $p$-dimensional Gaussian density function, $\phi_1(\cdot) $ is a unidimensional  Gaussian density function,  and $h(\cdot)$ is the density of a GIG distribution given in \eqref{eq:gig_density}.

We are now prepared to outline the calculations for our GEM algorithm for the MCGHDs. On the E-step, the expected value of the complete-data log-likelihood,$~\mathcal{Q}$,
is computed by replacing the sufficient statistics of the missing data by their expected values. For each component indicator label $z_{ig}$ and inner component label $u_{ig}$, for $i=1,\ldots, n$ and $g=1,\ldots,G$, we require the expectations
\begin{equation} \label{exweights}
\mathbb{E}\left[ Z_{ig}\mid \vecx_i \right] = \frac{\pi_gf_{\text{\tiny CGHD}}\left(\vecx\mid\vecmu_g,\gam_g,\vecPhi_g,\vecalpha_g,\vecomegaP_{g},\veclambda_{g},\omega_{0g},\lambda_{0g},\varpi_g\right)}{\sum_{h=1}^G\pi_hf_{\text{\tiny CGHD}}\left(\vecx\mid\vecmu_h,\gam_h,\vecPhi_h,\vecalpha_h,\vecomegaP_{h},\veclambda_{h},\omega_{0h},\lambda_{0h},\varpi_h\right)} \equalscolon \zig
\end{equation}
and 
 \begin{equation}\label{exweights}\begin{split}
&\mathbb{E}\left[ U_{ig}\mid \vecx_i, z_{ig}=1 \right]=\\
& \qquad \frac{\varpi _gf_{\text{\tiny GHD}}(\vecx\mid\vecmu_g,\gam_g\vecPhi_g\gam_g', \vecalpha_g,\omega_{0g},\lambda_{0g})}{\varpi _gf_{\text{\tiny GHD}}(\vecx\mid\vecmu_g,\gam_g\vecPhi_g\gam_g', \vecalpha_g,\omega_{0g},\lambda_{0g})+ (1-\varpi _g)f_{\text{\tiny MSGHD}}\left(\vecx\mid\vecmu_g,\gam_g,\vecPhi_g,\vecalpha_g,\vecomegaP_g,\veclambda_g\right)}\equalscolon\hat{u}_{ig} ,
\end{split}\end{equation}
where $f_{\tiny\text{CGHD}}$ is given in~\eqref{eq:cghd_dens}, $f_{\tiny\text{GHD}}$ is given in~\eqref{dist ghy} and $f_{\tiny\text{MSGHD}}$ is given in~\eqref{eq:GHNMVM2}. 
For the latent variable $W_{0ig}$, we use the expected value given in \cite{browne15}. The authors show that, given the density in~\eqref{eq:gig_density}, the following is true $$W_{0ig}\mid\vecx_i, z_{ig}=1, u_{ig}=1 \backsim\mathcal{GIG}\left( \omega_{0g}+\vecalpha_g'(\gam_g \vecPhi_g\gam_g')^{-1}\vecalpha_g,\omega_{0g}+\delta(\vecx_i,\vecmu_g\mid\gam_g \vecPhi_g\gam_g'),\lambda_{0g}-p/2\right).$$ For the MCGHDs, the maximization of $\mathcal{Q}$ requires the expected values of  $W_{0ig}$, $W^{-1}_{0ig}$ and $\log W_{0ig}$, i.e.,
 \begin{eqnarray}
&&{\mathbb E}[W_{0ig}~|~\vecx_i,z_{ig}=1,u_{ig}=1]= \sqrt{\frac{e_{ig}}{d_g}}\frac{K_{\lambda_{0g}-p/2+1}\left(\sqrt{d_ge_{ig}}\right)}{K_{\lambda_{0g}-p/2}\left(\sqrt{d_ge_{ig}}\right)}
\equalscolon a_{ig},\nonumber\\
&&{\mathbb E}[W_{0ig}^{-1}~|~\vecx_i,z_{ig}=1,u_{ig}=1]= \sqrt{\frac{d_g}{e_{ig}}}\frac{K_{\lambda_{0g}-p/2+1}\left(\sqrt{d_ge_{ig}}\right)}{K_{\lambda_{0g}-p/2}\left(\sqrt{d_ge_{ig}}\right)}-\frac{2\lambda_{0g}-p}{ e_{ig}}
\equalscolon b_{ig},\nonumber\\
&&{\mathbb E}[\log W_{0ig}~|~\vecx_i,z_{ig}=1,u_{ig}=1]=\log\sqrt{\frac{e_{ig}}{d_g}}+\left.\frac{\partial}{\partial v} \log \left\{K_{v}\left(\sqrt{d_ge_{ig}}\right)\right\}\right\vert_{v=\lambda_{0g}-p/2} \equalscolon c_{ig}\nonumber,
\end{eqnarray}
where  $d_g= \omega_{0g}+\vecalpha_g' (\vecGam_g\vecPhi_g\vecGam'_g)^{-1} \vecalpha_g$ and $e_{ig}=  \omega_{0g}+\delta(\vecx_i, \vecmu_g \mid \vecGam_g\vecPhi_g\vecGam'_g)$.

The maximization of $\mathcal{Q}$ also requires the expected values of the multidimensional weight variables $\vecDelta_{\vecw ig}$, $\vecDelta_{\vecw ig}^{-1}$, and $\log\vecDelta_{\vecw ig}$.  Given the density in~\eqref{eq:GHNMVM2}, 
it follows that $$W_{ijg}~|~\vecx_i, z_{ig}= 1, u_{ig} =0\backsim\mathcal{GIG}\left(  \omega_{jg} + \alpha_{jg}^2\Phi_{jg}^{-1},\omega_{jg} + ({ \left[\gam'\vecx\right]_j} -\vecmu_{gj} )^2/\phi_{jg}, \lambda_{jg}-1/2\right).$$ 
Each multidimensional weight variable is replaced by its 
expected value and so we need to compute
${\mathbf E}_{1ig}=\diag\{E_{1i1g}, \ldots, E_{1ipg}\}$, ${\mathbf E}_{2ig}=\diag\{E_{2i1g}, \ldots, E_{2ipg}\}$, and ${\mathbf E}_{3ig}=\diag\{E_{3i1g}, \ldots, E_{3ipg}\}$, where
\begin{equation}\begin{split}\label{eq:MSexp_vals}
&\mathbb{E}[W_{ijg}\mid\vecx_i,z_{ig}=1,u_{ig}=0] = \sqrt{\frac{\bar{e}_{ijg}}{\bar{d}_{jg}}}\frac{K_{\lambda_{jg}+1/2}\left( \sqrt{\bar{d}_{jg} \bar{e}_{ijg}}\right)}{K_{\lambda_{jg}-1/2}\left( \sqrt{\bar{d}_{jg} \bar{e}_{ijg}}\right)}\equalscolon E_{1ijg},\\
&\mathbb{E}[W_{ijg}^{-1}\mid\vecx_i,z_{ig}=1,u_{ig}=0] =  \sqrt{\frac{\bar{d}_{jg}}{\bar{e}_{ijg}}}\frac{K_{\lambda_{jg}+1/2}\left( \sqrt{\bar{d}_{jg} \bar{e}_{ijg}}\right)}{K_{\lambda_{jg}-1/2}\left( \sqrt{\bar{d}_{jg} \bar{e}_{ijg}}\right)} -\frac{2\lambda_{jg}-1}{\bar{e}_{ijg}}\equalscolon E_{2ijg},\\
&\mathbb{E}[\log W_{ijg}\mid\vecx_i,z_{ig}=1,u_{ig}=0] =  \log \sqrt{\frac{\bar{e}_{ijg}}{\bar{d}_{jg}}}+  \frac{\partial}{\partial v} \left. \log \left\{ K_{v}\left( \sqrt{\bar{d}_{jg} \bar{e}_{ijg}}\right)\right\} \right\vert_{v=\lambda_{jg}-1/2} \equalscolon E_{3ijg},
\end{split}\end{equation}
$\bar{d}_{jg} =  \omega_{jg} + \alpha_{jg}^2\Phi_{jg}^{-1}$ and $\bar{e}_{ijg} =\omega_{jg} + ( [\vecx_i -\vecmu_g]_j )^2/\phi_{jg}$. 
Let $n_g=\sum_{i=1}^n \hat z_{ig}$, $A_{g}=(1/n_g)\sum_{i=1}^n \hat z_{ig}a_{ig}$, $B_{g}=(1/n_g)\sum_{i=1}^n \hat z_{ig}b_{ig}$, $C_{g}=(1/n_g)\sum_{i=1}^n \hat z_{ig}c_{ig}$,
${\bar{E}}_{1jg}=(1/n_g)\sum_{i=1}^n \hat z_{ig}{ E}_{1ijg}$, ${\bar{E}}_{2jg}=(1/n_g)\sum_{i=1}^n \hat z_{ig}{ E}_{2ijg}$, and ${\bar{E}}_{3jg}=(1/n_g)\sum_{i=1}^n \hat z_{ig} { E}_{3ijg}$.

In the M-step, we maximize the expected value of the complete-data log-likelihood with respect to the model 
parameters. The mixing proportions and inner mixing proportions are updated via
$\hat{\pi}_g=n_g/n$ and $\hat{\varpi}_g={\sum_{i=1}^n \hat{u}_{ig} \hat{z}_{ig}}/{n_g}$, respectively.
The elements of the location parameter $\vecmu_g$ and skewness parameter $\vecalpha_g$ are replaced with
\begin{equation*}\begin{split}
\hat{\mu}_{jg} = \frac{ \sum_{i=1}^n \hat{z}_{ig}[\vecGam_g' \vecx_i]_j( \bar{s}_{1jg} s_{2ijg}-1)}{\sum_{i=1}^n \hat{z}_{ig} (\bar{s}_{1jg}s_{2ijg}-1)}
\quad\text{and}\quad
\hat{\alpha}_{jg} = \frac{ \sum_{i=1}^n \hat{z}_{ig}[\vecGam_g' \vecx_i]_j( \bar{s}_{2jg}- s_{2ijg})}{\sum_{i=1}^n \hat{z}_{ig} (\bar{s}_{1jg}s_{2ijg}-1)},
\end{split}\end{equation*}
respectively, where $[\vecGam_g' \vecx_i]_j$ is the $j$th element of the matrix $\vecGam_g' \vecx_i$,
$s_{1ijg}= \hat{u}_{ig}a_{ig}+\left(1- \hat{u}_{ig} \right) { E}_{1ijg}$,  $s_{2ijg} =\hat{u}_{ig}b_{ig}+\left(1- \hat{u}_{ig} \right) { E}_{2ijg}$, $\bar{s}_{1jg}=1/n_g\sum_{i=1}^n \hat{z}_{ig}s_{1ijg}$,$\bar{s}_{2jg}=1/n_g\sum_{i=1}^n \hat{z}_{ig}s_{2ijg}$.
The diagonal elements of the matrix  $\vecPhi_g$ are updated using
\begin{eqnarray*}
\hat{\phi}_{jg}  &=& \frac{1}{n_g}  \sum_{i=1}^n \left\{ \hat{z}_{ig} \hat{u}_{ig} \left[ b_{ig} \left(  [\vecGam_g' \vecx_i]_j - \hat{\mu}_{jg} \right)^2 -2 \left( [\vecGam_g' \vecx_i]_j - \hat{\mu}_{jg} \right) \hat{\alpha}_{jg} + a_{ig} \hat{ \alpha}_{jg}^2 \right]  \right. \\ 
& & \qquad\qquad\qquad \left. + \hat{z}_{ig}(1- \hat{u}_{ig} ) \left[  E_{2ijg} \left(  [\vecGam_g' \vecx_i]_j - \hat{\mu}_{jg} \right)^2-2 \left(  [\vecGam_g' \vecx_i]_j - \hat{\mu}_{jg} \right) \hat{\alpha}_{jg} + E_{1ijg}   \hat{\alpha}_{jg}^2 \right] \right\}.
\end{eqnarray*}
To update the component eigenvector matrices $\gam_g$, 
we wish to minimize the objective function
\begin{align}\label{eq:objfunc}
f&(\vecGam_g) = -\frac{1}{2} \tr \left\{ \zig \hat{\vecPhi}_{g}^{-1}\mathbf V_{ig}   \vecGam_g \vecx_i \vecx_i \vecGam_g'  \right\}  +  \tr \left\{ \zig \vecx_i \left(\mathbf V_{ig}  \hat{\vecmu}_g + \hat{\vecalpha}_g \right)' \hat{\vecPhi}_{g}^{-1} \vecGam_g  \right\} + C 
\end{align}
\noindent
with respect to $\gam_g$, where $\mathbf V_{ig} =  \hat{u}_{ig} b_{ig}\ident_p + (1-\hat{u}_{ig}){\bf E}_{2ig}$. 
We employ an optimization routine that uses two simpler majorization-minimization algorithms. Our optimization routine exploits the convexity of the objective function in~\eqref{eq:objfunc}, providing a computationally stable algorithm for estimating $\vecGam_g$. Specifically, we follow \cite{kiers02} and \cite{browne14a} and use the surrogate function
\begin{align}\label{eq:surofunc}
f(\vecGam_g)\leq C+\sum_{i=1}^n{\tr{\left\{\mathbf{F}_{rg}\vecGam_g\right\}}},
\end{align}
where $C$ is a constant that does not depend on $\vecGam_g$, $r\in\{1,2\}$ is an index, and the matrices $\mathbf{F}_{rg}$ are defined in~\eqref{eq:f1} and~\eqref{eq:f2}.

Therefore, on each $M$-step, we calculate either
\begin{equation}\label{eq:f1} 
\mathbf{F}_{1g} = \sum_{i=1}^n \zig\left[ - \vecx_i \left(\mathbf V_{ig}  \hat{\vecmu}_g + \hat{\vecalpha}_g \right)' \hat{\vecPhi}_{g}^{-1} 
+ \vecx_i \vecx_i'  \vecGam_g'  \hat{\vecPhi}_{g}^{-1}\mathbf V_{ig} - \alpha_{1ig}  \vecx_i \vecx_i'  \vecGam_g' 
\right]
\end{equation}
or
\begin{equation}\label{eq:f2}
\mathbf{F}_{2g} = \sum_{i=1}^n \zig\left[ -\vecx_i \left(\mathbf V_{ig}  \hat{\vecmu}_g + \hat{\vecalpha}_g \right)' \hat{\vecPhi}_{g}^{-1} 
+  \vecx_i \vecx_i'  \vecGam_g'  \hat{\vecPhi}_{g}^{-1}\mathbf V_{ig} - \alpha_{2ig} \mathbf V_{ig}  \hat{\vecPhi}_{g}^{-1} \vecGam_g' 
\right], 
\end{equation}
\noindent
where $\alpha_{1ig}$ is the largest eigenvalue of the diagonal matrix $ \vecPhi_{g}^{-1}\mathbf V_{ig}$, and $\alpha_{2ig}$ is equal to $\zig\vecx_i' \vecx_i$, which is the largest  eigenvalue of the rank-1 matrix $\zig\vecx_i \vecx_i'$. Following this, we compute the singular value decomposition of $\mathbf{F}_{rg}$ given by 
$$\mathbf{F}_{rg} =  \mathbf{P}\mathbf{B}\mathbf{R}'.$$
 It follows that our update for $\vecGam_g$ is given by
$$\hat{\gam}_g = \mathbf{R}\mathbf{P}'.$$ 

The  $p$-dimensional concentration and index parameters, i.e., $\vecomegaP_g$ and $\veclambda_g$, are estimated by maximizing the function
\begin{equation}
q_{jg}(\omega_{jg}, \lambda_{jg})=-\log K_{\lambda_{jg}}(\omega_{jg})+(\lambda_{jg} -1){\bar{E}}_{3jg}- \frac{\omega_{jg}}{2}({\bar{E}}_{1jg}+{\bar{E}}_{2jg}).
\end{equation}
This leads to
\begin{equation*}
\hat{\lambda}_{jg}= {\bar{E}}_{3jg}\lambda_{jg}^{\prev}\left[\left.\frac{\partial }{\partial v}\log K_{v}(\omega_{jg}^{\prev})\right\vert_{v=\lambda_{jg}^{\prev}}\right]^{-1}
\end{equation*}
and
\begin{equation*}
\hat{\omega}_{jg}= \omega_{jg}^{\prev}-\left[\left.\frac{\partial  }{\partial v}q_{jg}(v, {\hat\lambda_{jg}})\right\vert_{v=\omega_{jg}^{\prev}}\right]\left[\left.\frac{\partial^2  }{\partial v^2}q_{jg}(v, {\hat\lambda_{jg}})\right\vert_{v=\omega_{jg}^{\prev}}\right]^{-1},
\end{equation*}
{where the superscript ``prev'' denotes that the estimate from the previous iteration is used.}
The univariate parameters $\omega_{0g}$ and $\lambda_{0g}$ are estimated by maximizing the function
\begin{eqnarray}
q_{0g}(\omega_{0g}, \lambda_{0g})=-\log(K_{\lambda_{0g}}(\omega_{0g}))+(\lambda_{0g} -1)C_g- \frac{\omega_{0g}}{2}(A_g+B_g),
\end{eqnarray}
giving
\begin{equation*}
\hat\lambda_{0g}= C_g\lambda_{0g}^{\prev}\left[\left.\frac{\partial }{\partial v}\log K_{v}(\omega_{0g}^{\prev})\right\vert_{v=\lambda_{0g}^{\prev}}\right]^{-1}\qquad
\end{equation*}
and
\begin{equation*}
\hat \omega_{0g}= \omega_{0g}^{\prev}-\left[\left.\frac{\partial}{\partial v}q_{0g}(v, {\hat\lambda_{0g}})\right\vert_{v=\omega_{0g}^{\prev}}\right]\left[\left.\frac{\partial^2  }{\partial v^2}q_{0g}(v, {\hat\lambda_{0g}})\right\vert_{v=\omega_{0g}^{\prev}}\right]^{-1}. \nonumber
\end{equation*}

Our GEM algorithm is iterated until convergence, which 
is determined using the Aitken acceleration \citep{aitken26}. Formally, the Aitken acceleration is given by
\begin{eqnarray*}
a^{(k)}=\frac{l^{(k+1)}-l^{(k)}}{l^{(k)}-l^{(k-1)}},
\end{eqnarray*}
where $l^{(k)}$ is the value of the log-likelihood at the iteration $k$ and 
 \begin{eqnarray*}
l^{(k+1)}_\infty=l^{(k)}+\frac{1}{1-a^{(k)}}\left (l^{(k+1)}-l^{(k)}\right ),
\end{eqnarray*}
is an asymptotic estimate of the log-likelihood on iteration $k+1$. The algorithm can be considered to have converged when $l^{(k)}_{\infty}-l^{(k)}< \epsilon$, provided this difference is positive \citep{bohning94,lindsay95,mcnicholas10a}. Herein, we set $\epsilon=0.01$.
When the algorithm converges we compute the maximum \textit{a posteriori} (MAP) classification values using  the posterior $\zig$, where $\text{MAP}\left\{\zig\right\}=1$ if $g=\arg\max_h\left\{\hat{z}_{ih}\right\}$, and $\text{MAP}\left\{\zig\right\}=0$ otherwise.


\section{Quasi-Concavity of the cMSGHD}\label{app:convex}

In essence, we might want to consider only densities whose contours contain a set of points that are convex. Formally, such densities are quasi-concave. Extensive details on quasi-concavity, quasi-convexity, and related notions are given by \cite{niculescu06} and \cite{rockafellar09}. 

\begin{definition} A function $f(\vecx)$ is quasi-concave if each upper-level set $U_\alpha(f) = \{\vecx~|~f(\vecx) \geq \alpha\}$ is convex, for $\alpha\in\mathbb{R}$.
\end{definition}

\begin{definition}
A function $f(\vecx)$ is quasi-convex if each sub-level set $S_\alpha(f) = \{\vecx~|~f(\vecx) \leq \alpha\}$ is convex, for $\alpha\in\mathbb{R}$.
\end{definition}

\begin{lemma}
The class of elliptical distributions, whose density functions have the form $$f(\vecx) = \frac{1}{\sqrt{\lvert\matsig\rvert}}g\left( \delta\left(\vecx, \vecmu~|~\matsig\right) \right)$$ 
are quasi-concave if the generator function, $g$, is monotonic non-increasing.
\end{lemma}
\begin{proof}
Result follows from the fact that if the function $\delta\left(\vecx, \vecmu~|~\matsig\right) $ is convex since $ \matsig$ is positive definite and the function $g$ is monotonic non-increasing, then the function $f(\vecx)$ is quasi-concave.\qed
\end{proof}

\begin{theorem}\label{them:ghd}
The generalized hyperbolic distribution (GHD) is quasi-concave.
\end{theorem}
\begin{proof}
It is straightforward to show that the function
$$h(\vecx) = \sqrt{ a + b\; \delta\left(\vecx, \vecmu~|~\matsig\right)}$$ 
is convex, where $a$ and $b$ are positive constants, and $\delta\left(\vecx, \vecmu~|~\matsig\right)$ is the Malahanobis distance between $\vecx$ and $\vecmu$. Let $\tau =\lambda-p/2$. 
Then, the function 
$$k(z) = \tau \log z + \log K_\tau(z),$$ 
where $z\in\mathbb{R}^+$, and $K_\tau$ is the modified Bessel function of the third kind with index $\tau$, is monotonic decreasing (or non-increasing) because the first derivative
$$k'(z) = \frac{\tau}{z}  + \frac{  (\tau/z) K_{\tau}(z) - K_{\tau+1}(z) }{K_{\tau}(z)} = \frac{2\tau}{z}  - \frac{  K_{\tau+1}(z) }{K_{\tau}(z)}  =  -\frac{K_{\tau-1}(z)}{K_{\tau}(z)}$$ 
 is negative for all $\tau \in \mathbb{R}$ and $z >0$.  In addition to being monotonic decreasing, $k(z)$ is convex for $\tau < 1/2$, concave and convex (linear) for $\tau = 1/2$, and concave for $\tau > 1/2$. 
Because $k(z)$ is a monotonic function, it satisfies the criteria for quasi-convexity and quasi-concavity, so it is simultaneously  quasi-convex and quasi-concave. In this context, monotone functions are also known as quasi-linear or quasi-montone. 

Recall that if the function $U$ is quasi-convex and the function $g$ is decreasing, then the function $f(x) = g(U(x))$ is quasi-concave. It follows that the composition $k( h(\vecx) )$ is quasi-concave. Consider the skewness part of the GHD density function, i.e, $a(\vecx) = -\left(\vecx-\vecmu\right)'\matsig^{-1}\vecalpha$, which is a linear function. It follows that the function 
\begin{equation}\label{propto}
\exp\left\{k(h(\vecx))+a(\vecx)\right\}
\end{equation} 
is also quasi-concave, and the result follows from the fact that \eqref{propto} is proportional to the density of the GHD.
\qed
\end{proof}

\begin{theorem}
The convex multiple scaled generalized hyperbolic distribution (cMSGHD) is quasi-concave. In other words, the multiple scaled generalized hyperbolic distribution (MSGHD) is quasi-concave provided that $\lambda_j>1$ for all $j=1,\ldots,p$.
\end{theorem}
\begin{proof}
A $p$-dimensional multiple scaled distribution is a product of $p$ independent univariate densities. 
The density of the MSGHD has form 
$$g_p(x_1,x_2,\ldots,x_p) = g_1(x_1~|~\vectheta_1) g_1(x_2~|~\vectheta_2)\times\cdots\times g_1(x_p~|~\vectheta_p),$$
where $g_1(x_j~|~\vectheta_j)$ is the density of the univariate hyperbolic distribution with parameters $\vectheta_j$, $j=1,\ldots,p$. 
From Theorem~\ref{them:ghd}, $\log g_1(x_j~|~\vectheta_j)$ is a concave function for $\tau_j > 1/2$, i.e., for $\lambda_j > 1$ (because $p=1$). Therefore, the function $$\log g_p(x_1,x_2,\ldots,x_p) = \log g_1(x_1~|~\vectheta_1) + \log g_1(x_2~|~\vectheta_2)+\cdots+ \log g_1(x_p~|~\vectheta_p)$$ is concave provided that $\lambda_j>1$ for all $j=1,\ldots,p$. Therefore, the function 
$$g_p(x_1,x_2,\ldots,x_p) = g_1(x_1~|~\vectheta_1) g_1(x_2~|~\vectheta_2)\times\cdots\times g_1(x_p~|~\vectheta_p)$$ is quasi-concave provided that $\lambda_j>1$ for all $j=1,\ldots,p$. \qed
\end{proof}

Note that addition does not preserve quasi-convexity or quasi-concavity. The sum of two quasi-convex functions defined on different domains 
will be quasi-concave if they are additively decomposed \citep[see][]{debreu82}. \cite{debreu82} give necessary and sufficient conditions for the sum $f$ of a set of functions $f_1,\ldots,f_m$ to be additively decomposed. These conditions depend on the convexity index $c(f)$ in which $f$ is quasi convex if and only if either of the following hold:
(i) $c(f_i) \ge 0$ for every $i$, or 
(ii) $c(f_j) <0$ for some $j$, $c(f_i) >0$ for every $i\neq j$, and $\sum_{i=1}^m \frac{1}{c(f_i)} \le 0.$
For differentiable functions, the convexity index satisfies the inequality $f''(x)/[ f'(x) ]^2 \ge c(f)$. 

We have that a sufficient condition for the MSGHD to be quasi-concave is that all  $\lambda_j > 1$. Furthermore, a sufficient condition for the MSGHD not to be quasi-concave is that all $\lambda_j < 1$ and finite. 
Interestingly, {this means} the multiple scaled t-distribution cannot provide convex level sets for any finite degrees of freedom. For large degrees of freedom, the multiple scaled t-distribution will behave similarly to a normal distribution near the mode; however, as one moves away from the mode, non-convex contours will be encountered. Finally, note that \cite{debreu82} give necessary and sufficient conditions that suggest a quasi-concave MSGHD with some $\lambda_j$ positive and others negative is possible, but going this route would greatly complicate the estimation procedure.


\section{Finite Mixture Identifiability}\label{sec:PE}


In this section we consider the notion of identifiability for finite mixtures of MSGHDs and coalesced generalized hyperbolic distributions (CGHDs). Herein, we take the term identifiability to mean finite mixture identifiability.

\subsection{Background}

\cite{holzmann2006} prove identifiability of finite mixtures of elliptical distributions. They state that ``finite mixtures are said to be identifiable if distinct mixing distributions with finite support correspond to distinct mixtures''. 
A finite mixture of the densities $f_{p}(\vecx|\noisev_1),\ldots,f_{p}(\vecx|\noisev_G)$ is identifiable if the family $\left\{ f_{p}(\vecx|\noisev) : \noisev \in  \mathcal{A}^p \right\}$ is linearly independent. The founding work on finite mixture identifiability is by \cite{yakowitz1968}, who state that this linear independence is a necessary and sufficient condition for identifiability. 


The GHD can be expressed as a normal variance-mean mixture. The stochastic relationship of the normal variance-mean mixture is given by 
\begin{equation} \label{mean var mixture}
\vecX =\vecmu + W\vecalpha+ \sqrt{W} \mathbf{U},
\end{equation}
where $\mathbf{U} \backsim \mathcal{N}_p(\mathbf{0}, \matsig)$ and $W$, independent of $\mathbf{U}$, is a positive univariate random variable with density $h(w|\vectheta)$. 
\cite{browne15} proved identifiability for finite mixtures of GHDs through additivity of disjoint sets of identifiable distributions. 


\begin{definition}
In the present context, a finite mixture of the multiple scale distributions $f(\vecx | \vectheta_1),\ldots,f(\vecx | \vectheta_G)$ is
identifiable if 
\begin{equation} \label{def identifiable}
\sum_{g=1}^G \pi_g f\left({\vecx}|\vectheta_g \right) = \sum_{g=1}^G \pi_g^{\star} f\left({\vecx}|\vectheta_g^{\star} \right)
\end{equation}
for $\vecx \in \mathbb{R}^p$, where $G$ is a positive integer, $\sum_{g=1}^G \pi_g = \sum_{g=1}^G \pi_g^{\star} = 1$ and $\pi_g, \pi_g^{\star} > 0$ for $g=1,\ldots,G$, implies that there exists a permutation $\sigma$ such that $(\pi_g,\vectheta_g)=(\pi_{\sigma(g)}, \vectheta_{\sigma(g)})$ for all $g$. 
\end{definition}

\cite{browne15} prove identifiability for normal variance-mean mixtures, which includes the generalized hyperbolic. Here we view the results from a different vantage point to illustrate the concepts required for the identifiability of the multiple scaled distributions. We begin by noting the characteristic function for the generalized hyperbolic arises from the characteristic function of the normal variance-mean mixture,
\begin{equation} 
\varphi_{\vecX}( \vecv) =  \exp \left\{ i  \vecv'\vecmu_g \right\} 
M_{W} \left( \Beta_g' \vecv i -\frac{1}{2} \vecv' \matsig_g \vecv~\Bigg|~\gam_g \right), 
\end{equation}
where 
\begin{equation*}
M_{W} \left( u \right) 
 = \left[ \frac{\omega}{\omega -2u }  \right]^{ \frac{\lambda}{2} }
 \frac{ K_{\lambda} \left( \sqrt{ \omega(\omega-2u) } \right) }{   K_{\lambda} \left( \omega \right) }
  = \left[ 1 -2 \frac{u}{ \omega}  \right]^{- \frac{\lambda}{2} }
 \frac{ K_{\lambda} \left( \sqrt{ \omega(\omega-2u) } \right) }{   K_{\lambda} \left( \omega \right) }.
\end{equation*}
The characteristic function for the generalized hyperbolic is 
\begin{equation*}
\varphi_{\vecX}( \vecv ) = 
\exp\{ i  \vecv'\vecmu\} 
\left[ 1 + \frac{  \vecv' \matsig \vecv -2 i \Beta' \vecv   }{\omega}   \right]^{ -\frac{\lambda}{2} }
\frac{  K_{\lambda} \left( \sqrt{ \omega \left[\omega + ( \vecv' \matsig \vecv - 2 i \Beta' \vecv   )  \right] } \right)  }{  K_{\lambda} \left(  \omega \right) }.
\end{equation*}
In context of a coalesced distribution, with a eigen-decomposed scale matrix, the characteristic function is
\begin{equation*}
\varphi_{\vecX}( \vecv ) = 
\exp\{ i  \vecv'\vecmu\} 
\left[ 1 + \frac{  \vecv' \gam \vecPhi \gam' \vecv -2 i \Beta' \vecv   }{\omega}   \right]^{ -\frac{\lambda}{2} }
\frac{  K_{\lambda} \left( \sqrt{ \omega \left[\omega + ( \vecv' \gam \vecPhi \gam' \vecv - 2 i \Beta' \vecv   )  \right] } \right)  }{  K_{\lambda} \left(  \omega \right) }.
\end{equation*}
Now, we let  $\vecv = t \vecz$ and obtain
\begin{equation*}
\varphi_{\vecX}( \vecv = t \vecz) = 
\exp\{ i  t \vecz'\vecmu\} 
\left[ 1 + \frac{  t^2 (\vecz' \gam \vecPhi \gam' \vecz) -2 i t (\Beta' \vecz)   }{\omega}   \right]^{ -\frac{\lambda}{2} }
\frac{  K_{\lambda} \left( \sqrt{ \omega \left[\omega +  t^2 (\vecz' \gam \vecPhi \gam' \vecz) - 2 i t (\Beta' \vecz   )  \right] } \right)  }{  K_{\lambda} \left(  \omega \right) }.
\end{equation*}
To prove identifiability of the generalized hyperbolic we could now use the results from \cite{browne15} and Yakowitz and Spragins (1968,p. 211) that implies there exists $\vecz$ such that the tuple $(\vecz' \matsig_g \vecz, \Beta_g' \vecz, \vecz'\vecmu_g)$, where $\matsig_g= \gam_g \vecPhi_g \gam_g'$ is unique for all $g=1,\ldots, G$, allows a reduction to the univariate case. Now, we rewrite the term $\vecz'\matsig_g\vecz$ as 
\begin{equation*}
\vecz' \matsig_g \vecz = \vecz' \gam_g \vecPhi_g \gam_g' \vecz = \tr\left[ \vecz' \gam_g \vecPhi_g \gam_g' \vecz \right] = \tr\left[ \gam_g' \vecz \vecz' \gam_g \vecPhi_g  \right] = \sum_{j=1}^p \Phi_{jg} [\gam_g' \vecz ]_j^2, 
\end{equation*}
which implies the tuple 
\begin{equation*} \left(\vecz' \gam_g \vecPhi_g \gam_g' \vecz, \Beta_g' \vecz, \vecz'\vecmu_g\right) \equiv \left(\sum_{j=1}^p \Phi_{jg} [\gam_g' \vecz ]_j^2, \Beta_g' \vecz, \vecz'\vecmu_g\right) 
\end{equation*}
is unique for all $g=1,\ldots, G$. A similar argument indicates there exists a $\vecz$ such that the tuple
\begin{equation} \label{tuple condition}
\left(  \sum_{j=1}^p \Phi_{jg} | {[\gam_g' \vecz ]_j} | ,  \Beta_g' \vecz, \vecz'\vecmu_g\right)
\end{equation}
is unique. In fact, a more general statement indicates that there exists a $\vecz$ such that the tuple
\begin{equation*}
\left(  \sum_{j=1}^p \Phi_{jg} \varphi(   {[\gam_g' \vecz ]_j^2} ) ,  \Beta_g' \vecz, \vecz'\vecmu_g\right)
\end{equation*}
is unique for monotonic $\varphi : \mathbb{R}^+ \mapsto\mathbb{R}^+ $. Deriving this unique set of tuples facilitates the reduction to the univariate case. This is useful because the univariate generalized hyperbolic density is identifiable \citep[see][]{browne15}.



\subsection{Identifiability of a Finite Mixture of Multiple Scaled Distributions} 

For a multiple scaled distribution, we only need to find a single direction where the distribution is finite mixture identifiable because, as noted in Remark~2 of \cite{kent1983}, a distribution might be non-identifiable on a subset of $\mathbb{R}^p$ but identifiability can endure over $\mathbb{R}^p$. In other words, for a distribution to be non-identifiable, a linear combination has to be equal to zero for all $x \in \mathbb{R}^p$.  This is illustrated by the example given in \cite{kent1983}: \begin{quotation}``the polynomials $P(x_1, x_2) = 1$ and $P(x_1, x_2) = (x_1^2+ x_2)^3$, $x\in \mathbb{R}^2$, are equal on the unit circle, but are not the same on all of $\mathbb{R}^2$.'' \end{quotation} As a consequence, if a multivariate distribution is identifiable in some direction then it is identifiable over  $\mathbb{R}^p$.

To begin, consider that if there is at one least direction or column of $\gam_g$ that is equal across  $g=1,\ldots, G$, then the identifiability of a multiple scaled distribution follows from the identifiability of the univariate distribution. Whereas if one column of $\gam_g$ is unequal, that implies, by the nature of orthonormal matrices, that two columns of $\gam_g$ are unequal. We will now illustrate how the bivariate multiple scaled distribution is identifiable, which implies identifiability for finite $p$.

%

When $\gam_g$ differ, the identifiability of the multiple scaled distribution depends on the behaviour of the multiple scaled distribution's density and moment generating functions when we consider moving along directions other than the columns of $\gam_g$.  For example, a bivariate multiple scaled $t$-distribution behaves (by definition) like a $t$-distribution with $\nu_1$ and $\nu_2$ degrees of freedom along each of it's principal axes, but along any other direction, a bivariate multiple scaled $t$-distribution behaves asymptotically like a $t$-distribution with $\nu_1 + \nu_2$ degrees of freedom. 

Consider the following three orthonormal matrices in the context of an eigen-decomposition of a matrix;
 \begin{equation*}
\gam_1 = 
\left[\begin{array}{cc} 1 & 0 \\0 & 1\end{array}\right],
\quad\quad \mbox{} \quad\quad
\gam_2 = 
\left[\begin{array}{cc} 0 & 1 \\1 & 0\end{array}\right]
\quad\quad \mbox{and} \quad\quad
\gam_3 = 
\left[\begin{array}{cc} -1 & 0  \\ 0 &  1\end{array}\right] .
\end{equation*}
If we have equal eigenvalues then we cannot distinguish between  $\gam_1$ and $\gam_2$. In the same way, if we have the same distribution along the first and second axis, we cannot distinguish between them. However, if we have eigenvalue ordering we can distinguish between $\gam_1$ and $\gam_2$, but eigenvalue ordering will not allow us to distinguish between $\gam_1$ and $\gam_3$, since they yield the same basis or set of directions. Therefore, in general, $\gam$ is unique 
up to multiplication by
\begin{equation*}
\left[\begin{array}{cc} \pm 1 & 0 \\0  & \pm 1\end{array}\right] .
\end{equation*}
One way to establish uniqueness is to require the largest value of each column of $\gam$ to be positive. An equivalent requirement is for $\gam_1 \neq\gam_2$ which requires that 
\begin{equation} \label{orthonormal condition}
\gam_1' 
 \gam_2
 \neq 
\vecR
\quad\quad \mbox{or} \quad\quad
[  \gam_1'\vecz ]_j \neq - [  \gam_2' \vecz ]_j
\end{equation}
for $j=1,\ldots, p$, $\vecz \in \mathbb{R}^P$, $\vecz \neq \mathbf{0}_p$ and $\vecR$ is a set of diagonal matrices such that $\mbox{diag}{\left(\vecR\right)} = (\pm1,\ldots,\pm 1)$ excluding the identity matrix. Note that $[\veca]_j$ denotes the $j$th element of the vector $\veca$. However, if we had two orthonormal  matrices such that $\gam_1'  \gam_2  = \ident$, then $\gam_1 = \gam_2 $. If $\gam_1' \gam_2  = \vecR$, then our orthonormal condition amounts to $ \gam_1' =  \gam_2$ or equivalently, for all directions $\vecz \in \mathbb{R}^P$ and  $\vecz \neq \mathbf{0}_p$
 \begin{equation} \label{orthonormal condition}
| [  \gam_1'\vecz ]_j  | = | [  \gam_2' \vecz ]_j | 
\quad \mbox{for all} \quad  j =1,\ldots,p
\quad \mbox{then} \quad 
\gam_1 = 
 \gam_2.
\end{equation}
This prevents the $j$th column of $\gam_2$ from being in the opposite direction of the $j$th column of $\gam_1$. This form of the condition is easier to incorporate into the identifiability illustration. 


In the MSGHD, if we consider moving the amount $t$  in a direction $\vecz$, which entails setting $\vecx = t \vecz$, we can write the density as  
\begin{align} 
& f_{\text{MSGHD}}\left( \vecx = t \vecz \mid\vecmu,\gam,\vecPhi,\vecalpha,\vecomegaP,\veclambda\right) \nonumber \\
&=\prod_{j=1}^p\left[\frac{\omega_j+ \Phi_j^{-1}\left( t \left[\gam' \vecz\right]_j-\mu_j\right)^{2}}{\omega_j+ \alpha_j^2 {\Phi_j}^{-1}} \right]^{\frac{\lambda_j-\frac{1}{2}}{2}}
\frac{K_{\lambda_j-\frac{1}{2}}\bigg(\sqrt {[\omega_j+\alpha_j^2 {\Phi_j}^{-1}]\left[\omega_j+ \Phi_j^{-1}\left(t \left[\gam' \vecz\right]_j-\mu_j\right)^{2}\right]}\bigg)}
 {(2\pi)^{\frac{1}{2}}{\Phi_j}^{\frac{1}{2}}K_{\lambda_j}(\omega_j)\exp{\{-(t \left[\gam' \vecz\right]_j-\mu_j){\alpha_j} \Phi_j^{-1}\}}},
\end{align}
Note, if $\vecz$ is equal to the $k$th eigenvector, which is the $k$th column of $\gam$, then the density reduces to 
\begin{equation*} 
c_k \left[\frac{\omega_k+ \Phi_k^{-1}\left( t -\mu_k\right)^{2}}{\omega_k+ \alpha_k^2 {\Phi_k}^{-1}} \right]^{\frac{\lambda_k-\frac{1}{2}}{2}}
\frac{K_{\lambda_k-\frac{1}{2}}\bigg(\sqrt {[\omega_k+\alpha_k^2 {\Phi_k}^{-1}]\left[\omega_k+ \Phi_k^{-1}\left(t -\mu_k\right)^{2}\right]}\bigg)}
 {(2\pi)^{\frac{1}{2}}{\Phi_k}^{\frac{1}{2}}K_{\lambda_k}(\omega_k)\exp{\left\{-\left(t-\mu_k\right){\alpha_k} \Phi_k^{-1} \right\}}},
\end{equation*}
where
\begin{equation*} 
c_k = \prod_{j=1, j\neq k}^p\left[\frac{\omega_j+ \Phi_j^{-1} \mu_j^{2}}{\omega_j+ \alpha_j^2 {\Phi_j}^{-1}} \right]^{\frac{\lambda_j-\frac{1}{2}}{2}}
\frac{K_{\lambda_j-\frac{1}{2}}\bigg(\sqrt {[\omega_j+\alpha_j^2 {\Phi_j}^{-1}]\left[\omega_j+ \Phi_j^{-1}\mu_j^{2}\right]}\bigg)}
 {(2\pi)^{\frac{1}{2}}{\Phi_j}^{\frac{1}{2}}K_{\lambda_j}(\omega_j)\exp{\left\{ \mu_j \alpha_j\Phi_j^{-1} \right\}}}.
\end{equation*}
Therefore, the density is simply proportional to 
\begin{equation*} 
\propto \left[\frac{\omega_k+ \Phi_k^{-1}\left( t -\mu_k\right)^{2}}{\omega_k+ \alpha_k^2 {\Phi_k}^{-1}} \right]^{\frac{\lambda_k-\frac{1}{2}}{2}}
\frac{K_{\lambda_k-\frac{1}{2}}\bigg(\sqrt {[\omega_k+\alpha_k^2 {\Phi_k}^{-1}]\left[\omega_k+ \Phi_k^{-1}\left(t -\mu_k\right)^{2}\right]}\bigg)}
 {(2\pi)^{\frac{1}{2}}{\Phi_k}^{\frac{1}{2}}K_{\lambda_k}(\omega_k)\exp{\left\{-\left(t-\mu_k\right){\alpha_k} \Phi_k^{-1} \right\}}} .
\end{equation*}

First, note that if the parameterizations are one-to-one, then if one parameterization is shown to be identifiable, the others are identifiable as well. Similar to  \cite{browne15}, we let $\delta_j = \beta_j / \Phi_j$, $\alpha_j = \sqrt{ \omega_j/\Phi_j + \beta_j^2/\Phi_j^2 }$ and $\kappa_j = \sqrt{\Phi_j \omega_j} $, where $\alpha_j \ge | \delta_j | $. Under this reparameterization, we now have 
\begin{equation} \label{reparameterization}
\Phi_j = \frac{\kappa_j}{\sqrt{\alpha_j^2-\delta_j^2}}, \quad \omega_j =  \kappa_j\sqrt{\alpha_j^2-\delta_j^2} \quad \mbox{ and } \quad\beta_j = \frac{\delta_j \kappa_j}{\sqrt{\alpha_j^2-\delta_j^2}}.
\end{equation}  

For large $z$, the Bessel function can approximated by 
\begin{equation*} 
K_{\lambda} (z) = \sqrt{ \frac{ \pi }{2 z} } e^{-z} \left[ 1+ O\left(\frac{1}{z}\right)\right],
\end{equation*} 
which yields, using the alternative parameterization,
\begin{align} \label{ density large t } 
f( t \mid\vectheta ) 
&\propto  
\left[ 1 + \frac{(t-\mu_j)^2}{\kappa_j^2} \right]^{\lambda_j/2}\exp\left\{ - \alpha_j |t-\mu_j|+\delta_j \left(t-\mu_j\right)
\right\} .
\end{align}

If $\vecz$ is not equal to  the $k$th eigenvector, than, using the reparameterization given in \eqref{reparameterization}, we have
\begin{align} \label{ density large t 2 } 
f( t \mid\vectheta )
&\propto 
\exp\left\{ - \sum_{j=1}^p \alpha_j \left| t \left[\gam' \vecz\right]_j - \mu_j \right|  + \sum_{j=1}^p \delta_j \left( t  \left[\gam' \vecz\right]_j - \mu_j \right)   \right\}
\prod_{j=1}^p\left[ 1 +  \frac{\left( t \left[\gam' \vecz\right]_j-\mu_j\right)^{2}}{\kappa_j^2} \right]^{\frac{\lambda_j-\frac{1}{2}}{2}}  \nonumber\\
&\propto 
\exp\left\{ - \sum_{j=1}^p \alpha_j \left| t \left[\gam' \vecz\right]_j - \mu_j \right|  + \sum_{j=1}^p \delta_j \left( t  \left[\gam' \vecz\right]_j - \mu_j \right)   \right\}
t^{2  \sum_{j=1}^p I\left( \left[\gam' \vecz\right]_j \neq 0 \right) \frac{\lambda_j-\frac{1}{2}}{2}  }  \nonumber\\
 &\propto 
\exp\left\{ \sum_{j=1}^p  \left[ - \alpha_j \left| t \left[\gam' \vecz\right]_j - \mu_j \right|  + \delta_j \left( t  \left[\gam' \vecz\right]_j - \mu_j \right)  + 2 I\left( \left[\gam' \vecz\right]_j  \neq 0 \right) \frac{\lambda_j-\frac{1}{2}}{2} \log (t)  \right] \right\} \nonumber\\
&\propto 
 \prod_{j=1}^p \exp\left\{ - \alpha_j \left| t \left[\gam' \vecz\right]_j - \mu_j \right|  + \delta_j \left( t  \left[\gam' \vecz\right]_j - \mu_j \right)  + 2 I\left( \left[\gam' \vecz\right]_j  \neq 0 \right) \frac{\lambda_j-\frac{1}{2}}{2} \log (t)   \right\}. 
\end{align}

The characteristic function for a multiple scaled distribution can be written as 
\begin{equation*}\begin{split}
\varphi_{\vecX}( \vecv ) = \prod_{j=1}^P
\exp\{ i  | {[\gam' \vecv ]_j}| \mu_j \} 
&\left[ 1 + \frac{ \Phi_j | {[\gam' \vecv ]_j}|^2 -2 \beta_j | {[\gam' \vecv ]_j}| i  }{\omega_j}   \right]^{ -\frac{\lambda_j }{2} }\\&\times
\frac{  K_{\lambda_j } \left( \sqrt{ \omega_j  \left[\omega_j  + ( \Phi_j  | {[\gam' \vecv ]_j}|^2 - 2\beta_j | {[\gam' \vecv ]_j}| i  )  \right] } \right)  }{  K_{\lambda_j } \left(  \omega_j  \right) },
\end{split}\end{equation*}
which, under the alternative parameterization from equation \eqref{reparameterization},  becomes
\begin{equation}\begin{split}
\varphi_{\vecX}( \vecv ) = \prod_{j=1}^P \exp\{ i  | {[\gam' \vecv ]_j}| \mu_j \} 
&\left[ 1 + \frac{ | {[\gam' \vecv ]_j}|^2 - 2 \delta_j | {[\gam' \vecv ]_j}| i }{\alpha_j^2 - \delta_j^2} \right]^{ -\frac{\lambda_j}{2} }\\& \times
\frac{  K_{\lambda_j} \left( \sqrt{  \kappa_j^2 \left[  | {[\gam' \vecv ]_j}|^2 - 2 \delta_j | {[\gam' \vecv ]_j}| i + \alpha_j^2 - \delta_j^2   \right] } \right)  }{  K_{\lambda_j} \left(  \kappa_j \sqrt{\alpha_j^2 - \delta_j^2} \right)}.
\end{split}\end{equation}
Now if we consider moving $t$ in the direction $\vecz$ 
\begin{equation*} \begin{split}
\varphi_{\vecX}( \vecv = t \vecz ) = \prod_{j=1}^P \exp\{ i  t  | {[\gam' \vecz ]_j}| \mu_j \} 
&\left[ 1 + \frac{ t^2 | {[\gam' \vecz ]_j}|^2 - 2 \delta_j t | {[\gam' \vecz ]_j}| i }{\alpha_j^2 - \delta_j^2} \right]^{ -\frac{\lambda_j}{2} }\\
&\times\frac{  K_{\lambda_j} \left( \sqrt{  \kappa_j^2 \left[ t^2 | {[\gam' \vecz ]_j}|^2 - 2 \delta_j t | {[\gam' \vecz ]_j}| i + \alpha_j^2 - \delta_j^2   \right] } \right)  }{  K_{\lambda_j} \left(  \kappa_j \sqrt{\alpha_j^2 - \delta_j^2} \right)},
\end{split}\end{equation*}
and, for large $t$, the characteristic function is 
\begin{align*} 
\varphi_{\vecX}( \vecv = t \vecz ) &\propto  \exp\left\{ i  t  \sum_{j=1}^P | {[\gam' \vecz]_j}| \mu_j - t \sum_{j=1}^P \kappa_j | {[\gam' \vecz ]_j} |  - \log(t) \sum_{j=1}^P \lambda_j I\left(  |{[\gam' \vecz ]_j}| \neq 0 \right)  + O(1)  \right\} \\
& \propto  \exp\left\{ i  t  \; \vecz' \gam \vecmu - t \sum_{j=1}^P \kappa_j |  {[ \gam' \vecz ]_j} |  - \log(t) \sum_{j=1}^P \lambda_j I\left(  | {[\gam' \vecz ]_j}| \neq 0 \right)  + O(1)  \right\} .
\end{align*}
Therefore, from the condition given in \eqref{tuple condition}, there exists $\vecz$ such that the tuple $( \sum_{j=1}^P \kappa_j \left| [ \gam' \vecz ]_j \right|,   \vecz' \gam \vecmu  )$ is unique for all $g=1,\ldots, G$ and reduces to the univariate hyperbolic distribution, which is identifiabile.

\subsection{Identifiability of the Coalesced Generalized Hyperbolic Distribution}

To prove the identifiability of the CGHD we only need to show that two sets of distributions, the multiple scaled and the generalized hyperbolic distribution are disjoint. Consider moving along the $k$th eigenvalue such that $(\lambda_{k}, \kappa_{k})$ is distinct from $(\lambda_{0}, \kappa_{0})$ and the proof easily follows from the identifiability of the univariate generalized hyperbolic distribution.


\clearpage
\section{Figures}\label{app}
\begin{figure}[!th]
\hspace{-0.2in}\begin{tabular}{ccc}
\includegraphics[width=0.32\textwidth,height=2.1in]{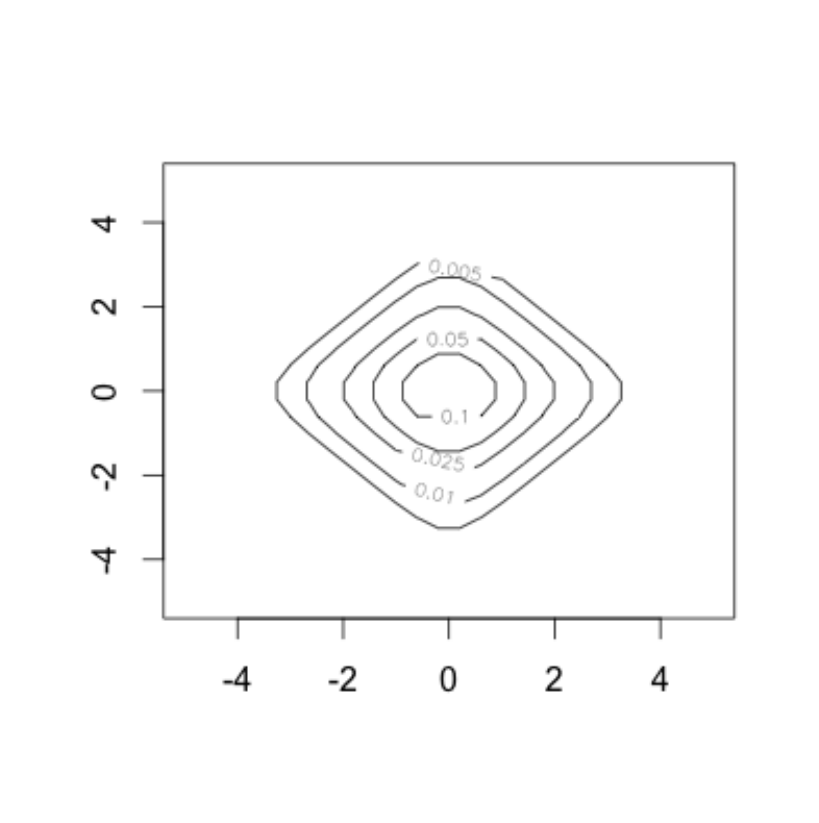}&
\includegraphics[width=0.32\textwidth,height=2.1in]{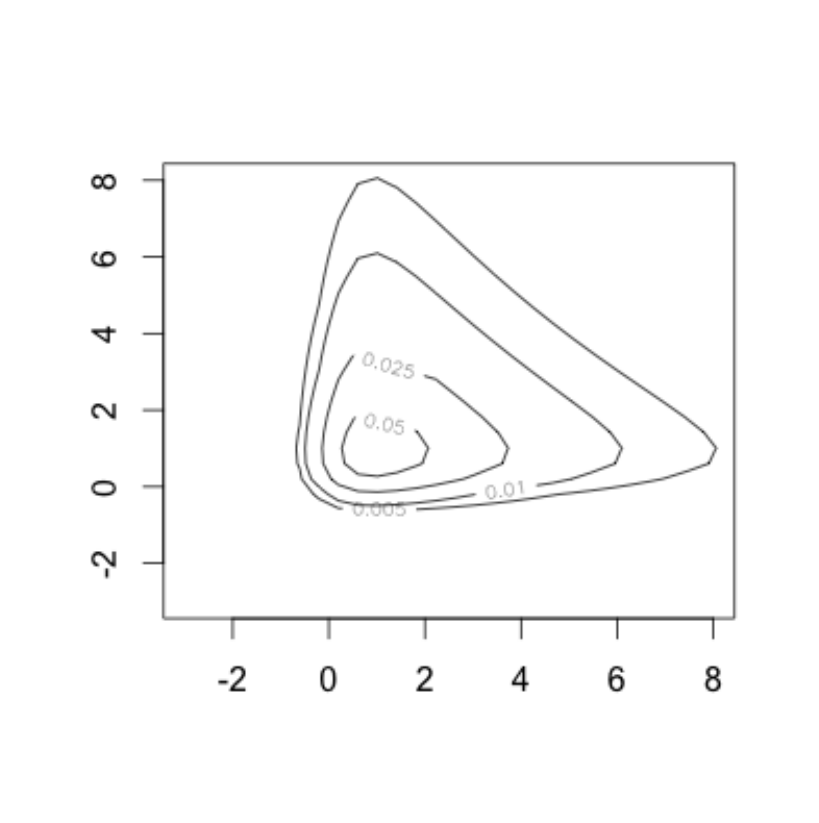}&
\includegraphics[width=0.32\textwidth,height=2.1in]{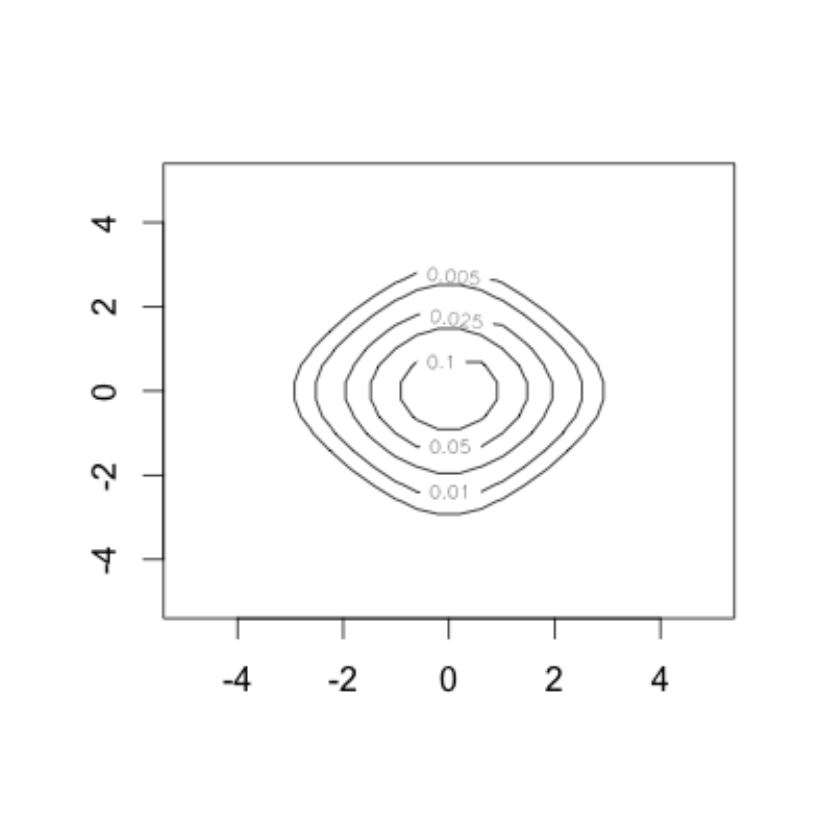}\\[-20pt]
{\scriptsize$\matsig=\diag(1,1)$, $\vecalpha=(0,0)'$,}&
{\scriptsize$\matsig=\diag(1,1)$, $\vecalpha=(2,2)'$,}&
{\scriptsize$\matsig=\diag(1,1)$, $\vecalpha=(0,0)'$,}\\
{\scriptsize$\vecomegaP=(1,1)'$, $\veclambda=(0,0)'$}&
{\scriptsize$\vecomegaP=(1,1)'$, $\veclambda=(0,0)'$}&
{\scriptsize$\vecomegaP=(3,3)'$, $\veclambda=(0,0)'$}\\[-1pt]
\includegraphics[width=0.32\textwidth,height=2.1in]{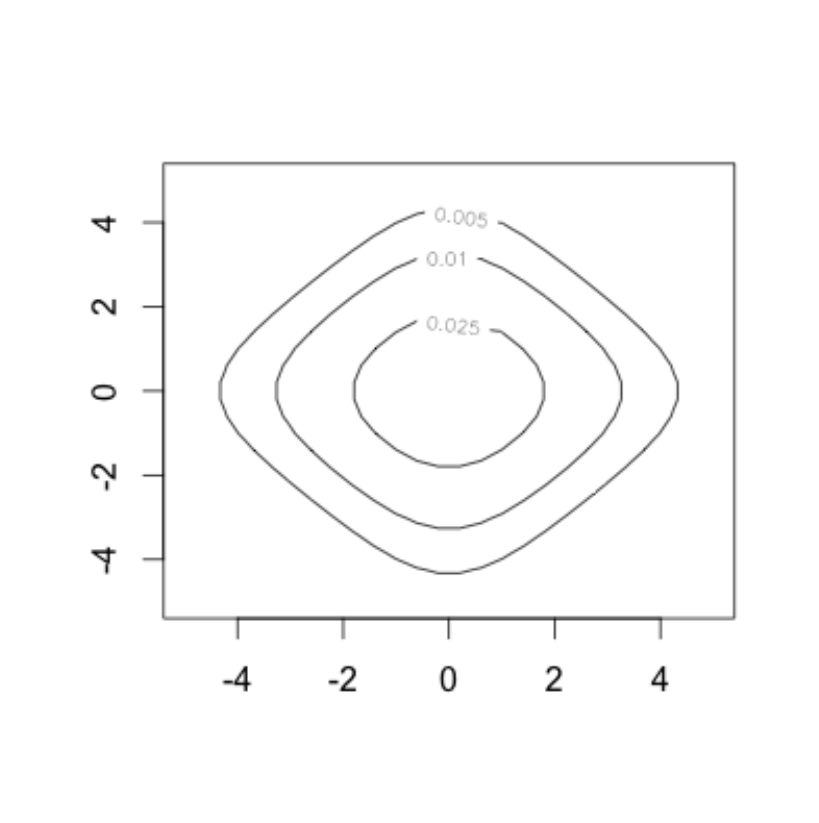}&
\includegraphics[width=0.32\textwidth,height=2.1in]{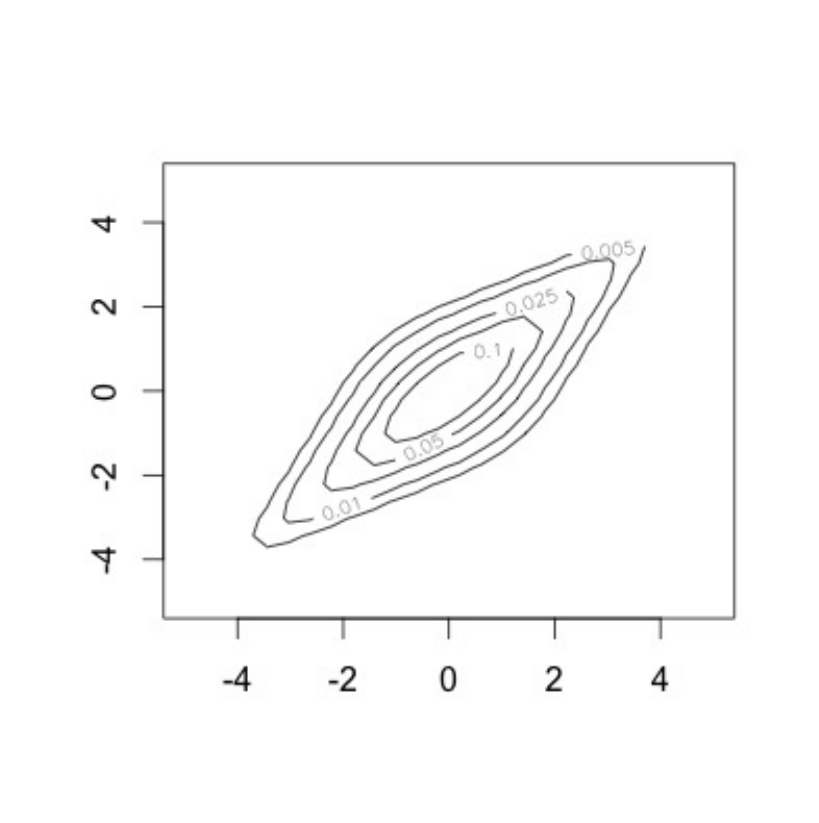}&
\includegraphics[width=0.32\textwidth,height=2.1in]{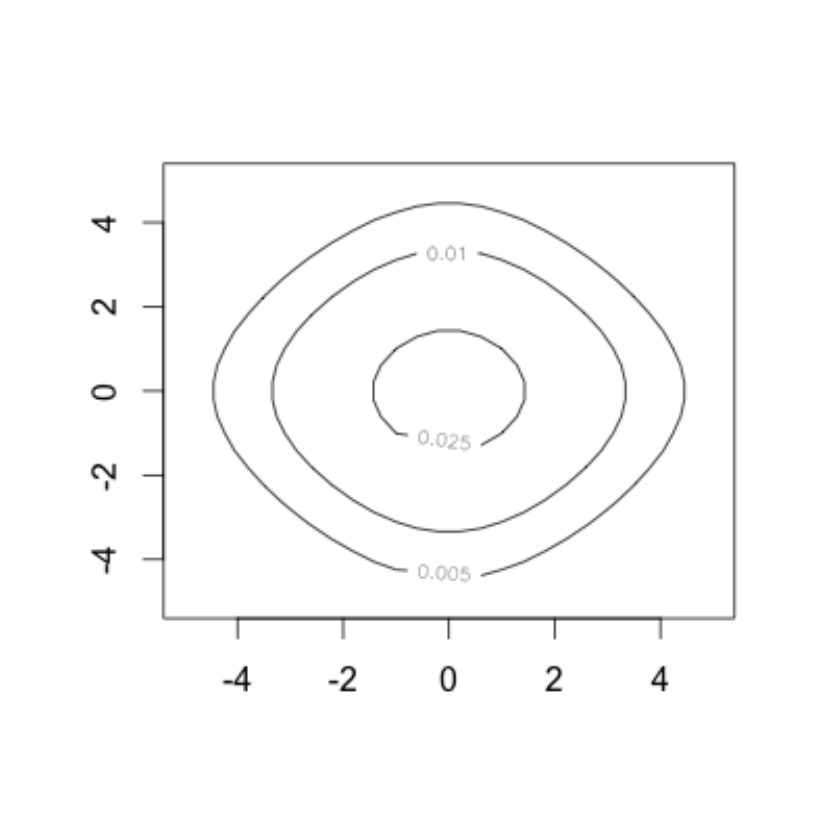}\\[-20pt]
{\scriptsize$\matsig=\diag(2,2), \vecalpha=(0,0)'$,}&
{\scriptsize$\matsig=\diag(1,1)$ plus off-diagonal $0.5$, }&
{\scriptsize$\matsig=\diag(1,1), \vecalpha=(0,0)'$,}\\
{\scriptsize$\vecomegaP=(1,1)'$, $\veclambda=(0,0)'$}&
{\scriptsize$\vecalpha=(0,0)'$, $\vecomegaP=(1,1)'$, $\veclambda=(0,0)'$}&
{\scriptsize$\vecomegaP=(1,1)'$, $\veclambda=(3,3)'$}\\[-1pt]
\includegraphics[width=0.32\textwidth,height=2.1in]{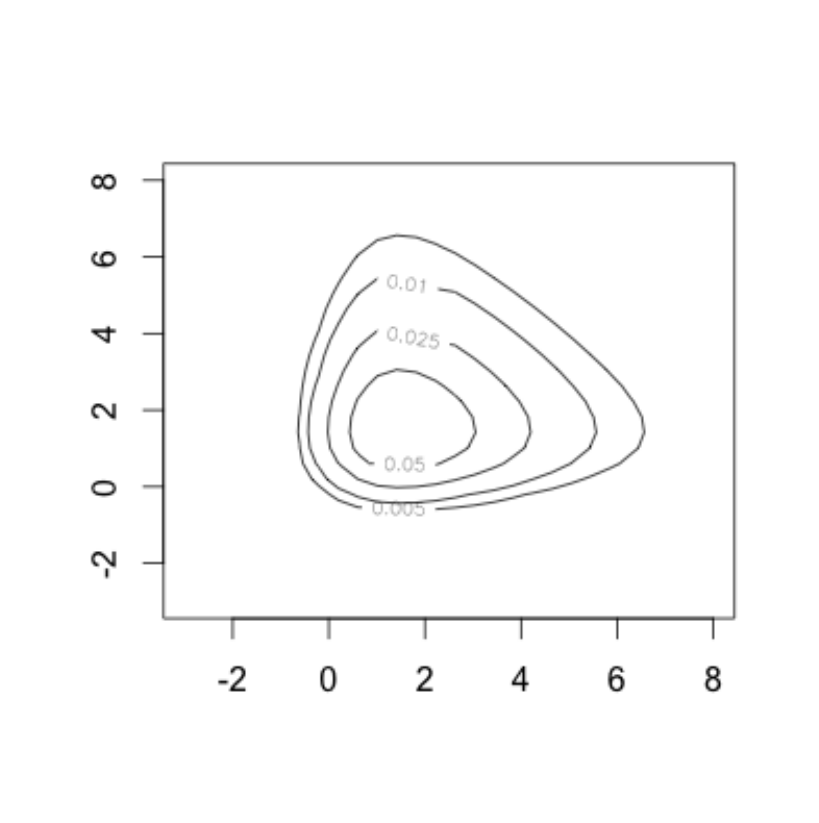}&
\includegraphics[width=0.32\textwidth,height=2.1in]{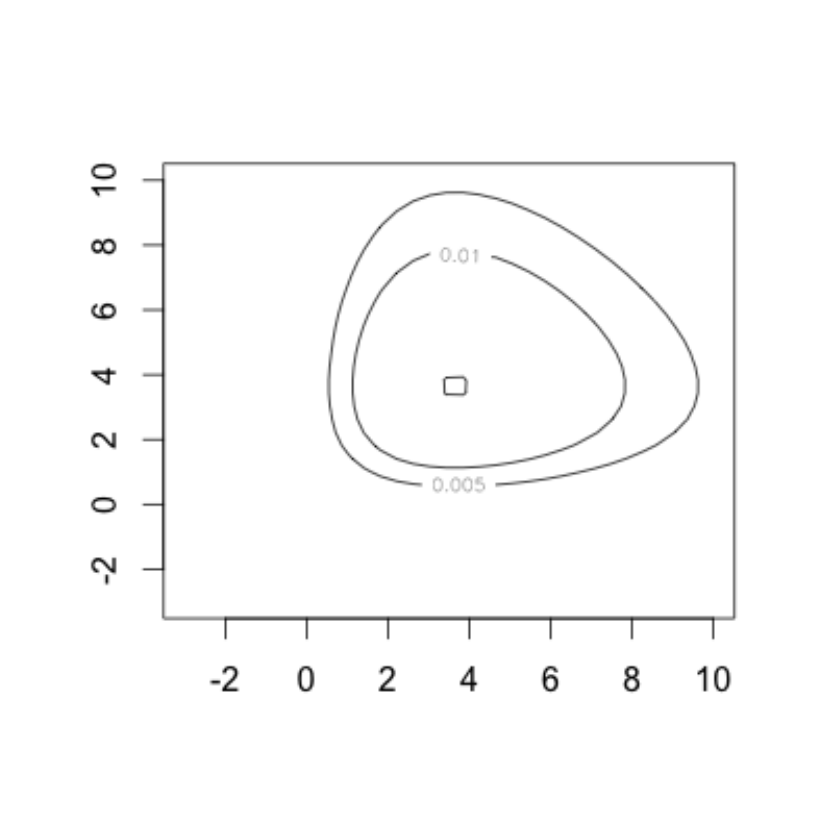}&
\includegraphics[width=0.32\textwidth,height=2.1in]{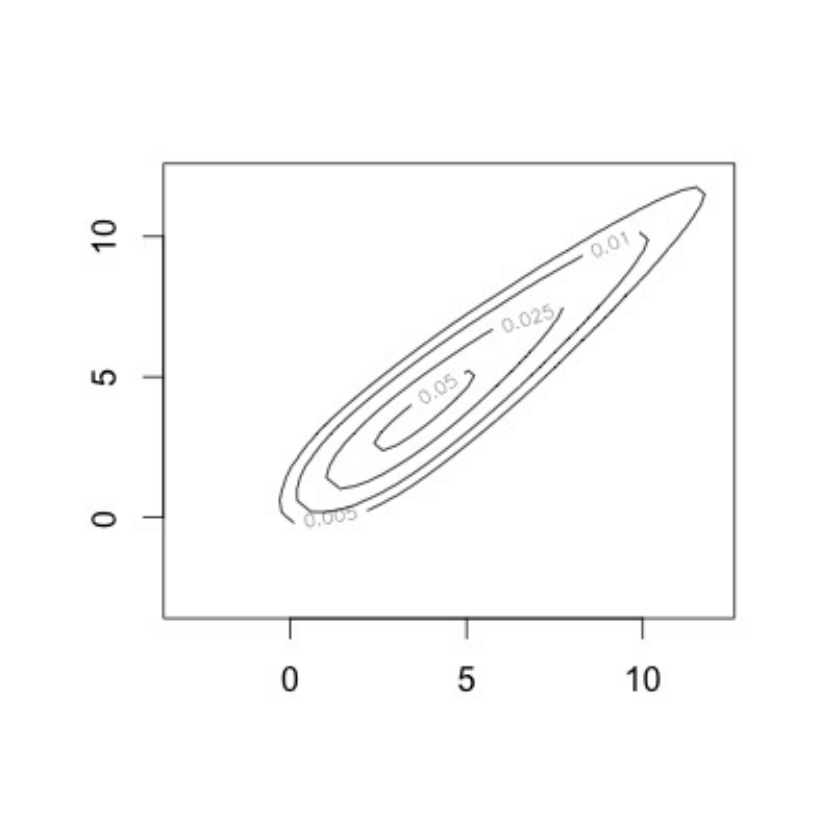}\\[-20pt]
{\scriptsize$\matsig=\diag(1,1)$, $\vecalpha=(2,2)'$,}&
{\scriptsize$\matsig=\diag(1,1)$, $\vecalpha=(2,2)'$,}&
{\scriptsize$\matsig=\diag(1,1)$ plus off-diagonal $0.5$,}\\
{\scriptsize$\vecomegaP=(3,3)'$, $\veclambda=(0,0)'$}&
{\scriptsize$\vecomegaP=(3,3)'$, $\veclambda=(3,3)'$}&
{\scriptsize$\vecalpha=(2,2)'$, $\vecomegaP=(3,3)'$, $\veclambda=(3,3)'$}
\end{tabular}
\caption{Bivariate contour plots of the MSGHD density with $\vecmu=(0,0)'$  and varying $\matsig$, $\vecalpha$, $\vecomegaP$, and $\veclambda$. \label{MSContours}}
\end{figure}

\newpage

\begin{figure}[!ht]
\begin{tabular}{ccc}
\includegraphics[width=0.32\textwidth]{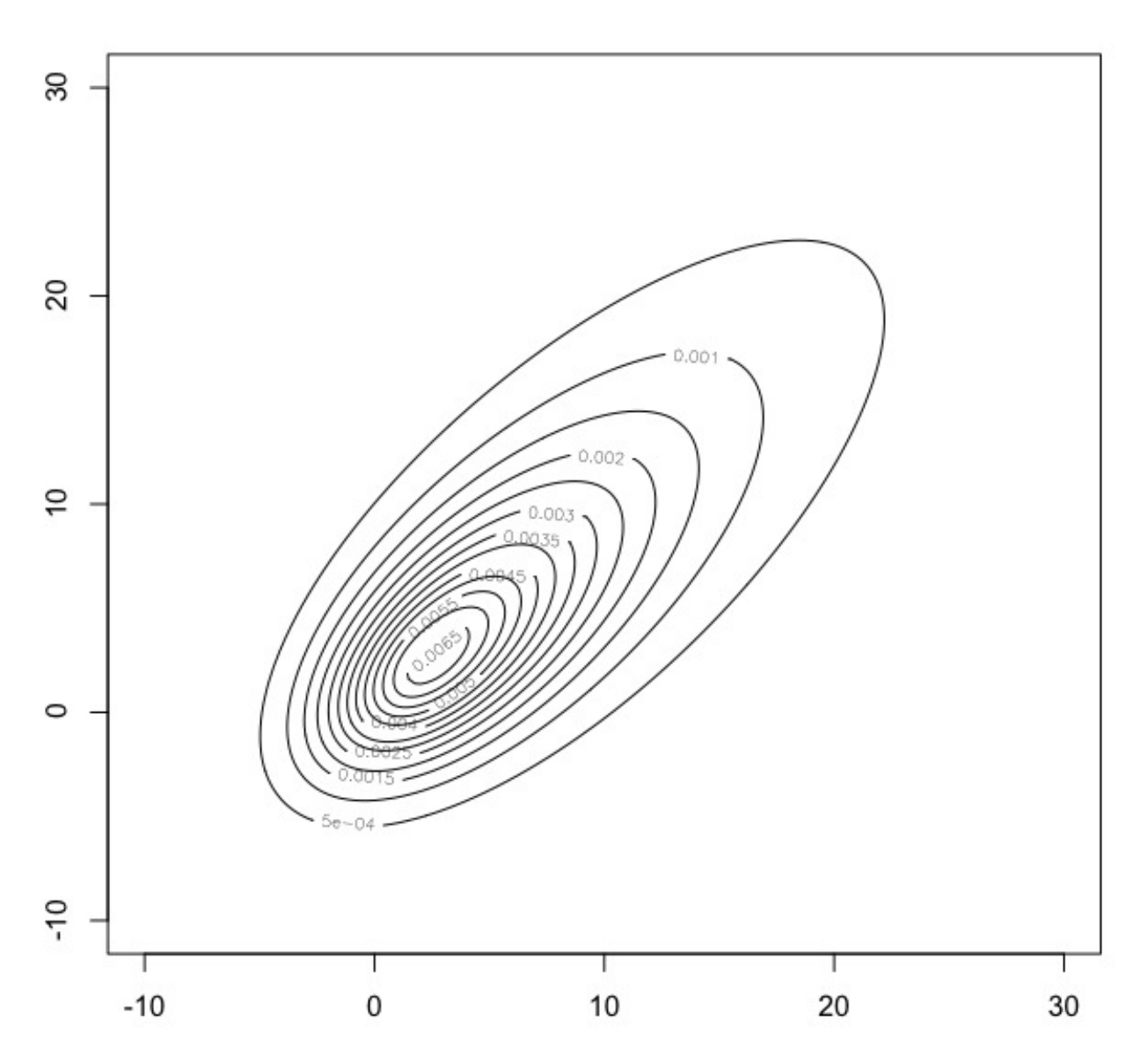} &
\includegraphics[width=0.32\textwidth]{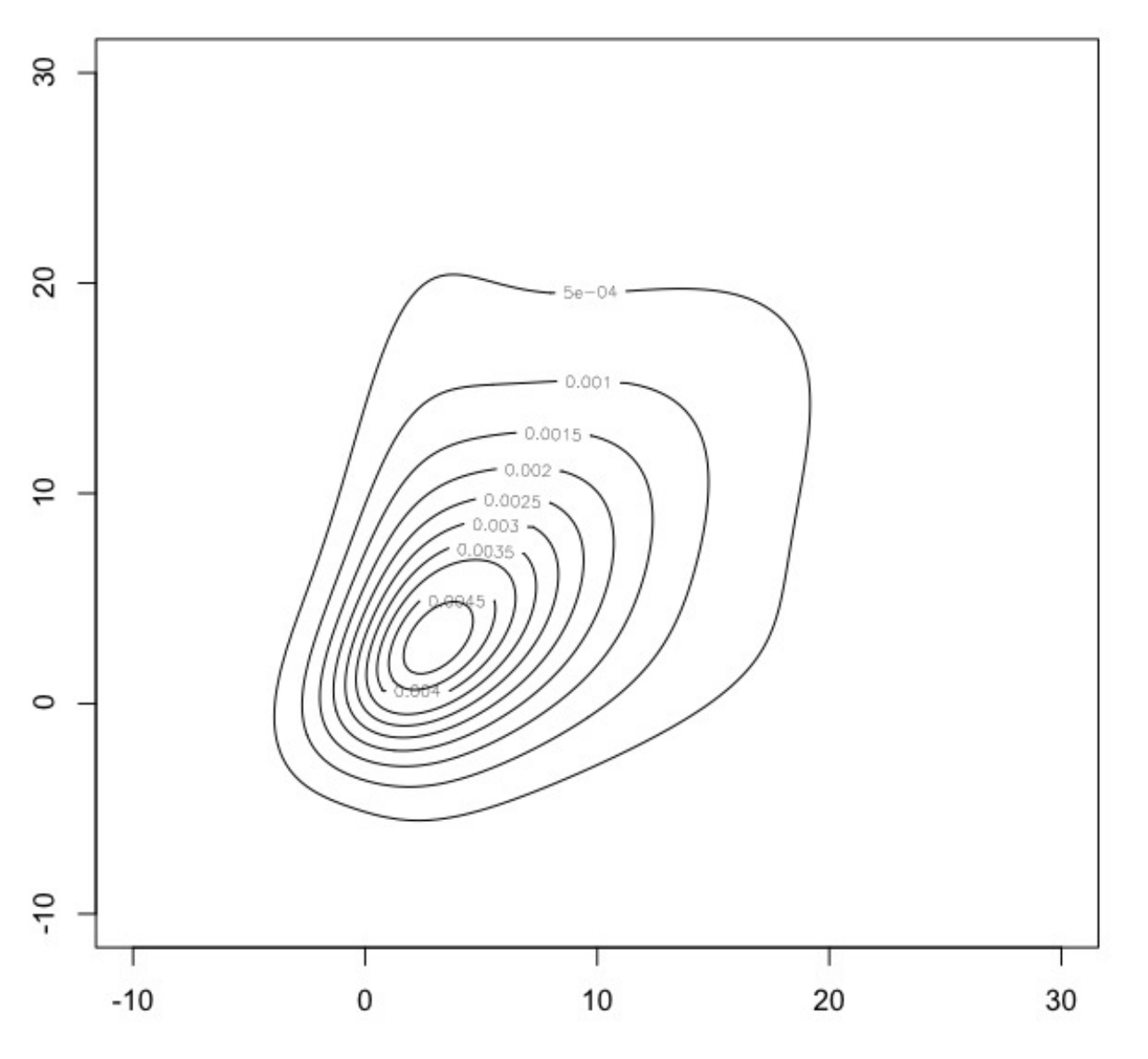} &
\includegraphics[width=0.32\textwidth]{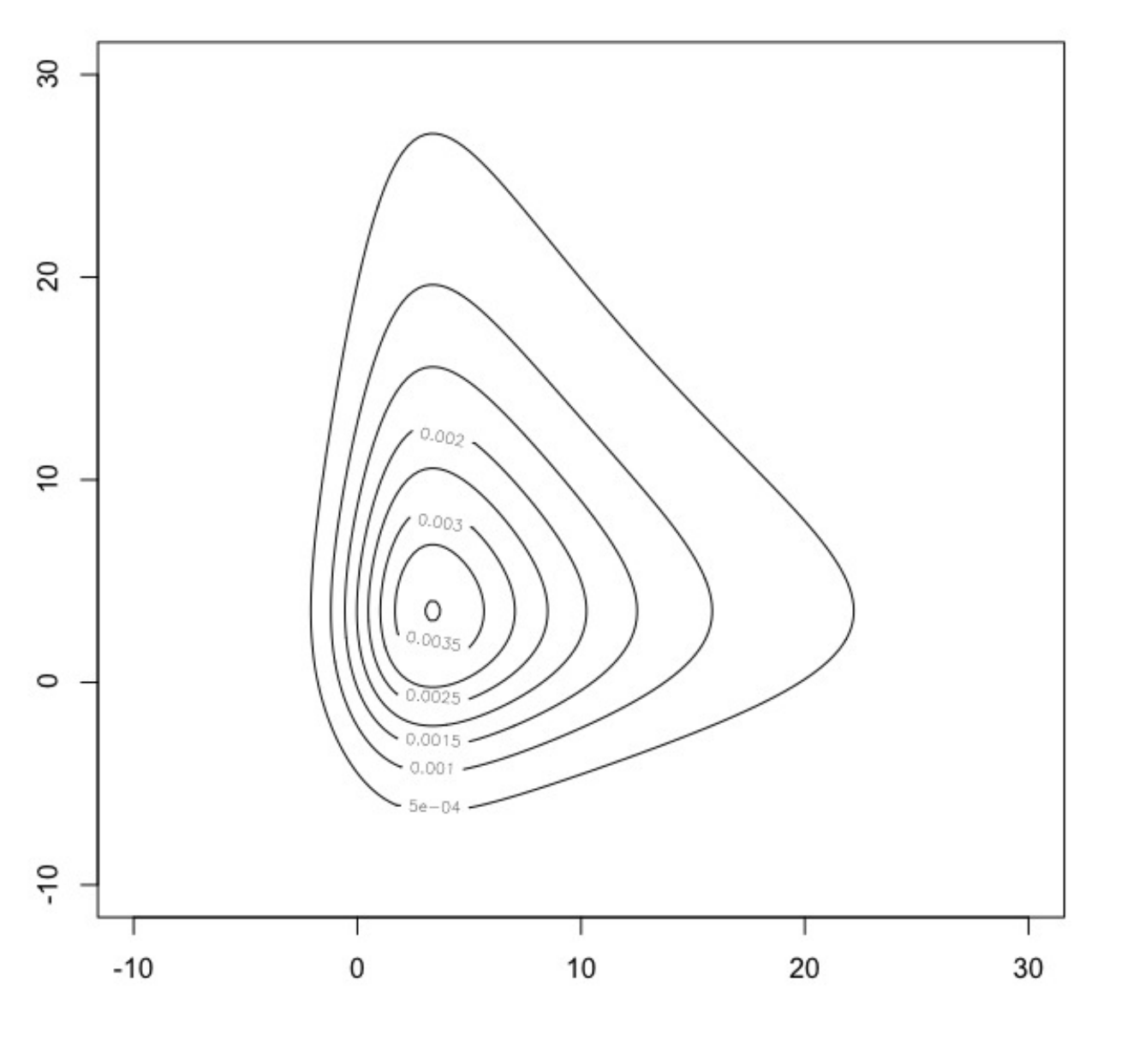}\\[-2pt]
{\scriptsize$\matsig=\diag(1,1)$, $\vecalpha=(2,2)'$, $\varpi=1$}&
{\scriptsize$\matsig=\diag(1,1)$, $\vecalpha=(2,2)'$, $\varpi=0.5$}&
{\scriptsize$\matsig=\diag(1,1)$, $\vecalpha=(2,2)'$, $\varpi=0$}\\[+4pt]
\includegraphics[width=0.32\textwidth]{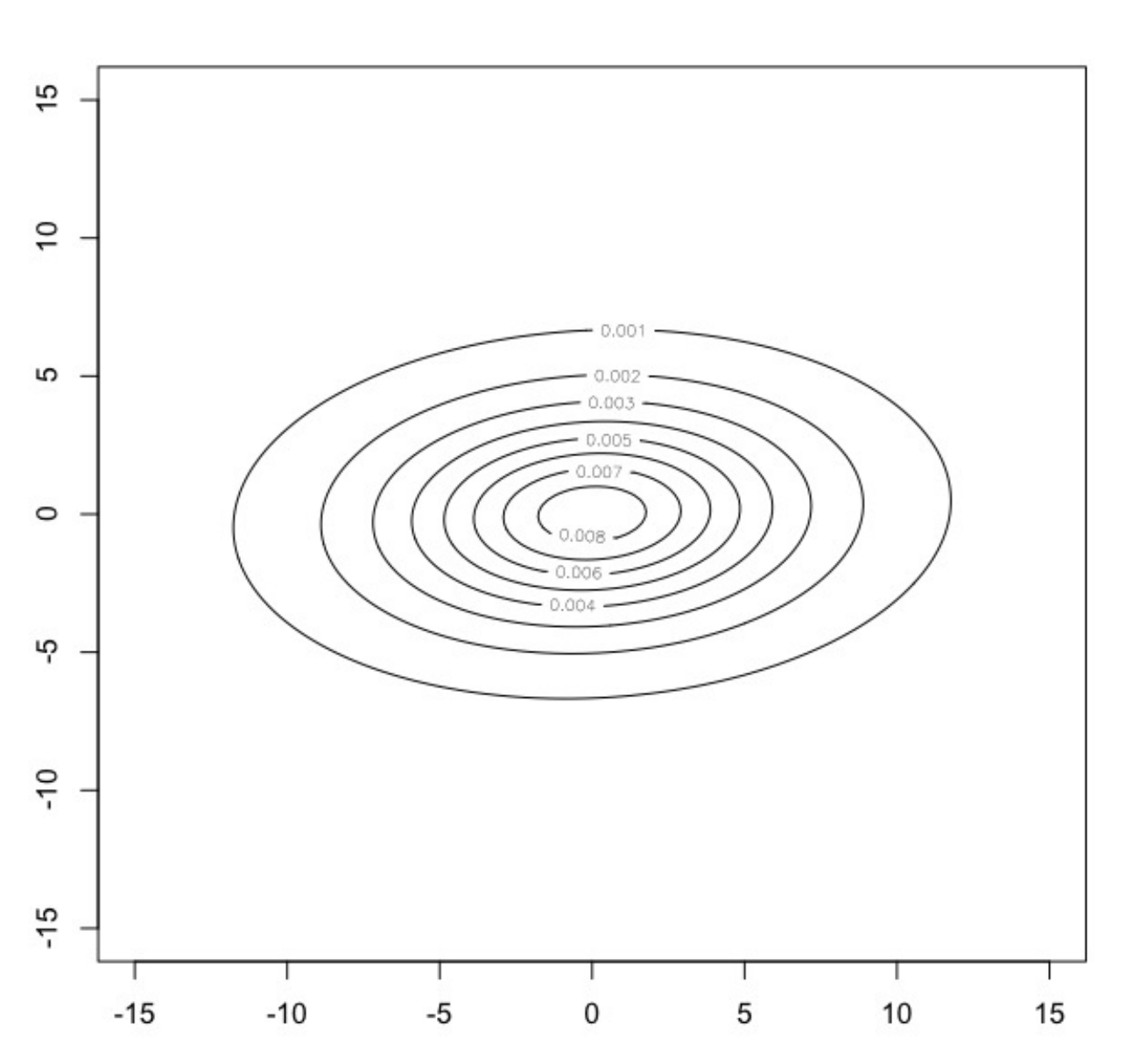}&
\includegraphics[width=0.32\textwidth]{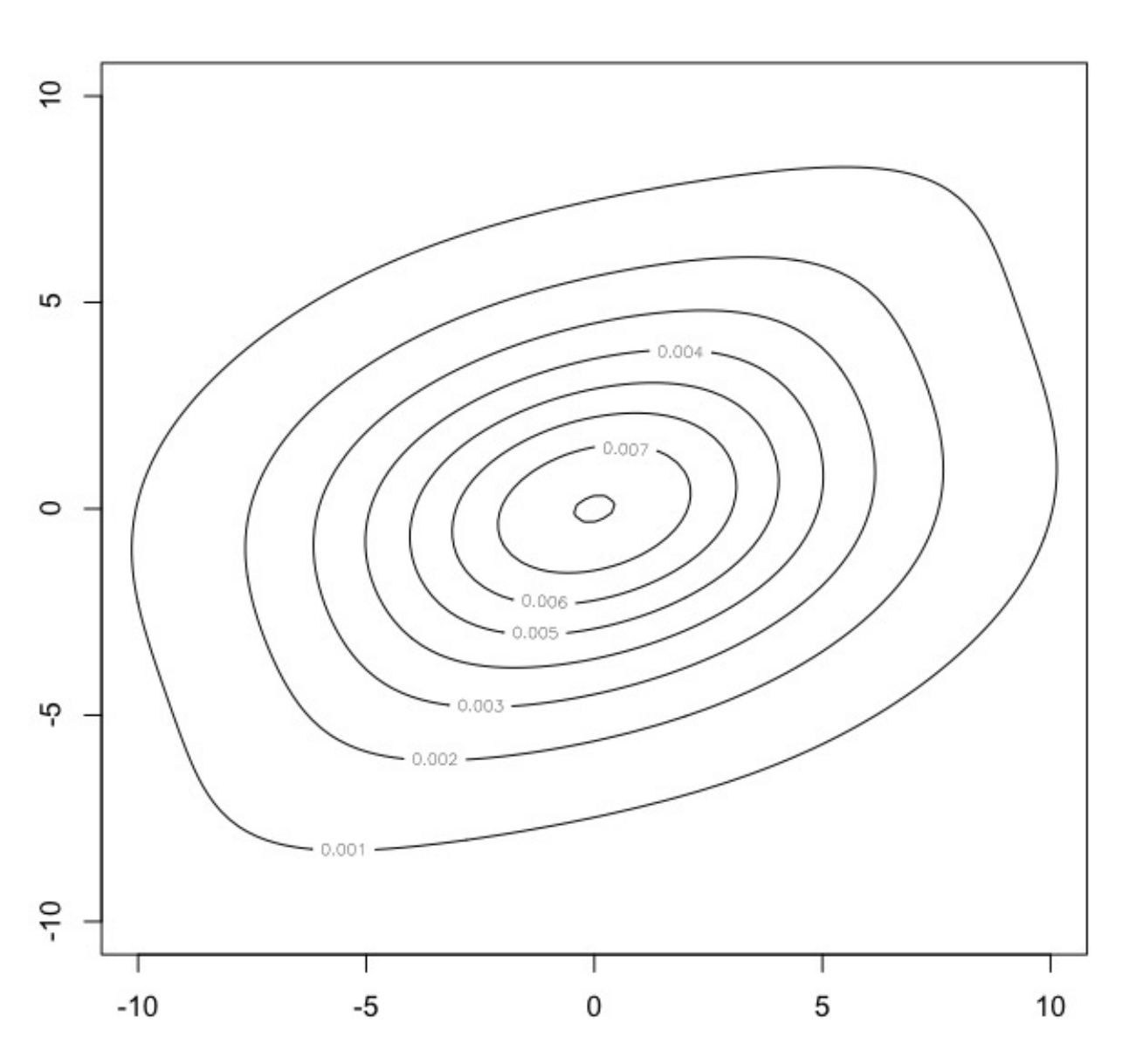}&
\includegraphics[width=0.32\textwidth]{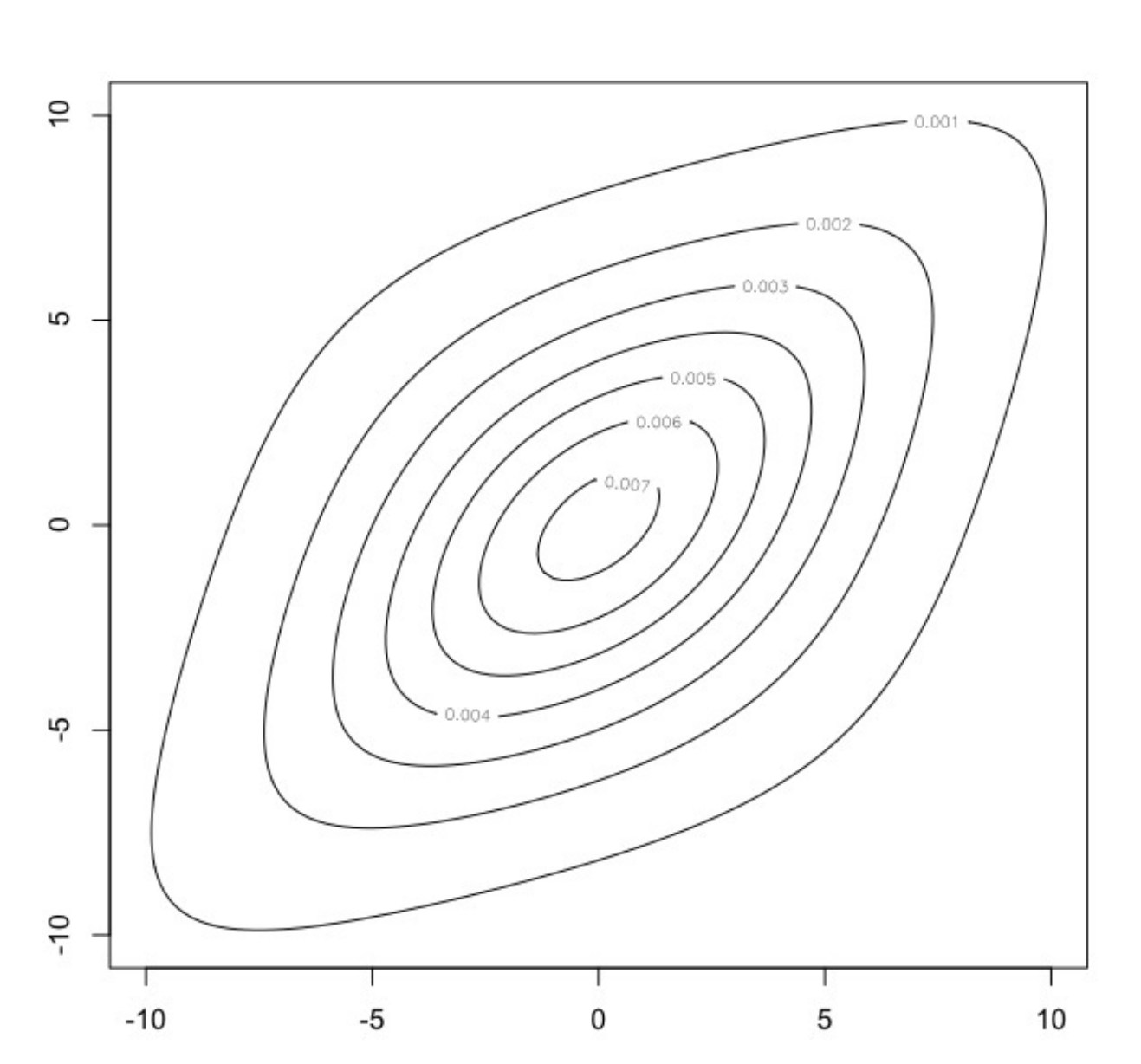}\\[-2pt]
{\scriptsize$\matsig=\diag(1,1)$ plus off-diagonal $0.8$,}&
{\scriptsize$\matsig=\diag(1,1)$ plus off-diagonal $0.8$,}&
{\scriptsize$\matsig=\diag(1,1)$ plus off-diagonal $0.8$,}\\
{\scriptsize$\vecalpha=(0,0)'$, $\varpi=1$}&
{\scriptsize$\vecalpha=(0,0)'$, $\varpi=0.5$}&
{\scriptsize$\vecalpha=(0,0)'$, $\varpi=0$}\\[+4pt]
\includegraphics[width=0.32\textwidth]{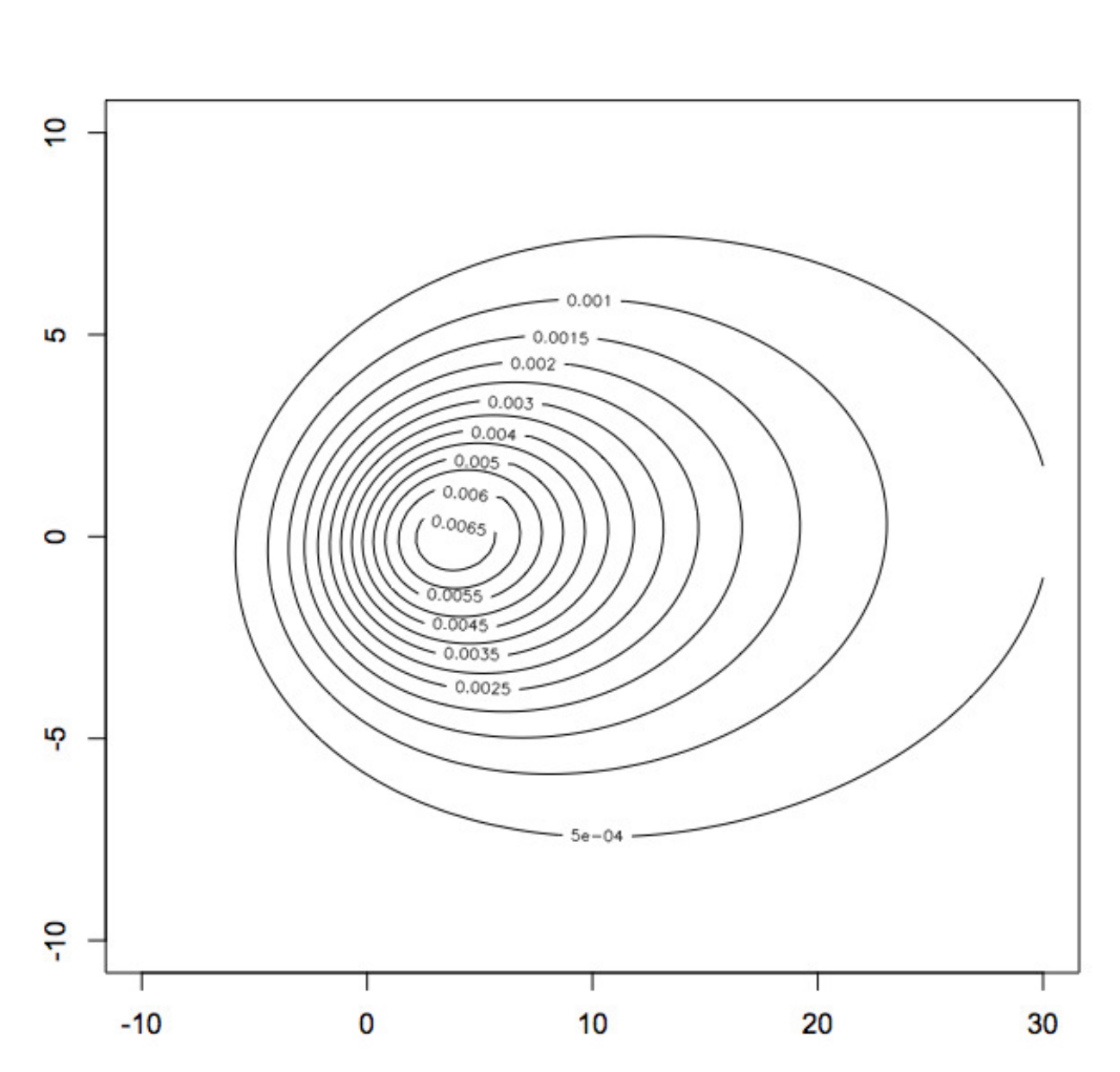}&
\includegraphics[width=0.32\textwidth]{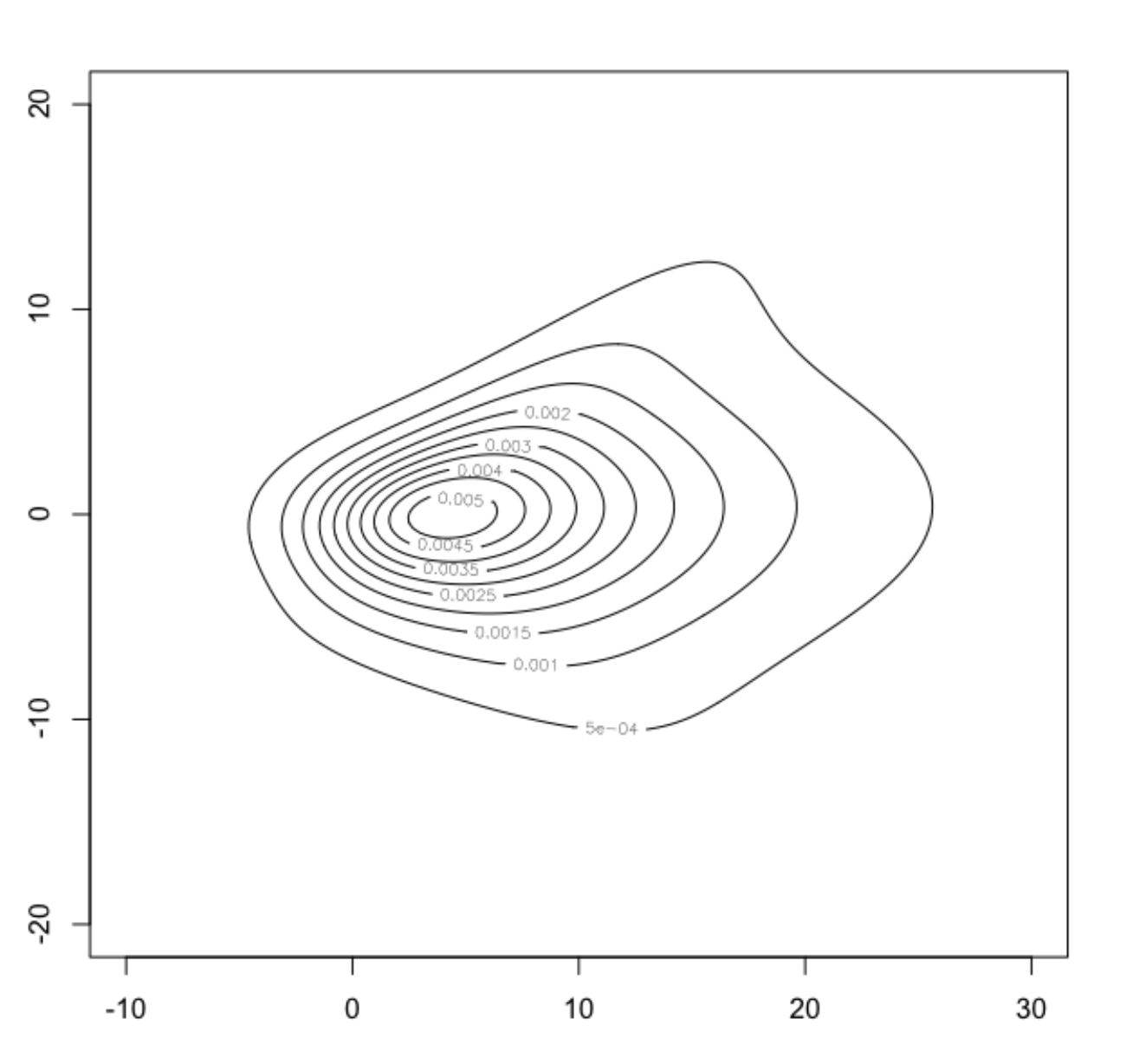}&
\includegraphics[width=0.32\textwidth]{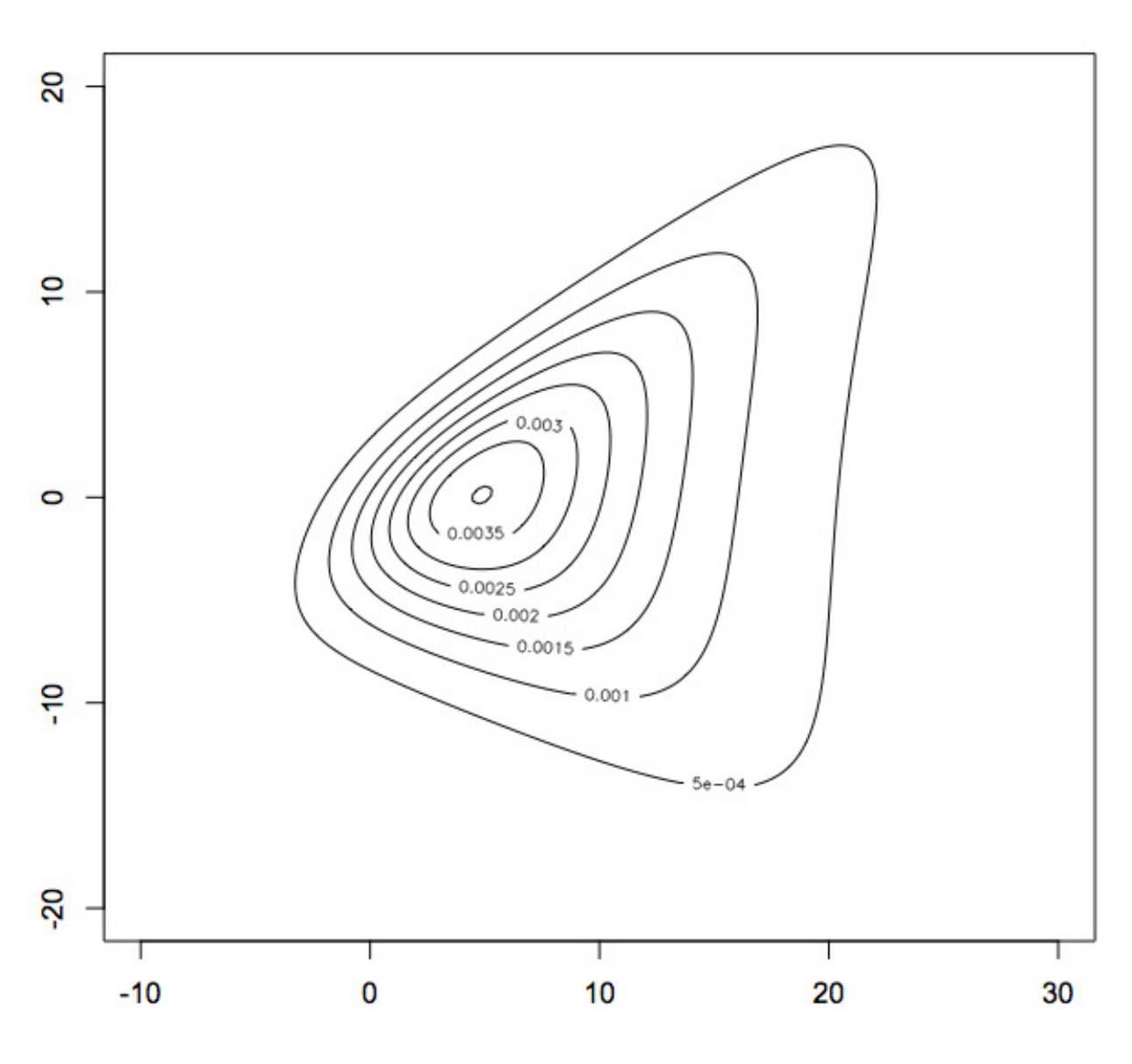}\\[-2pt]
{\scriptsize$\matsig=\diag(1,1)$ plus off-diagonal $0.8$,}&
{\scriptsize$\matsig=\diag(1,1)$ plus off-diagonal $0.8$,}&
{\scriptsize$\matsig=\diag(1,1)$ plus off-diagonal $0.8$,}\\
{\scriptsize$\vecalpha=(2,2)'$, $\varpi=1$}&
{\scriptsize$\vecalpha=(2,2)'$, $\varpi=0.5$}&
{\scriptsize$\vecalpha=(2,2)'$, $\varpi=0$}
\end{tabular}
\caption{Bivariate contour plots of the MCGHD density varying $\matsig$, $\vecalpha$, and $\varpi$. \label{MCContours}}
\end{figure}
}

\end{document}